%% file: paper_core.tex
\documentclass[11pt]{article}

\usepackage{verbatim}
\usepackage{amsmath}
\usepackage{amsfonts}
\usepackage{amssymb}
\usepackage{bbm}
\usepackage{latexsym}
\usepackage{graphics}
\usepackage{lscape} 
\usepackage{epsfig}
\usepackage{color}

\usepackage{amscd}
\usepackage{cite}

\renewcommand{\d}{\mathrm{d}}

\newcommand{\captn}[1]{\vspace{-3ex}\caption{\small #1}}

\newcommand{\Ra}{\Rightarrow}

\DeclareMathSymbol{\mg}{\mathrel}{symbols}{"1D}

\newcommand{\ga}{\alpha}
\newcommand{\gb}{\beta}
\renewcommand{\gg}{\gamma}
\newcommand{\gd}{\delta}
\renewcommand{\ge}{\epsilon}

\newcommand{\gf}{\phi}
\newcommand{\gvf}{\varphi}

\newcommand{\gx}{\xi}

\newcommand{\gk}{\kappa}
\newcommand{\gl}{\lambda}
\newcommand{\gr}{\rho}
\newcommand{\gth}{\theta}

\newcommand{\gs}{\sigma}
\newcommand{\gt}{\tau}

\newcommand{\gp}{\pi}
\newcommand{\gps}{\psi}

\newcommand{\gG}{\Gamma}

\newcommand{\gF}{\Phi}

\newcommand{\gX}{\Xi}
\newcommand{\gL}{\Lambda}
\newcommand{\gS}{\Sigma}
\newcommand{\gTh}{\Theta}

\newcommand{\gPs}{\Psi}

\newcommand{\cC}{{\cal C}}
\newcommand{\cD}{{\cal D}}

\newcommand{\cL}{{\cal L}}
\newcommand{\cM}{{\cal M}}
\newcommand{\cN}{{\cal N}}
\newcommand{\cO}{{\cal O}}

\newcommand{\cT}{{\cal T}}

\newcommand{\cV}{{\cal V}}

\newcommand{\ua}{{\underline a}}

\newcommand{\tq}{{\tilde q}}

\newcommand{\tv}{{\tilde v}}

\newcommand{\tD}{{\widetilde D}}

\newcommand{\tN}{{\widetilde N}}

\newcommand{\dga}{{\dot \alpha}}

\newcommand{\lra}{\longrightarrow}

\newcommand{\ra}{\rightarrow}

\newcommand{\der}{\partial}

\newcommand{\inv}{^{-1}}
\newcommand{\dsp}{\displaystyle}

\newcommand{\undr}[1]{{\underline{#1}}}

\newcommand{\Kh}{K\"{a}hler}
\newcommand{\beq}{\begin{equation}}
\newcommand{\eeq}{\end{equation}}
\newcommand{\barr}{\begin{array}}
\newcommand{\earr}{\end{array}}
\newcommand{\equ}[1]{\begin{gather} #1 \end{gather}}
\newcommand{\equa}[1]{\begin{align} #1 \end{align}}
\newcommand{\items}[1]{\begin{itemize} #1 \end{itemize}}
\newcommand{\enums}[1]{\begin{enumerate} #1 \end{enumerate}}
\newcommand{\tabu}[2]{\begin{tabular}{#1} #2 \end{tabular}}
\newcommand{\arry}[2]{\begin{array}{#1} #2 \end{array}}

\newcommand{\non}{\nonumber}
\newcommand{\sfrac}[2]{\mbox{$\frac{#1}{#2}$}}
\newcounter{oldcounter}
\newenvironment{subeqns}{
\addtocounter{equation}{1}
\setcounter{oldcounter}{\value{equation}}
\setcounter{equation}{0}
\renewcommand{\theequation}{\arabic{oldcounter}
\alph{equation}}
}
{
\setcounter{equation}{\value{oldcounter}}
\renewcommand{\theequation}{\arabic{equation}}\vspace{-10pt}\\
}

\newcommand{\bder}{\bar\partial}
\newcommand{\bb}{{\bar b}}

\newcommand{\bg}{{\bar g}}
\newcommand{\bh}{{\bar h}}

\newcommand{\bk}{{\bar k}}
\newcommand{\bl}{{\bar l}}

\newcommand{\bn}{{\bar n}}

\newcommand{\br}{{\bar r}}
\newcommand{\bs}{{\bar s}}
\newcommand{\bt}{{\bar t}}

\newcommand{\bz}{{\bar z}}
\newcommand{\bA}{{\overline A}}

\newcommand{\bD}{{\overline D}}

\newcommand{\bF}{{\overline F}}

\newcommand{\bK}{{\overline K}}

\newcommand{\bM}{{\overline M}}

\newcommand{\bgf}{{\bar\phi}}
\newcommand{\bgvf}{{\bar\varphi}}

\newcommand{\bgl}{{\bar\lambda}}

\newcommand{\bgth}{{\bar\theta}}
\newcommand{\bgs}{{\bar\sigma}}

\newcommand{\bgps}{{\bar\psi}}

\newcommand{\bgG}{{\overline\Gamma}}

\newcommand{\bgX}{{\overline\Xi}}
\newcommand{\bgL}{{\overline\Lambda}}
\newcommand{\bgS}{{\overline\Sigma}}
\newcommand{\bgTh}{{\overline\Theta}}

\newcommand{\bgPs}{{\overline\Psi}}

\newcommand{\bcC}{{\overline{\cal C}}}
\newcommand{\bcD}{{\overline{\cal D}}}

\newcommand{\bcL}{{\overline{\cal L}}}

\newcommand{\bcT}{{\overline{\cal T}}}

\newcommand{\tga}{{\tilde \alpha}}

\newcommand{\Intr}{\mathbb{Z}}
\newcommand{\Cplx}{\mathbb{C}}

\newcommand{\ba}[2]{\[\begin{array}{#2}\label{#1}}
\newcommand{\ea}{\end{array}\]}
\newcommand{\be}{\begin{equation}}
\newcommand{\ee}{\end{equation}}
\newcommand{\bea}{\begin{eqnarray}}
\newcommand{\eea}{\end{eqnarray}}

\newcommand{\Spin}[1]{\mathrm{Spin(#1)}}

\newcommand{\hc}{\text{h.c.}} 

\newcommand{\brkt}[2]{\bigl[ ^{#1}_{#2} \bigr]}

\newcommand{\rep}[1]{\mathbf{#1}}
\newcommand{\crep}[1]{\overline{\rep{#1}}}

\newcommand{\sm}{{\,\mbox{-}}}

\newcommand{\ztwo}{\Intr_2\!\times\!\Intr_2}

\begin{document}

\thispagestyle{empty}

\begin{flushright}
LMU-ASC 103/10
\\
\end{flushright}
\vskip 2 cm
\begin{center}
{\Large {\bf Heterotic orbifold resolutions as (2,0) gauged linear sigma models} 
}
\\[0pt]

\bigskip
\bigskip {\large
{\bf Stefan Groot Nibbelink\footnote{
E-mail: Groot.Nibbelink@physik.uni-muenchen.de},
\bigskip }}\\[0pt]
\vspace{0.23cm}
${}$ {\it 
Arnold Sommerfeld Center for Theoretical Physics,\\
~~Ludwig-Maximilians-Universit\"at M\"unchen, 80333 M\"unchen, Germany
 \\} 
\bigskip
\end{center}

\subsection*{\centering Abstract}

In this work we attempt to bridge the gap between heterotic orbifold models and Calabi-Yau compactifications using gauged linear sigma models (GLSMs) with (2,0) worldsheet supersymmetry. 
We associate a specific GLSM to a heterotic orbifold model with twisted states that have non-vanishing vacuum expectation values (VEVs): 
The charges of the GLSM superfields are essentially determined by the shifted momenta of these states. 
When a twisted state contains an oscillator excitation, a fermionic gauging is introduced on the worldsheet, inducing a non-Abelian gauge bundle, e.g.\ the standard embedding. 
However, irrespectively of whether the twisted states contain oscillators or not, they can be interpreted as blow-up modes, as their VEVs are correlated with sizes of exceptional cycles in the resolved geometry. 
We show that the GLSM anomaly cancellation conditions ensure that the Bianchi identities are fulfilled for all possible triangulations of a resolution. 
By considering marginal deformations of a GLSM in the large volume limit we are able to directly determine its effective four dimensional spectrum. 
In the cases considered the spectra coincide with those computed via index theorems.

\newpage 
\setcounter{page}{1}

\section{Introduction and motivation}


String theory provides a fascinating arena to learn about possible physics beyond the Standard Model (SM) of Particle Physics, because it is capable of describing both gravity and gauge theories in a unified framework. To show that string theory is really the ultimate theory of nature, a string theoretical realization of the SM has to be uncovered. In recent years progress has been made in achieving this goal within the heterotic string. 


Soon after the invention of the heterotic string~\cite{Gross:1985fr,Gross:1985rr} it was realized that its E$_8\times$E$_8$ version naturally leads to a Grand Unified Theory (GUT) when compactified on a six dimensional internal Calabi-Yau (CY) manifold preserving $\cN=1$ supersymmetry in four dimensions~\cite{Candelas:1985en}. With the simplest gauge embedding, the so-called standard embedding, the largest GUT group E$_6$ is realized. But since it is not clear how to break this huge gauge group, the search was on for more complicated gauge embeddings which by the Donaldson-Uhlenbeck-Yau theorem~\cite{Donaldson:1985,Uhlenbeck:1986} are characterized as stable vector bundles with vanishing first Chern class. Unfortunately smooth CY spaces and such bundles are very complicated objects, so progress in finding the SM or its supersymmetric version, the MSSM, has been slow. Ongoing efforts of refs.~\cite{Donagi:1999ez,Donagi:2000zs,Donagi:2000fw} have resulted in MSSM-like candidates~\cite{Braun:2005ux,Braun:2005nv,Bouchard:2005ag,Bouchard:2008bg}. 


As smooth CY are difficult spaces, orbifolds were considered as simple -- though geometrically singular -- realizations of them. In particular $T^6/\Intr_N$ orbifolds define exactly solvable string models, as they are described as free worldsheet theories~\cite{Dixon:1985jw,Dixon:1986jc}. Because of modular invariance the $\Intr_N$ orbifold twist has to be extended to act on the gauge degrees of freedom. In this work a heterotic orbifold is therefore always specified by a certain orbifold twist and its gauge shift embedding. Such heterotic orbifolds allow for a systematic search for phenomenologically viable models. In particular in the mini-landscape search on the $\Intr_{6\text{-II}}$ orbifold over a hundred MSSM realization were found~\cite{Lebedev:2006tr,Buchmuller:2006,Buchmuller:2006ik,Lebedev:2006kn,Lebedev:2007hv} based on the idea of string theoretical orbifold GUTs~\cite{Kobayashi:2004ya,Kobayashi:2004ud,Forste:2004ie}. Recently another interesting MSSM realization was constructed on a $\Intr_2\times\Intr_2$ orbifold as a $SU(5)$ GUT that was broken down to the MSSM using a freely acting involution~\cite{Blaszczyk:2009in,Blaszczyk:2010db}.  


True MSSM realizations obtained from orbifold compactification require for both, phenomenological and consistency reasons, that a certain number of scalars attain non-vanishing Vacuum Expectation Values (VEVs).\footnote{The presence of moduli, i.e.\ such VEVs, poses a serious obstacle to fully satisfactory phenomenology. In this work we do not address the question of moduli stabilization further.} All known orbifold models contain exotic states that are not part of the SM yet charged under it. When these VEVs are for SM singlets, this  results in a Higgs mechanism that preserves the SM group but nevertheless decouples many vector-like exotics. In addition, the one--loop correction to the $D$-term~\cite{Fischler:1981zk} of the so--called anomalous $U(1)$ requires that at least one scalar charged under this $U(1)$ has a non-zero VEV~\cite{Dine:1987xk,Atick:1987gy,Dine:1987gj}. If the VEVs occur for twisted states, i.e.\ string states localized at the orbifold fixed points, this means that some of these fixed points of the orbifold with curvature singularities get blown up, i.e.\ smoothed out.\footnote{We will see that happens irrespectively of whether the twisted mode which gets a VEV has oscillator excitations.} If this would happen to all fixed points one would end up with a smooth CY. To understand the properties of the resulting CY and its gauge backgrounds one needs to systematically study the blow-up process. In refs.~\cite{Douglas:1997de,Denef:2004dm,Lust:2006zh,Reffert:2007im} the topology of the smooth CY resulting from orbifold resolutions were discussed using toric geometry methods~\cite{Fulton,Hori:2003ic}. There has been recent progress in realizing gauge fluxes (line bundles) on both non-compact and compact resolutions~\cite{Honecker:2006qz,Nibbelink:2007pn,Nibbelink:2008tv,Nibbelink:2009sp}. In some particular non-compact orbifolds even the metric and Abelian and non--Abelian gauge backgrounds were explicitly constructed~\cite{Nibbelink:2007rd,GrootNibbelink:2007ew,Nibbelink:2008qf}.

Unfortunately, the above description of the blow-up procedure has some severe limitations. The main problem is that one uses different frameworks in different regimes: On the one hand, one has an orbifold theory which can be investigated using conformal field theory (CFT) techniques. On the other hand, one investigates heterotic supergravity on smooth CY spaces using topological methods like index theorems. This means one has two very different descriptions that can only be compared on the level of the chiral spectrum. Unfortunately, it is unclear whether an overlapping region exists in which both descriptions can be trusted: An exact CFT description requires that all VEVs are absent, while a supergravity description on smooth CYs is valid when the volumes of exceptional divisors are large. Moreover, it is unlikely that the $D$-- (and $F$--)flatness and the decoupling of the exotics are realized in such regions even if they would exist. Therefore, a formalism that both applies in the orbifold regime and allows for a supergravity description would be very helpful.

In addition, there is the practical complication that often a single orbifold can be related to many smooth CYs which are distinguished by their intersection numbers: Even though a toric resolution of an orbifold singularity is well understood mathematically, it often does not offer an uniquely determined geometry, because the resolution allows for various triangulations. For the $T^6/\Intr_{6\text{-II}}$ orbifold this gives us millions of smooth CYs to consider~\cite{Nibbelink:2009sp}; and there are many orders more resolutions of the orbifold $T^6/\Intr_2\times\Intr_2$ \cite{Blaszczyk:2010db}. In light of this, it would be extremely useful to have a framework that treats all resolutions of a single orbifold on equal footing.


Two dimensional gauged linear sigma models (GLSMs) seems to provide a formalism which has all these features. Witten introduced GLSMs as a concrete setting for investigating stringy geometries and their gauge bundles~\cite{Witten:1993yc}. This gives a particularly convenient description of CY hyper surfaces in (weighted) projected spaces and complete intersection CYs. When the standard embedding is employed, this gives rise to a $(2,2)$ supersymmetric gauge theory on the worldsheet, while for other gauge embeddings $(2,0)$ supersymmetry remains. (2,2) worldsheet supersymmetry ensures that the compactification space is a K\"ahler manifold; (2,0) supersymmetry only requires that the target space is complex with a trivial canonical bundle \cite{Hull:1985jv,Hull:1986hn,Strominger:1986uh}. One important ingredient of a $(2,0)$ GLSM is the Fayet-Iliopoulos (FI) parameter, which has the geometrical interpretation of a K\"ahler parameter. When this parameter tends to minus infinity, the worldsheet gauge symmetry is dynamically broken to a finite discrete subgroup, i.e.\ in this limit the GLSM describes a Landau-Ginzburg orbifold, the spectrum of which can be computed exactly~\cite{Witten:1993yc,Kachru:1993pg}. The opposite limit leads to a large volume description, which can be compared with a supergravity treatment. In addition, GLSMs are capable of describing vector bundles and their deformations, even when they develop singularities~\cite{Distler:1993mk,Distler:1995mi,Distler:1996tj,Chiang:1997kt}. Investigations of the type--II and heterotic strings using (2,2) and (2,0) GLSMs have recently been revived, see e.g.\ \cite{Melnikov:2005hq,Aspinwall:2010ve,Kreuzer:2010ph,McOrist:2010ae}.


In this work we use GLSMs as an explicit framework which is able to interpolate between singular heterotic orbifold CFTs and smooth non-compact CY compactifications with gauge bundles. 
Concretely we study GLSM descriptions of resolutions of non--compact singularities $\Cplx^n/G$ with $n=2,3$ and $G\subset SU(n)$ a discrete Abelian group, i.e.\ $G=\Intr_p$ or $\Intr_p\times \Intr_p$.  
Let us emphasize that our purpose is not just to give some examples of such GLSM resolutions. (It is well--known how to write down consistent (2,2) GLSMs associated to orbifold resolutions using the standard embedding.) But rather we aim to identify the unique (2,0) GLSM, that is induced when certain states in the heterotic orbifold CFT spectrum take non--vanishing VEVs. In particular, these VEVs should characterize which gauge bundle arises on the resolution. In more detail our proposal entails the following:

From the target space perspective the VEVs of twisted states of the orbifold CFT generate the blow-up. We therefore expect that the twisted state VEVs specifies which GLSM should be used. We give a precise recipe how to associate a GLSM to an heterotic orbifold model with certain blow-up modes switched on: In a nutshell, the charges of the superfields in the GLSM are dictated by the shifted momenta that characterize the twisted blow-up modes of the orbifold model. This identification is inspired by our recent findings that the vectors that characterize line bundle embeddings are identical to the shifted momenta of certain twisted states in the orbifold spectrum \cite{Nibbelink:2009sp}. In the blow down limit the original orbifold theory is recovered. In the opposite -- large volume -- limit the four dimensional charged spectra agree with those computed using topological techniques on toric resolutions. To this end we propose a novel technique to read off the four dimensional chiral charged spectrum directly from the GLSM.\footnote{Even though we use non--holomorphic kinetic terms on the worldsheet to do so, which are not protected against renormalization, the spectrum computation seems to be remarkably reliable in the cases examined.} (For other methods see e.g.\ \cite{Beccaria:2010yp} and references therein.) The relations between the orbifold CFT, the GLSM and the supergravity descriptions are summarized in Figure \ref{fg:NCHO<>GLSM}.

Finally, let us stress that GLSMs as such do not represent true string models, because the conformal symmetry is lost; the gauge coupling in two dimensions has the dimension of mass. The basic motivation why GLSMs may nevertheless be relevant for string theory is that GLSMs defined in the ultraviolet (UV) are assumed to flow to a CFT in the infrared (IR). Some evidence in this direction has been provided in ref.~\cite{Silverstein:1995re,Silverstein:1994ih,Basu:2003bq,Beasley:2003fx}. Since the renormalization group equation (RGE) flow is governed by a continuous parameter, the renormalization scale, any quasi-topological quantity is presumably protected and can be reliably computed in the UV using GLSMs. In this paper we take a pragmatic approach, and simply take GLSMs as convenient tools to characterize the stringy geometry of orbifold resolutions and their gauge bundles. We realize that we should consider  gauged non-linear sigma models (GNLSMs) to incorporate the full string dynamics. However, for most purposes of this paper the linear approximation is sufficient; only for the computation of the charged spectrum in target space we do need to take non-linear effects into account.

\begin{figure}
\begin{center} 
\scalebox{.5}{\input{Limits}}
\end{center} 
\captn{The dashed line schematically indicates that by selecting some blow-up modes $|p_r, P_r\rangle$ within the twisted orbifold CFT spectrum, we can define a specific GLSM with given gauging of the chiral and chiral-Fermi multiplets encoded by the charges $(q_r, Q_r)$. When taking the blow down limit ($b_r \ra -\infty$) of this GLSM we recover the orbifold theory back, while the large volume limit ($b_r \ra \infty$) gives describes a non-compact CY with a certain (line) bundle. 
\label{fg:NCHO<>GLSM}}
\end{figure}

\subsection*{Paper setup}

To this end we have structured the manuscript as follows: In Section \ref{sc:(2,0)susy} we give a brief account of the necessary ingredients of (2,0) supersymmetric GLSMs in two dimensions. Holomorphic marginal deformations are introduced, and it is recalled how the supersymmetric minima of the worldsheet $D$-term potential describes the effective target space geometry. Finally for later use, we  summarize the GLSM consistency requirements. In Section \ref{sc:Orbifolds} we review the basic construction of non-compact heterotic orbifold models, and recall how to determine their  massless spectra. Section \ref{sc:ResGLSM} describes our proposal of how switching on twisted states as blow-up modes leads to a specific GLSM on the worldsheet. Section \ref{sc:Examples} the general discussion is illustrated by describing the blow-up procedure of the orbifolds $\Cplx^3/\Intr_3$ and $\Cplx^3/\Intr_4$. In Subsection \ref{sc:Z2Z2top} the $\Cplx^3/\Intr_2\!\times\!\Intr_2$ orbifold is used to demonstrate how changes in topology are treated in an entirely smooth fashion by the GLSM. Finally, Section \ref{sc:Spectrum} gives a proposal for how to determine the charged chiral spectrum in the effective four dimensional theory by considering marginal deformations of the kinetic terms of the chiral-Fermi multiplets. We check that this method reproduces the charged spectra of line bundle models on the resolution of the $\Intr_3$ singularity. Our main conclusions and outlook are collected in Section \ref{sc:concl}.

The (2,2) and (2,0) superspace conventions are collected in Appendix \ref{sc:superspace}. In Appendix \ref{sc:(2,2)theories} we briefly recall (2,2) supersymmetric theories, and in Appendix \ref{sc:(2,0)theories} we review (2,0) theories. The reduction of (2,2) theories in two dimensions to (2,0) models is recollected in Appendix \ref{sc:Reduction}.



\section{Gauged linear sigma models}
\label{sc:(2,0)susy}


This section provides a review of  gauged linear sigma models (GLSMs) described in terms of (2,0) supersymmetric field theories in two dimensions. Further details can be found in refs.\ \cite{Witten:1993yc,Distler:1987ee,Distler:1992gi,Distler:1995mi}. Here we mainly focus on the physical content of such theories; the precise details have been collected in Appendix \ref{sc:(2,0)theories}. Worldsheet theories with (2,0) supersymmetry are conveniently described in terms of superfields. Such superfields are functions of (2,0) superspace, which are spanned by the worldsheet coordinate $\gs, \bgs$ and the fermionic variables $\gth^+,\bgth^+$. The relation of (2,0) superspace in two dimensions and the maybe more familiar four dimensional $N=1$ superspace is reviewed in Appendix \ref{sc:superspace}.

The field content of a generic GLSM can be encoded in a number of superfields: the gauge, chiral, chiral-Fermi and Fermi-gauge multiplets. As in four dimensions, these superfields contain physical degrees of freedom and often also non-dynamical auxiliary fields. The basic superfields together with their physical and off-shell components have been collected in Table~\ref{tb:LinearSigma}. As can be seen from Table \ref{tb:LinearSigma} a chiral superfield $\gPs^a$ contains a complex scalar $z^a$ and a holomorphic (or right-moving) fermion $\gps^a$ (i.e.\ see \eqref{compChiral}). A chiral-Fermi superfield $\gL^\ga$ only contains a physical anti-holomorphic (or left-moving) fermion $\gl^\ga$ (i.e.\ see \eqref{compChiralFermi}). The worldsheet theory of the free heterotic string has labels $a = 0,1,2,3$ and $\ga=1,\ldots,16$ in light-cone gauge, respectively. (Recall that in light-cone gauge one only describes the coordinates transverse to the light-cone, so that $z^0 = x^2+i x^3, \ldots, z^3 = x^8+ix^9$.) Their worldsheet action reads 
\equ{
S_\text{het free}  = \int \d^2\gs\d^2\gth^+\, \Big\{ 
\frac i2 \, \bgPs_a \bder \gPs^a 
- \frac 12 \, \bgL_\ga \gL^\ga
\Big\}~, 
\label{FreeHetAc} 
}
in the (2,0) superspace language, see Appendix \ref{sc:(2,0)theories}.

\begin{table}
\begin{center}
\renewcommand{\arraystretch}{1.2} 
\tabu{| c c || c | c | c c | c c c | }{
\hline 
\multicolumn{2}{|c||}{superfield} &   &  & \multicolumn{2}{|c|}{bosonic DOF} & \multicolumn{3}{|c|}{fermionic DOF}
\\
type & notation & dimension & charge & on & off & & on     & off  
\\ \hline\hline 
chiral & $\gPs^{a}$ & 0 & $q_I^a$ & $z^a$ & - & & $\gps^a$ & -
\\ 
chiral-Fermi & $\gL^{\ga}$ & 1/2 & $Q_I^\ga$ & - & $h^\ga$ & & $\gl^\ga$ & -  
\\ 
\hline 
gauge & $(V, A)^{I}$ & (0,1) & $0$ & $A_\gs^I, A_\bgs^I$ & $\tD^I$ & & $\gf^I$ & -  
\\ 
Fermi-gauge & $\gS^{i}$ & 1/2 & $0$ & $s^i$ & - & &$\gvf^i$ & -  
\\ \hline 
}
\renewcommand{\arraystretch}{1} 
\end{center} 
\captn{The superfield content of a gauged linear sigma model and their physical on- and off-shell degrees of freedom (DOF). 
\label{tb:LinearSigma}}
\end{table}

In two dimensions we can distinguish two types of gaugings: bosonic and fermionic. One introduces gauge multiplets $(V,A)^I$ for the former and Fermi-gauge multiplets $\gS^i$ for the latter. (For their precise gauge transformations and their components see \eqref{(2,0)gauge}, \eqref{compVectorM} and \eqref{FermionicGauge}, \eqref{compFermiGauge} of Appendix \ref{sc:(2,0)superfields}, respectively.) Such gaugings remove some degrees of freedom from the worldsheet theory. 
Therefore, in order to keep the number of physical degrees of freedom the same, one needs to add an equal number of new (or exceptional) chiral or chiral-Fermi superfields to the theory. The charges of the superfields $\gPs^a$ and $\gL^\ga$ are denoted by $q_I^a$ and $Q_I^\ga$, respectively. The fermionic gaugings 
\equ{
\gL^\ga \ra \gL^\ga + M^\ga{}_i(\gPs)\, \gX^i~, 
\qquad 
\gS^i \ra \gS^i + \gX^i~, 
\label{gXgauge} 
}
are characterized by holomorphic functions $M^\ga{}_i(\gPs)$ and chiral superfield gauge parameters $\gX^i$.


In addition to the gauge symmetries GLSMs often possess various discrete gauge symmetries: These lead e.g.\ to the various spin structures and are therefore of crucial importance, as they define the heterotic string. We call them $L$- or $R$-symmetries depending on whether the left-moving fermions $\gl^\ga$ or the right-moving fermions $\gps^a$ transform under them, respectively. In this paper we will be only concerned with the $SO(32)$ heterotic string, therefore we assign $L$-charge of $1/2$ to each chiral-Fermi superfield $\gL^\ga$. The Grassmann coordinate $\gth^+$ has an $R$-charge of $-1/2$, while the chiral superfields $\gPs^a$ are neutral, so that the right-moving fermions $\gps^a$ carry $R$-charge $1/2$.

\subsection{Holomorphic deformations} 
\label{sc:HoloDefs}

The standard parts of the action for a gauged linear sigma model consists of kinetic terms, displayed in Appendix \ref{sc:(2,0)actions}. The relevant and marginal holomorphic deformations can be represented by chiral superspace ($\int\d\gth^+$ integral) contributions 
\equ{
S_\text{deform} =  \int\d^2\gs\d\gth^+\, \Big\{
W(\gPs,\gL) + \gr_I(\gPs) \, F_I 
\Big\} + \hc~,
\label{SuperfieldAction}
}
where $F_I$ is the super fieldstrength \eqref{(2,0)superfieldstrengths} defined in terms of the fields displayed in Table~\ref{tb:LinearSigma}. The deformations are parameterized by a gauge invariant fermionic superpotential $W$ of mass dimension 3/2 and  by gauge invariant dimensionless holomorphic functions $\gr_I$. It follows that the functions $\gr_I$ constitute marginal deformations. They have the target space interpretation as K\"ahler deformations, while the superpotential $W$ encodes complex structure deformations.

Given the dimension of the chiral superfields $\gPs^a$ and $\gL^\ga$, the most relevant superpotential terms are given by  
\equ{
W(\gPs,\gL)  =  m\, N_\ga(\gPs)\, \gL^\ga~,
\label{superpotential} 
}
with mass $m$. Since the superpotential is holomorphic and fermionic, it gives the most general $W(\Psi,\Lambda)$ consistent with worldsheet Lorentz invariance with operators of mass dimension 1/2. When the GLSM has fermionic gauge transformations, one needs to require that the superpotential is gauge invariant, i.e.\ 
\equ{
N_{\ga}(\gPs) \, M^\ga{}_i(\gPs) = 0~. 
\label{FermionicWinvariance}
}

At this stage it is important to make some comments concerning the relation between GLSMs and the CFTs that govern the string dynamics. A GLSM is a supersymmetric field theory which does contain dimensionful parameters, whereas a CFT by definition only has dimensionless parameters. Indeed, in two dimensions the gauge couplings $e_I$ and $e_i$ (see \eqref{(2,0)kinactions} of Appendix \ref{sc:(2,0)theories}) for the bosonic and fermionic gaugings have the dimension of mass. Similarly, also the superpotential \eqref{superpotential} is controlled by a dimensionful parameter $m$. Hence both gauging and adding a superpotential constitute relevant deformations of the theory.  This means that one needs to take a conformal limit of a GLSM before one can give it a string theoretical interpretation. In particular for the theories discussed here, we are only really interested in the limit where the dimensionful couplings are all sent to infinity simultaneously. Consequently, introducing a (bosonic or fermionic) gauging or a certain superpotential structure is not a small deformation, but rather changes the dynamics of the theory in a very significant manner.

\subsection{Phases of a GLSM -- Effective target space geometry}
\label{sc:Phases}

A GLSM can be used to study changes of topology \cite{Aspinwall:1993xz,Witten:1993yc,Aspinwall:1994ev}. A change of topology, and in particular of the intersection numbers, is described as a phase transition induced by varying some of its parameters. The zero modes of the scalars $z^a$ in a given background are interpreted as the target space coordinates of the effective geometry. Such a background is characterized by scalars that acquire VEVs, $\langle z^a \rangle$ minimizing the scalar potential. In order to preserve (2,0) supersymmetry the scalar potential needs to vanish identically. Different configurations of VEVs that are necessarily non-zero correspond to different phases of the model. Not all phases of a GLSM have a smooth target space interpretation: Only deep in the interior of a K\"ahler cone
(where all K\"ahler parameters are positive and sufficiently large in our parameterization), can the model be treated using supergravity techniques. A GLSM may have various additional phases for which a smooth CY interpretation is not available.

Let us describe the relation between GLSM and topology of its target space in more detail. Using the consequences of the fermionic gauge invariance \eqref{FermionicWinvariance} of the superpotential \eqref{superpotential} the scalar potential takes the form
\equ{
V = \sum_I\, \frac {e_I^2}2 \, 
\Big( \sum_a q_I^a\, |z^a|^2 - b_I \Big)^2 
+ |m|^2\,  
\sum_\ga\, 
\Big| N_\ga(z) \Big|^2 
+ \sum_{i,j}\, \bs_i \, (\cM^2)^i{}_j \, s^j~, 
\label{ScalarPot}
} 
where $b_I$ are the real parts of the complexified K\"ahler parameters $\gr_I| =b_I+i\gb_I$. (The $\gb_I$ appear in the expansion of the Kalb-Ramond two-form and behave in the effective four dimensional theory as axions.) The mass matrix squared 
\(
(\cM^2)^i{}_j = \sum_\ga\, \bM^i{}_\ga(\bz) M^\ga{}_j(z) 
\) 
has generically maximal rank when the range of the indices $\ga$ is larger than that of $i$. In this generic case the scalars $s^i$ are massive and their VEV lies at zero.

As the scalar potential is a sum of squares, each of the squares has to vanish individually for the potential to vanish as a whole. The first two terms are obtained by eliminating the auxiliary fields $h^\ga$ and $\tD_I$ contained in $\gL^\ga$ and $A^I$, respectively. These terms vanish provided that the VEVs of the complex scalar can be chosen such that 
\equ{
\sum_a q_I^a\, \big|\langle z^a\rangle\big|^2 = b_I~, 
\qquad 
N_{\ga}\big(\langle z\rangle\big) = 0~, 
\label{HyperSurfaces}
}
for all $I$ and $\ga$.

\subsubsection*{Divisors, curves and intersections}

A given complex geometry may possess a certain number of divisors and curves. Its topology is specified in part by their intersection numbers. (The information on divisors and curves fails to capture the $H^{1,2}$--part of the topological data of the geometry.)  The volume of the various cycles is set by the K\"ahler form $J$, which can be expanded as 
\equ{ 
J = \sum_D a_D \, D~, 
\label{KahlerForm} 
}
in terms of a basis of independent divisors $\lbrace D \rbrace$. The volumes of curves $C$ or divisors $D$ are determined according to
\equ{ 
\text{Vol}(C) = \int_C J~, 
\qquad 
\text{Vol}(D) = \frac 12\,  \int_D J^2~, 
\label{VolCurvesDivisors} 
}
respectively. The volumes of curves depend on genuine intersections and double-self-intersections, while the volumes of divisors even contain triple-self-intersections.

This geometrical information can be determined in any given phase of a GLSM. However, one should realize that some divisors, curves or intersections may exist in one phase, while being absent in another. It is straightforward to investigate whether a given divisor (or curve) exists: One simply checks whether the scalar potential can be made to vanish when the corresponding complex coordinates are set to zero. If this is possible, the divisor (or curve) exists, otherwise it does not. Using the same procedure one can also decide whether a certain intersection of three distinct divisors exists. Its intersection number equals unity, unless the vacuum is not uniquely defined: When a discrete remnant of the gauge symmetry is not fixed by the VEVs, the intersection number equals one over the order of this discrete gauge group.

Provided that a curve $C$ exists in a given phase, its volume is  determined by the integral 
\equ{
\text{Vol}(C) = \frac 1{\gp} \int_{V\big|_C = 0} 1~, 
\label{Vol_Curve} 
}
over the region where the potential $V$ vanishes when restricted to the curve $C$. The integral is normalized to the unit disc. When this integral diverges, it signals that the curve $C$ is non-compact. The computation of the volume of divisor $D$ proceeds via an analogous integral
\equ{
\text{Vol}(D) = \frac 1{\gp^2} \int_{V\big|_D=0} 1~. 
\label{Vol_Divisor} 
}
By comparing these results with the expressions \eqref{VolCurvesDivisors} for volumes of the various curves and divisors after inserting the expansion of the K\"ahler form \eqref{KahlerForm}, one can determine a large number of intersections (including self-intersections) in a given GLSM phase.

\subsection{Fermionic zero modes -- Gauge bundle}

As we have analyzed in the previous section, the zero modes of the scalars inside the chiral multiplets characterize the target space geometry. Similarly the zero (or constant) modes of the fermions, $\gl_0^\ga$, inside the chiral-Fermi multiplet determine the properties of the vector bundle. There are at least three ways of obtaining non-Abelian bundles: (i) non-Abelian gauging on the worldsheet, (ii) fermionic gauging and (iii) left-moving fermion constraints from the superpotential.

In the present work we do not consider non-Abelian gaugings, so we restrict ourselves to the other constructions. The fermionic gauging, introduced in \eqref{gXgauge}, lead to the transformation of the left-moving zero mode $\gl_0$
\equ{
\gd_\gX \gl_0^\ga = M_i{}^\ga(\langle z\rangle)\, \gx~,  
\label{FermiGaugings} 
}
where $\gx = \gX|$ is the lowest component of the chiral-Fermi super gauge parameter $\gX$. The constraints on the left-moving fermion zero modes $\gl_0^\ga$ arise as the lowest component of the equations of motion of the superfields $\gPs^a$ lead to the linear constraints 
\equ{ 
N_{\ga,a}(\langle z\rangle)\, \gl_0^\ga = 0~, 
\label{FermiConstraints}
}
on the fermion zero modes $\gl^\ga_0$ in the limit $m \ra \infty$.

The fermionic gauge transformations \eqref{FermiGaugings} and the constraints \eqref{FermiConstraints} have the geometric interpretation of a complex of line bundles: Let $\cO_\ga$, $\cO_\gr$ and $\cO_r$ be the line bundles whose transition functions act on the Fermi multiplets $\gL^\ga$, $\gG^\gr$ and the chiral superfields $\gF^\gk$, respectively. The holomorphic mappings 
$M = \big(M^\ga{}_i(z) \big)$ and 
$N = \big(N_{\ga}(z)\big)$ define a complex of vector bundles \cite{Witten:1993yc} 
\equ{
0 
\quad\lra\quad 
\cO^{N_\gS} 
\quad\stackrel{M}{\dsp \lra}\quad 
\bigoplus\limits_{\ga}^{} \cO_\ga 
\quad\stackrel{N}{\dsp \lra}\quad  
\bigoplus\limits_{r}^{} \cO_r
\quad\lra\quad 
 0~,
}
where $N_\gS$ denotes the number of fermionic gaugings. This complex generically determines a holomorphic vector bundle via $\mathbb{V} = \text{Ker}(N)/\text{Im}(M)$. Under certain circumstances the mappings $M$ or $N$ may degenerate and the resulting object is a sheaf rather than a bonafide vector bundle. Provided that the resulting singularities are not too severe the GLSM description still makes sense perturbatively \cite{Distler:1996tj}.

\subsubsection*{Standard embedding -- enhancement to a (2,2) subsector} 


The models presented so far had (2,0) worldsheet supersymmetry. Under certain conditions a subsector of the theory can possess a higher amount of worldsheet supersymmetry. This happens when the GLSM describes the standard embedding where the worldsheet theory has (2,2) supersymmetry.

It is not difficult to understand the necessary conditions for this to happen: As reviewed in Appendix \ref{sc:Reduction} (2,2) multiplets can be decomposed in two (2,0) superfields \cite{Witten:1993yc}. A (2,2) chiral superfield decomposes into a chiral and a chiral-Fermi superfield of (2,0) supersymmetry. And a (2,2) vector multiplet decomposes into a gauge and a Fermi-gauge superfield. Thus a minimal requirement for a sector of a (2,0) theory to possess (2,2) supersymmetry is that the number of chiral and chiral-Fermi superfields are equal, as well as, the number of gauge and Fermi-gauge superfields.

For a sector to be (2,2) supersymmetric, not only the degrees of freedom have to match, but also the interactions need to respect the higher amount of supersymmetry. From the results in the Subappendix \ref{sc:ReductionAction} we see that the (2,0) superpotential and FI-terms can be lifted directly to a (2,2) superpotential and twisted-superpotential. The kinetic terms of the gauge and Fermi-gauge multiplets are (2,2) supersymmetric when the couplings $e_I$ and $e_i$ are equal. The kinetic terms of the chiral and chiral-Fermi superfields possess (2,2) supersymmetry provided the bosonic and fermionic gaugings are compatible, i.e.\ the function $M^\ga_i$ has to be of the form 
\equ{
M^a{}_I = (Q_I)^a{}_b \, \gPs^b~.
}
When this choice is made the sigma model describes what is conventionally called the standard embedding.

\subsection{Consistency requirements for GLSM} 
\label{sc:Consistency}

A generic GLSM does not constitute a consistent model, because it suffers from gauge, $L$- or $R$-symmetry anomalies. Furthermore, some of its physical {\em finite} parameters, like the K\"ahler parameters $\gr_r$, may receive infinite quantum corrections. 


Gauge anomalies are problematic because then the GLSM does not define a well-defined quantum worldsheet theory. Consistency of this model requires that all pure and mixed gauge anomalies cancel. This leads to the conditions on the charges 
\equ{
\sum_\ga Q_I^\ga \,Q_J^\ga =\sum_a q_I^a \,q_J^a~, 
\label{gauge_anomalies} 
}
for all $I,J$. These consistency requirements (for $J=I$) may be interpreted as the Bianchi identities from the target space perspective: The left- and right-hand-sides constitute the second Chern classes $c_2(\mathbb{V})$ and $c_2(\mathbb{T})$ of the gauge bundle $\mathbb{V}$ and the tangent bundle $\mathbb{T}$, respectively.


Anomalies in the $L$/$R$-symmetries are not problematic for the worldsheet theory by itself, but impair the string theoretical interpretation: The existence of the $L$/$R$-symmetries at the quantum level is necessary to ensure that the proper GSO projections, i.e.\ spin-structures, that define the heterotic string, can be implemented. As the $L$- and $R$-symmetries are only discrete gauge symmetries, we do not have to require that their anomalies vanish identically; it is sufficient that 
\equ{ 
R \cdot q_I = \frac 12 \, \sum_a q_I^a \equiv 0~, 
\qquad 
L \cdot Q_I = \frac 12\, \sum_\ga Q_I^\ga \equiv 0~, 
\label{R/L_anomalies}
}
modulo integers (indicated by "$\equiv$'').


In order that we may interpret $\gr_I$ as a {\em finite} K\"ahler deformation, it is necessary that this function does not receive any infinite quantum corrections. This means that no divergent FI-term should be generated at the one loop level for the auxiliary field $\tD_I$, given in Table \ref{tb:LinearSigma}. As they are contained in superfields $A_I$, they only couple to the chiral multiplets, but not to the chiral-Fermi multiplets, see Appendix \ref{sc:(2,0)theories}. Therefore, this requirement reads 
\equ{
\sum_a q_I^a  = 0~.
\label{tadpole_chiral}
}
An alternative way to understand this requirement follows from four dimensional $\cN=1$ supersymmetry: It implies that the supposed (2,0) CFT must have a non--anomalous $U(1)_R$ symmetry, therefore its GLSM counterpart should be anomaly free as well. By comparing with \eqref{R/L_anomalies} we conclude that this condition is stronger than the requirement that $R$-symmetry anomalies are absent. This condition has the geometrical interpretation as the CY condition that the first Chern class of the tangent bundle $\mathbb{T}$ vanishes: $c_1(\mathbb{T}) = 0$.

Similarly, a tadpole for a $A_{I\gs}$ is avoided when $\sum_\ga Q_I^\ga = 0$. When this condition fails, the gauge field component $A_{I\gs}$ receives a strongly renomalization dependent VEV. From the worldsheet perspective this might seem as a nuisance, but does not impede the target space interpretation. However, as noted above to ensure the proper GSO projection, we need to require that the $L$-symmetry is not anomalous \eqref{R/L_anomalies}, i.e.\ $\sum_\ga Q_I^\ga = 0$ modulo even integers. From a target space perspective this means that the vector bundle $\mathbb{V}$ has $c_1(\mathbb{V}) = 0 \mod 2$.

\section{Non-compact heterotic orbifolds}
\label{sc:Orbifolds}

In this Section we briefly recall some basics of heterotic orbifold model building, details can be found in the refs.\ \cite{Dixon:1985jw,Dixon:1986jc,Ibanez:1988pj,Vaudrevange:2008sm}.  We consider non-compact orbifolds $\Cplx^3/\Intr_N$ defined by the action 
\equ{
z^a \ra e^{2\gp i v^a}\, z^a~, 
\qquad 
\gps^a \ra e^{2\gp i v^a}\, \gps^a~, 
}
for $a =0,1,2,3$ on the coordinates $z^a$ of $\Cplx^4$ and their superpartners $\gps^a$ with $N\, v^a \in \Intr$ and $\sum_i v^a /2 \equiv 0$. (As explained above equation \eqref{FreeHetAc} the coordinate $z^0$ refers to the non--compact directions in light--cone gauge.) 
The former requirement states that the action is generically a $\Intr_N$ action, and the latter that the spinor $(\frac 12^4)$ is left invariant, hence ($\cN=1$) target space supersymmetry is preserved. We can make a specific choice for $v$ given by 
\equ{
v = \big(0, \frac mN, \frac nN, -\frac{m+n}N \big)~, 
\qquad 
0 \leq m \leq  n \leq  \frac{N-m}2~, 
\quad m \leq \frac N3~. 
}
(When $m=0$ we produce a two dimensional complex orbifold $\Cplx^2/\Intr_N$.) The heterotic orbifold is fully specified by the additional $\Intr_N$ action on the gauge fermions $\gl^I$ parameterized by the shift vector $V = (V^1,\ldots, V^{16})$ as 
\equ{
\gl^\ga \ra e^{2\gp i V^\ga}\, \gl^\ga~, 
\qquad \ga= 1, \ldots, 16~.
}
These entries are required to satisfy $N\, V \in \gL_{16}$, with $\gL_{16}$ the $\Spin{32}/\Intr_2$ lattice.

\subsection{Partition function}

We consider the quantization of the orbifold theory on the worldsheet torus and define the path integral 
\equ{
Z\brkt{g}{\bg}(\gt) = \int \cD \gPs \cD \gL \, e^{-S_\text{free}}~, 
\label{PIsector} 
}
as a function of the Teichm\"uller parameter $\gt$; the complex structure of the worldsheet torus. Here $g=(r,s,t)$ parameterizes the various worldsheet boundary conditions: The parameter $r=0,\ldots, N-1$ labels the different orbifold sectors of the theory; the parameters $s=0,1$ and $t=0,1$ label the spin--structures for the right-- and left--moving fermions, respectively. On the worldsheet superfields these boundary conditions read: 
\begin{subequations} 
\equ{
\gPs^a(\gs+1,\gth^+) = e^{2\gp i\, r v^a} \,
\gPs^a(\gs,e^{2\gp i\, \frac s2} \gth^+)~, 
\\[2ex]  
\gL^\ga(\gs+1,\gth^+) = e^{2\gp i( r V^\ga+ \frac t2)} \,
\gL^\ga(\gs,e^{2\gp i\, \frac s2} \gth^+)~,  
}
\end{subequations} 
 and similar $\bg=(\br,\bs,\bt)$ for the other torus boundary condition (that relates $\gs-\gt$ to $\gs$.) These boundary conditions implement the orbifold action and the various spin structures of the heterotic string.

An important constraint is that the wordsheet theory is modular invariant under the transformations 
\equ{
T:~\gt \ra \gt +1~, 
\qquad
S:~ \gt \ra - \frac 1\gt~. 
}
The partition functions \eqref{PIsector} transform covariantly under the modular transformations, hence we need to combine the various sectors together to form a modular invariant partition function
\equ{
Z(\gt) =  \frac 1{4 N} \sum_{g,\bg} C\brkt{g}{\bg} \, Z\brkt{g}{\bg}(\gt)~, 
\qquad 
C\brkt{g}{\bg} = e^{\dsp 2\pi i\, \sfrac 12 \big\{ V_g \cdot V_\bg  - v_g\cdot v_\bg \big\}}~, 
\label{TotPI} 
}
where 
\equ{ 
V_g = V_r + \sfrac t2 \, e_{16}~, 
\qquad 
v_g = v_r + \sfrac s2\, e_4~, 
}
with $e_n = (1,\ldots, 1)$ with $n$ entries $1$ and $v_r = r\, v$ and $V_r = r\, V$. For a $\Intr_N$ orbifold applying the $T$ transformation $N$ times takes the path integral \eqref{PIsector} for each of the sectors back to itself up to a phase, hence modular invariance requires that 
\equ{
\frac {N}2 \, V^2 \equiv \frac {N}2\, v^2~. 
}

\subsection{Massless (twisted) orbifold states}

The massless spectrum of the orbifold in target space is obtained by expanding the partition function \eqref{TotPI} up to constant terms in $\gt$. This decomposes the spectrum in sectors characterized by $r = 0,\ldots, N-1$. We define $\tv_r \equiv v_r$ such that all entries in $\tv_r$ satisfy $0 \leq \tv_{r\,a} < 1$. A sector $r$ is independent provided that 
\equ{ 
 \sum_a \tv_{r}^{a} = \tv_r \cdot e_4 = 1~.
 \label{independent_sector} 
}
When this condition is not satisfied, the sector is conjugated to another sector that does satisfy it. From now on we only restrict to the independent sectors that fulfill this condition. Each twisted state $|T_r \rangle = |p_r, P_r\rangle, \tga^a_{-\tv_r}|p_r, P_r\rangle, \tga^\ua_{-\tv_r}|p_r, P_r\rangle,$ etc., is characterized by shifted left- and right-moving momenta 
\equ{ 
p_{r} = p + v_r~, 
\qquad 
P_r = P + V_r~, 
}
with $r \neq 0$. Here $p$ takes values in the vector lattice $\gL_4$ of $SO(8)$ and $P$ in the direct sum $\gL_{16}$ of the root and spinor lattices of $SO(32)$. In addition, to specify a twisted state fully, the numbers of left-moving oscillators, $\tga_{-\tv_r}^a$ or $\tga_{-\tv_r}^\ua$, that hit the vacuum state $|p_r; P_r \rangle$, has to be given. (Here $\ua$ denotes the index that is complex conjugated to that of $a$.) The twisted number operators $n_r^a$ and $\bn_r^\ua$ count number of left-moving oscillators $\tga_{-\tv_r}^a$ and $\tga_{-\tv_r}^\ua$, respectively. The twisted number operator 
\equ{
\tN_r = \sum_a (n_r^a - \bn_r^\ua)\, \tv_{r\,a}
}
counts these excitations weighted by $\tv_{r\, a}$, e.g.\ the oscillator state $|\tga^a_{-\tv_r}|p_r, P_r\rangle$ has $\tN_r$ equal to $\tv_{r\,a}$.

The sums over the orbifold parameter $r$ and the spin structures $s,t$ in the total partition function \eqref{TotPI} lead to the orbifold projection 
\equ{
V_\br \cdot P_r - v_\br \cdot k_r - \sfrac 12 (V_\br \cdot V_r - v_\br \cdot v_r) \equiv 0~,
\label{OrbiProj} 
}
where $k_r = p_r - n_r + \bn_r$.

The masses of the states $|T_r\rangle$ are determined by the following formulae. We define the zero point off-set 
\equ{
\gd c_r = \frac 12 \, \tv_{r} \cdot (e_4-\tv_r) = \frac 12 -\frac 12\, \tv_r^2~. 
}
The right-moving mass is given by 
\equ{
M_R^2 = \frac 12\, p_{r}^2 - \frac 12 + \gd c_r~. 
\label{MR2} 
}
The two formulae above combined imply that the massless twisted state $|T_r\rangle$ has $p_r^2 = \tv_r^2$. The level matching condition says that right-moving mass needs to be equal to the left-moving mass 
\equ{
M_L^2 = \frac 12\, P_{r}^2 - 1 + \gd c_r + \tN_r~. 
\label{ML2} 
}
This implies that the shifted left- and right-moving momenta squared are related via 
\equ{
P_r^2 = 1 + p_r^2 - 2 \, \tN_r~.
\label{LevelMatch}
} 

\section{Resolutions as GLSMs}
\label{sc:ResGLSM}

In this section we investigate how to associate a GLSM to a non-compact heterotic orbifold. The procedure we propose is as follows: From the massless spectrum of the non-compact heterotic orbifold theory we select one or more twisted states that acquire non-vanishing VEVs. We describe how these twisted states characterize specific bosonic and fermionic gaugings in the GLSM.

As we have seen a given twisted state $|T_r\rangle$ is characterized by shifted right-- and left--moving momenta $p_r$ and $P_r$. The idea is to promote these momenta to the charges $q_r^a$ and $Q_r^\ga$ of chiral and chiral--Fermi superfields $\gPs^a$, $a=0,1,2,3$, and $\gL^\ga$, $\ga=1,\ldots,16$, under a bosonic gauge symmetry. This entails the introduction of a vector multiplet $(V,A)^r$, which removes one chiral superfield from the worldsheet spectrum. Therefore, in order to keep the number of degrees of freedom the same as in the free heterotic theory, the gauging requires the introduction of a new (or exceptional) charged chiral superfield $\gPs^{\sm r}$. This new superfield is essential to ensure that the FI-tadpole cancels, and that the model has an orbifold limit. When the twisted state involves oscillator excitations, anomaly cancellation further requires the introduction of an extra (or exceptional) chiral-Fermi multiplet $\gL^{\!\sm r}$. As this changes the number of the degrees of freedom again, the introduction of this exceptional chiral-Fermi multiplet has to be accompanied by a fermionic gauging with a Fermi-gauge superfield $\gS^r$. Below we describe these proposals in detail.

\subsection{Characterization of a GLSM by blow-up modes}

In the target space picture of the orbifold theory, the VEV $\langle T_r\rangle$ of a twisted state $|T_r\rangle$ generates a deformation of the orbifold theory. Hence in the GLSM description of this deformation process we should be able to identify a complex function associate with this VEV. For any Abelian gauge multiplet $(V,A)^r$ one can write down a Fayet-Iliopoulos(FI)-action
\equ{
S_{\langle T_r\rangle}  =  \dsp 
\int\d^2\gs \d \gth^+ \, \gr_r(\gPs) \, F^r  + \hc = 
\int \d^2\gs \Big\{ 
b_r(z) \, \tD^r - 
\gb_r(z)\, F^r_{\gs\bgs}
 \Big\}~, 
 \label{FI} 
}
see \eqref{super&FI} in Appendix \ref{sc:(2,0)theories}, 
where the superfield strength $F^r$ defined in \eqref{(2,0)superfieldstrengths}. Here is $\gr_r(\gPs)$ is a neutral holomorphic function of the chiral superfields $\gPs^a$. Its lowest component is related to the VEV of the twisted state $|T_r\rangle$ as 
\equ{ 
\gr_r| = b_r + i\, \gb_r \sim \ln \Big\{\frac{\langle T_r \rangle}{M_r}\Big\} + \ldots~, 
\label{logdependence}
}
where $M_r$ is a mass parameter of the order of the string scale. To justify the appearance of the logarithm of the twisted state VEV we notice the following: The pullback $B_{\gs\bgs}$ of the Kalb-Ramond tensor $B_{MN}$ to the worldsheet has the same index structure and symmetry properties as the gauge field strength $F_{\gs\bgs}$. In fact, by a gauge transformation the Kalb-Ramond field can absorb this field strength. This is the sigma model realization of the expansion of the $B_{MN}$ in terms of exceptional divisors in target space. The expansion coefficients can be interpreted as twisted axions, because they shift under Abelian gauge transformation. In refs.~\cite{GrootNibbelink:2007ew,Nibbelink:2008tv} it was realized that these axions are the phases of the four dimensional twisted chiral superfields that take VEVs in the blow-up procedure. Combining these pieces of information motivates the $\ln \langle T_r\rangle$ in \eqref{FI}. However, these arguments do not tell us what the precise value is of the mass $M_r$ is which sets the scale for the twisted VEV.

As noticed above the bosonic gauging requires the introduction of a new chiral superfield $\gPs^{\sm r}$ to restore the right number of degrees of freedom. The charge of this superfield is $-1$, because the sum of all chiral superfield charges needs to vanish  \eqref{tadpole_chiral}: 
\equ{
q_r^{\sm r} + \sum_{\ga=0,1,2,3} q_r^\ga = 0
\qquad\Ra\qquad
q_r^{\sm r} = - e_4 \cdot \tv_r = -1~. 
}

To understand the target space interpretation of the worldsheet FI-term \eqref{FI} we study the resulting scalar potential  
\equ{
V \supset \frac {e_r^2}2 \Big( 
\sum_{a=1,2,3} q_r^a |z^a|^2 - |z^{\sm r}|^2 - b_r
\Big)^2~,
\label{DtermPot} 
}
obtained from \eqref{ScalarPot}. To preserve supersymmetry on the worldsheet, this potential has to vanish.

In the limit where the gauge sigma model should reduce to the orbifold CFT, the twisted state VEV goes to zero: $\langle T_r\rangle \ra 0$. Because of the logarithm in \eqref{logdependence} this means that the K\"ahler parameter $b_r$ tends to minus infinity. Since the charges $q^a$ of $z^a$ are all positive, this implies that 
\equ{ 
|\langle z^{\sm r} \rangle| = \sqrt{ - b_r}
}
needs to become very large. As the charge of $z^{\sm r}$ is $-1$ while the charges $q^a$ of the $z^a$ are fractional, the gauge transformations that preserve $\langle z^{\sm r} \rangle$ generate a residual $\Intr_{n_r}$ gauge action on $z^a$. (Here $n_r = N/\gcd(r,N)$ is the order of the twisted sector $r$. In particular, when $r$ and $N$ are relatively prime $n_r=N$.) In this limit the gauge superfields decouple, since they form a massive vector multiplet.  As removing some states from the spectrum could result in anomalies, in this process some vacuum phases could arise \cite{Distler:1994hs,Adams:2009tt}. (We leave this interesting effect for future study.)

This GLSM also gives a description of the exceptional divisor that arises in the blow--up process: When the parameter $b_r > 0$ the zero locus of the potential \eqref{DtermPot} can be written as 
\equ{
\sum_{a=1,2,3} q_r^a |z^a|^2 = b_r + |z^{\sm r}|^2~. 
}
This shows that at least one $z^a$ needs to take a non-vanishing VEV in order that the potential is zero. Moreover, if we are located on the divisor $E_r$, i.e.\ set $z^{\sm r} = 0$, this equation defines a (deformed) five-sphere $S^5$. On this five-sphere the $U(1)$ gauge symmetry still acts, hence the GLSM describes a four--cycle $S^5/U(1)$ of ``radius'' $\sqrt{b_r}$ for $b_r > 0$. This justifies to interpret $b_r$ as a \Kh\ parameter (or a blow-up parameter) that measures the size of the exceptional four--cycle. Hence it makes sense to call a twisted state $|T_r\rangle$ a blow-up mode when it takes a non-vanishing VEV, irrespectively of whether it contains any oscillators.

In the following subsections we first describe how to associate a GLSM to  a single non-oscillatory blow-up mode, then blow-up modes that have oscillator excitations, and finally consider some restrictions and consequences of using multiple blow-up modes simultaneously.

\subsection{Single non-oscillatory blow-up mode GLSM}

We consider a single massless non-oscillatory twisted state $|T_r\rangle = |p_r, P_r\rangle$ as blow-up mode that survives the orbifold projection \eqref{OrbiProj}. According to the procedure above we would like to take the  $U(1)$ charges of chiral superfields $\gPs^a$ and Fermi multiplets $\gL^\ga$ equal to the shift momenta of $|T_r\rangle$, i.e.\ $q_r^a=p_r^a =\tv_r^a$ and $Q_r^\ga=P_r^\ga$, respectively. As noticed above the bosonic gauging requires the introduction of a new chiral superfield $\gPs^{\sm r}$ with charge $-1$ to restore the right number of degrees of freedom and avoid a divergent FI-tadpole on the worldsheet.

In this case no new chiral-Fermi multiplets need to be introduced as the worldsheet gauge anomaly vanishes automatically: The sum of charges squared of all chiral multiplets $\gPs^a$, $a=0,\dots,3;\sm r$ equals 
\equ{
q_r^2 = \sum_{a=0,\ldots,3;\sm r} (q_r^a)^2 = 
1+ p_r^2~. 
\label{SumChiralCharges} 
}
This is equal to the same sum for the Fermi multiplets 
\equ{
Q_r^2 = \sum_\ga (Q_r^\ga)^2 = P_r^2 = 1 + p_r^2~, 
}
because of the level matching condition \eqref{LevelMatch} with $\tN_r=0$ (since $|T_r\rangle$ does not have any oscillator modes).  To set the notation for later use, we collectively write all chiral superfields $\gPs = (\gPs^0,\ldots,\gPs^3;\gPs^{\sm r})$ and their charges $q_r = (q_r^0,\ldots,q_r^3;q_r^{\sm r}=-1)$ separated by a semi colon ";" to distinguish the exceptional chiral superfields from the ordinary ones.

In ref.\ \cite{Nibbelink:2009sp} it was found that the shifted momenta of twisted states without oscillators can be directly identified with bundle vectors that specify how line bundles on the resolution are embedded in the Cartan of the gauge group. In particular, it was found that certain Bianchi identities are related to mass--shell conditions of the corresponding twisted states. The description presented here gives the GLSM realization of these findings: The role of the Bianchi identities in the GLSM is played by the anomaly conditions. And precisely when we take $Q_r = P_r$ the level matching condition ensures the absence of the gauge anomaly.

\subsection{Oscillator blow-up mode GLSM}
\label{sc:OscillatorBlowup}

Next we consider a twisted state that has one oscillator excitation, e.g.: $|T_r\rangle = \tga_{-\tv_r}^a | p_r, P_r \rangle$. Since $p_r$ still defines the charges of the chiral superfield $\gPs^a$, $a=0,\ldots,3$, we still need to introduce an exceptional chiral superfield $\gPs^{\sm r}$ with charge $-1$, as discussed above. Therefore a twisted state with oscillators can also be interpreted as a blow-up mode, because the GLSM has again an FI-term on the worldsheet that is related to a K\"ahler parameter of an exceptional cycle.

However, because now $\tN_r = \tv_r^a$ we violate the anomaly cancellation constraint if we simply set $Q_r^\ga$ equal to $P_r^\ga$, as  
\(
P_r^2 = 1 + \tv_r^2 - 2\tv_r^a  \neq 1 + \tv_r^2~.  
\)
Hence unless we modify the identification between the charges $Q_r^\ga$ and the shifted momentum $P_r^\ga$, we are not able to associate a GLSM to an oscillatory state $|T_r\rangle$. We propose to add some integral entries to the momentum $P_r$ to define the charge vector $Q_r$ and insert an extra exceptional chiral-Fermi multiplet $\gL^{\!\sm r}$ with charge $-1$ in the GLSM. The reason that we are free to modify the charges of $\gL^\ga$ by integral amounts is that this does not modify the transformation properties of the orbifold state $|T_r\rangle$ in the blow down limit. For the same reason an integral charged extra chiral-Fermi superfield $\gL^{\!\sm r}$ is invisible in the orbifold limit, provided we also introduce a fermionic gauging to balance the number of degrees of freedom. Hence for certain shifted momenta $P_r$ we can construct charge vectors $Q_r$ that characterize anomaly free GLSMs.

Here we describe one such case, which we will be using in various examples throughout the remainder of this work. Suppose that the state with an oscillator excitation has an entry of $P_r$ that equals $P_r^\ga =  \pm(\tv_r^a-1)$ for some $\ga \in \{1,\ldots,16\}$ and $a\in \{0,\ldots, 3\}$.  In this case we may take 
\equ{
Q_r^{\gb\neq \ga} = P_r^\gb~, 
\qquad 
Q_r^\ga = \pm (P_r^\ga +1) = \pm \tv_r^a~. 
}
Then in order to obtain an anomaly free GLSM,  we introduce an additional chiral-Fermi multiplet $\gL^{\!\sm r}$ with charge $Q_r^{\sm r} = -1$. The anomaly cancellation then works out fine: 
\equ{
Q_r^2 
= 
\sum_{\gb \neq \ga} (Q_r^\gb)^2 + (Q_r^\ga)^2 + (Q_r^{\sm r})^2 
= 
P_r^2 - (\tv_r^a-1)^2 + (\tv_r^a)^2 + (-1)^2 
= P_r^2 +2 \tv_r^a = 1 + \tv_r^2~, 
}
using the level matching condition \eqref{LevelMatch} with $\tN_r = \tv_r^a$, which equals to the sum of chiral superfield charges \eqref{SumChiralCharges}. When the twisted state contains more oscillators, one proceeds in the same fashion to find an anomaly free charge vector $Q_r$. It turns out that for all twisted states with oscillator excitations in the same twisted sector, one finds the same charge vector.

When we consider twisted states with more than one oscillator, e.g.\ 
 $\tga_{-\tv_r}^a \tga_{-\tv_r}^b| p_r, P_r\rangle$, we can proceed in a similar fashion, but now changing two or more entries of the left--moving shifted momentum $P_r$ before interpreting it as the charge vector $Q_r$ of the chiral--Fermi multiplets. It turns out that this charge $Q_r$ is the same as that obtained for twisted states with a single oscillator. Hence, it seems that the twisted states with oscillator excitations are indistinguishable within the GLSM description. However, as we explain next these various oscillatory states lead to different contributions in a fermionic gauging:

As already alluded to above, in order to keep the number of worldsheet degrees of freedom the same, we need to introduce a fermionic gauging with a Fermi-gauge superfield $\gS_r$. The fermionic gauging is required to be holomorphic, see \eqref{gXgauge}. In addition, only superfields with the same bosonic gauge charges can transform into each other under the fermionic transformation. Hence we find that this gauging takes the generic form
\equ{
\gd_{r} \gL^\ga = \sum_{Q_r^\ga=q_r^a} 
 \gb_r^\ga{}_a\, \gPs^a \, \gX_r  + 
 \sum_{Q_r^\ga=q_r^a+q_r^b} 
 \gb_r^\ga{}_{ab}\, \gPs^a \gPs^b\, \gX_r  
    + \ldots~, 
\qquad
\gd_{r} \gL^{\sm r} = - \gPs^{\sm r}\, \gX_r~. 
\label{gXgauging} 
}
We have fixed the complex coefficient in the last equation to $-1$, using that we can always absorb a free complex parameter in $\gX_r$. The coefficients $\gb_r^\ga{}_a$ and $\gb_r^\ga{}_{ab}$, etc., some complex parameters, can be identified with the twisted states with one, two, or more, oscillator excitations. The standard embedding is a special case of this where these parameters take specific values.

\subsection{Multiple blow-up modes}

We now consider the situation with multiple blow-up modes. We distinguish two cases: (i) in each independent twisted sector we select a single blow-up mode, and (ii) in some sectors we have more than one blow-up mode.

\subsubsection*{Single blow-up mode in each different twisted sector}

In case (i) each twisted state $|T_r\rangle$ in each independent twisted sector is represented in the GLSM by its own extra chiral superfield $\gPs^{\sm r}$ and a bosonic gauging by a vector multiplet $(V,A)^r$. Not all possible configurations of twisted state VEVs have a GLSM realization, since one has to ensure that all mixed anomalies \eqref{gauge_anomalies} vanish as well. This gives a rather non-trivial set of consistency requirements. When we employ states that survive the orbifold projections \eqref{OrbiProj}, these conditions turn out to be satisfied in all cases examined in this work. This suggests that the orbifold projections \eqref{OrbiProj} and the mixed anomaly conditions \eqref{gauge_anomalies} of a GLSM are closely related.

Reversely, one may wonder whether a twisted state that is projected out  can ever be related to a consistent GLSM. Indeed, in general the mixed anomaly conditions are not fulfilled. Yet, these conditions may be satisfied for some projected-out states, if they contains oscillators. However, this generically leads to a non-holomorphic fermionic gauging which is incompatible with the (2,0) worldsheet supersymmetry. Hence we conclude that it is generically impossible to associate a GLSM with twisted states that got projected out. We illustrate these findings for a specific $\Intr_4$ resolution model where states got projected out from the second twisted sector, see Subsection \ref{sc:Z4ResModel}.

\subsubsection*{Multiple VEVs in a single twisted sector}

In case (ii) we have selected more than one twisted state in the same twisted sector, say two, $|T_r\rangle$ and $T_r'\rangle$, with shifted momenta $p_r, P_r$ and $p_r',P_r'$, respectively. (As we have seen above all states in a given twisted sector with oscillators can be treated simultaneously using the same bosonic gauging; their distinction is only made in the fermionic gauge transformation.) Since we have two distinct blow-up modes, we introduce two bosonic gaugings $(V,A)$ and $(V',A')$ and two extra chiral superfields $\gPs^{\sm r}$ and $\gPs^{\prime \sm r}$ with charges $(-1,0)$ and $(0,-1)$ w.r.t.\ the first and second gauging, respectively. Since the blow-up modes come from the same twisted sector, $p_r = p_r'= \tv_r$, hence it follows that the charges of the chiral superfields $\gPs^a$, $a=0,\ldots,3$ are equal. The full charge assignment reads
\equ{
\arry{| c || c | c | c |}{ \hline 
\gPs & \gPs^a & \gPs^{\sm r} & \gPs^{\prime \sm r} 
\\ \hline \hline 
q_r & \tv_r^a & -1 & 0 
\\ \hline 
q'_r & \tv_r^a & 0 & -1
\\ \hline 
}
}

It might sound surprising that a single twisted sector induces two exceptional superfields. But this is necessary in order to keep the number of degrees of freedom equal, when each of the two blow-up modes introduces a gauging. Nevertheless, from a target space perspective there is just a single exceptional divisor. The reason is that the charge basis is not unique. Any change of basis of the $U(1)$ charges still satisfies the anomaly constraints provided that the original basis was free of all pure and mixed anomalies. To ensure that the quantization conditions are not modified, the change of basis has to be an $SL(2; \Intr)$ element. By performing the change of basis, $\tq_{r} = q_r$ and $\tq_{r-} = q_r - q_r'$, we obtain the $D$-term potential in the form  
\equ{
V \subset \frac{e_{r}{}^2}2 
\Big( \sum_a \tv_r^a\, |z^a|^2 - |z^{\sm r}|^2  - b_{r}\Big)^2 
+ 
\frac {e_{r-}{}^2}2 \Big( |z^{\sm r}|^2 - |z^{\prime \sm r}|^2 - b_{r-} \Big)^2~. 
}
Because the $z^a$ are neutral w.r.t.\ $q_{r-}$,  the coordinates $z^a$ do not appear in the second square. This implies that the expectation values of $z^{\sm r}$ and $z^{\prime \sm r}$ are always correlated. This shows that from the target space perspective there is effectively just a single exceptional divisor with a single K\"ahler parameter.

We should therefore interpret the parameter $b_{r-} = b_r - b'_r$ as a vector bundle modulus rather than a K\"ahler modulus: This parameter can distinguish between different gauge bundles and therefore different unbroken gauge groups in the effective four dimensional theory. Depending on the field content of the twisted sector in question various situations may occur, for example (we do not aim to give an exhaustive list here): 
\enums{ 
\item If both VEVs are in the same vector representation of an $SU(n)$ gauge group, then by a simple transformation these two VEVs can be transformed to a single VEV of a given component of this vector.  In this case the parameter $b_{r-}$  seems to be redundant. 
\item If the VEVs are in two representations of two different gauge groups, then this parameter only measures which representation has the larger VEV.
\item If both VEVs are in the same rank--two anti--symmetric tensor representation of $SU(n)$, then interesting effects may happen. An anti--symmetric tensor can be block--diagonalized. So if the two VEVs correspond to two of its eigenvalues, then their relative values determine the effective unbroken four dimensional gauge group, e.g.: 
\items{ 
\item the gauge group is $SU(2)\times SU(2)\times SU(n-4)$, when both non--zero VEVs are different, 
\item 
 the gauge group is instead $Sp(4)\times SU(n-4)$, when both VEVs are non--zero but equal.  
}
} 
We have studied such effects in detail in ref.\ \cite{Nibbelink:2008qf} on the resolution of $\Cplx^2/\Intr_3$. There we were able to associate such different VEVs to different configurations of instantons on Eguchi--Hanson spaces. The arguments given here used an effective four dimensional field theoretical language, how to precisely describe these effects in an GLSM formulation we leave for future study.

\subsection{VEVs of untwisted states}

In the discussion so far we have investigated what kind of GLSMs are generated when twisted states take non-vanishing VEVs. From a target space perspective it is natural and often even necessary that also charged untwisted states take non-zero VEVs. As untwisted states correspond to the internal component of the ten dimensional gauge field, such untwisted VEVs result in continuous Wilson lines.

We can reach the same interpretation from the GLSM perspective. Untwisted states reside by definition in the 0th twisted sector, and can therefore be described by the same mass formulae \eqref{MR2} and \eqref{ML2} with $\gd c_0 = 0$. The solutions are are given by 
\equ{
p_0 = \big(0, \undr{1\, 0^2} \big)~, 
\qquad 
P_0 = \big( \undr{\pm 1^2\, 0^{14}} \big)~, 
}
for $\tN_0 = 0$. (For $\tN_0 = 1$ we have $P_0 = (0^{16})$ describing the graviton and the anti-symmetric Kalb-Ramond tensor.) Of course these momenta are still subject to the orbifold invariance condition \eqref{OrbiProj}. To ensure that both the gauge anomalies \eqref{gauge_anomalies} and the $\tD$-tadpole \eqref{tadpole_chiral} vanishes, one is forced to introduce a new chiral superfield $\gPs^{\prime 0}$ with charge $-1$.

From the analysis of the resulting vacuum manifold we conclude that the zero locus of the potential reads
\equ{
|z^1|^2 = b_0 + |z^{\prime 0}|^2~,
}
for say $p_0 = (0,1,0,0)$ when the parameter $b_0$ is positive. For the minimum value of $\langle z^{\prime 0}\rangle = 0$ this equation defines a circle $S^1$ of radius $\sqrt{b_0}$. But given the Abelian gauge transformation, this merely defines a point in the target space. This shows that an untwisted VEV does not result in an Abelian gauge flux and hence is not subject to any quantization condition. Hence, also within the GLSM description one concludes that an untwisted VEV can be interpreted as a continuous Wilson line.

\section{Some explicit examples}
\label{sc:Examples}

In this section we would like to illustrate some of the general proposals that were put forward in the previous Sections to construct GLSMs that represent resolutions of heterotic orbifolds. To be concrete, we consider the heterotic $SO(32)$ theory on certain non-compact orbifolds $\Cplx^3/\Intr_N$. These orbifold models were classified e.g.\ in \cite{Choi:2004wn,Nilles:2006np}.

\subsection{Resolutions of some $\boldsymbol{\Cplx^3/\Intr_3}$ orbifold models}
\label{sc:Z3resolutions}

The $\Intr_3$ orbifold is the simplest six dimensional orbifold, because it only has a single twisted sector. The orbifold twist reads 
\equ{
v = \big(0, \sfrac 13^2, -\sfrac 23 \big)~. 
}
The classification of the $\Intr_3$ models is straightforward and one can show that there are six modular invariant theories distinguished by the gauge shifts 
\equ{
V_n = \big( \sfrac 13^{2n}, -\sfrac 23^{n}, 0^{16-3n}\big)~,
\label{Z3class}
}
with $n = 0,1,\ldots, 5$. The resulting gauge group is $U(3n)\times SO(32-6n)$. The twisted states and the spectra can be found e.g.\ in Tables 1 and 2 of Ref.~\cite{Choi:2004wn}. In the following we will use $n=0,1, \ldots, 5$ to distinguish these six different $\Intr_3$ orbifold models.

\subsubsection*{The $\Intr_3$ orbifold standard embedding model ($\boldsymbol{n=1}$)}

We first describe GLSMs that can be associated with the model, which may be called orbifold standard embedding model, because the orbifold twist and shift are chosen equal. This model has $n=1$ and its gauge group reads $U(3)\times SO(26)$. The complete non--compact orbifold $\Cplx^3/\Intr_3$ spectrum reads 
\equ{
\frac 1{27} \, \Big[ \rep{3_R}(\rep{3},\rep{26})_1 + \rep{3_R}(\rep{3},\rep{1})_{\sm 2} \Big] + \rep{3_R}(\crep{3},\rep{1})_0 + (\rep{1},\rep{26})_1 + (\rep{1},\rep{1})_{\sm 2}~. 
}
Here the states with $\rep{3_R}$ transform as a triplet under the unbroken $SU(3)_R\subset SO(6)$ part of the internal Lorentz group. The twisted states have integral multiplicity, while the states with fractional $1/27$ multiplicity are the untwisted states \cite{Gmeiner:2002es,GrootNibbelink:2003gb,GrootNibbelink:2003gd}: On the compact orbifold $T^6/\Intr_3$ the untwisted states live everywhere on the orbifold. At one of the $27$ fixed points of this orbifold the untwisted states give a contribution of $1/27$. (These fractional multiplicities persist on resolutions of non--compact orbifolds as well \cite{Nibbelink:2007rd,Nibbelink:2007pn}.  But then only fractions of $1/9$ appear as the $SU(3)_R$ becomes contained in the $SU(3)$ holonomy group, hence for the untwisted states we get $3\cdot 1/27 = 1/9$.)

\begin{table}
\[
\renewcommand{\arraystretch}{1.3} 
\arry{|c|| c c c | c c | }{
\hline 
\text{state} & p & P & osc. & q & Q 
\\ \hline\hline 
\multicolumn{6}{|c|}{\text{twisted}} 
\\\hline 
(\rep{1},\rep{1})_{\sm 2} & \big(0,\frac 13^3\big) & \big(\sm \frac23^3,0^{13} \big) & &  \big(0,\frac 13^3;\sm 1, 0^4 \big) & \big(- \frac23^3,0^{13} \big) 
\\
(\rep{1},\rep{26})_1 &  \big(0,\frac 13^3\big) & \big(\frac 13^3,\undr{\pm1,0^{12}}\big) & &  \big(0,\frac 13^3;0,\sm 1,0^3\big) & \big(\frac 13^3,\undr{\pm1,0^{12}}\big)  
\\
\rep{3_R}(\crep{3},\rep{1})_0 &  \big(0,\frac 13^3\big) & \big(\undr{\frac 13^2,\sm \frac 23},0^{13}\big) & \tga_{-1/3}^i &  \big(0,\frac 13^3;0^2,\sm 1,0^2\big) & \big(\frac 13^3,0^{13}; \sm 1\big) 
\\\hline 
\multicolumn{6}{|c|}{\text{untwisted}} 
\\ \hline 
\rep{3_R}(\rep{3},\rep{26})_1 & \big(0,\undr{1, 0^2}\big) & \big(\undr{1, 0^2}, \undr{\pm 1, 0^{12}}\big) & &  \big(0,\undr{1, 0^2};0^3,\sm 1,0\big) & \big(\undr{1, 0^2}, \undr{\pm 1, 0^{12}}\big)
\\
\rep{3_R}(\rep{3},\rep{1})_{\sm 2} & \big(0,\undr{1, 0^2}\big) & \big(\undr{\sm 1^2, 0},  0^{13}\big) & &  \big(0,\undr{1, 0^2};0^4,\sm 1\big) & \big(\undr{\sm 1^2, 0}, 0^{13}\big)
\\ \hline
}
\renewcommand{\arraystretch}{1} 
\]
\captn{$\Intr_3$ (un)twisted states in the orbifold standard embedding model ($n=1$) and the $U(1)$ charges $q$ and $Q$ in the corresponding GLSM.
\label{tb:Z3orbiSE}}
\end{table}

We now analyze the possible GLSMs we can obtain from this spectrum. In Table \ref{tb:Z3orbiSE} we have indicated the shifted momenta that define the untwisted and twisted states of this orbifold model, and the associated gauge charge of the possible GLSMs. Because we associate with each of these states a different bosonic worldsheet gauge symmetry, the charges $q$ have five entries after the semi-column ";''. Consequently there are also five exceptional chiral superfields, because  we need to preserve the number of spacetime dimensions. The charges $q$ and $Q$ for each of the five types of states in that Table have been chosen such that the pure $U(1)$ anomaly cancels. The last two states in that Table are untwisted states, hence they generate continuous Wilson lines. Each of the other twisted states define a GLSM with a single blow-up mode.

Only the last twisted state in Table \ref{tb:Z3orbiSE}, $\rep{3_R}(\crep{3},\rep{1})$, has an oscillator excitation. Therefore, in the corresponding GLSM we need to introduce an additional Fermi multiplet $\gL^{\!\sm 1}$ with charge $-1$, and the fermionic gauging\footnote{Here we have employed the following index conventions: $i =1,2,3$ denote the complex internal space indices; $I=1,2,3$ denote the fundamental $U(3)$ indices of the chiral-Fermi superfields $\gL^I$, the remaining indices of the chiral-Fermi multiplets $\gL^\ga$ have the range $\ga = 1,\ldots, 13$.} 
\equ{
\gd \gL^I = \gb^I{}_i \, \gPs^i\, \gX~, 
\qquad 
\gd \gL^{\!\sm 1} = - \gPs^{\sm 1} \, \gX~. 
}
The complex parameters $\gb^I{}_i$ carry the same index structure as the twisted mode $\rep{3_R}(\crep{3},\rep{1})$ of the orbifold spectrum. When we take them to be equal to $\gb^I{}_i = \frac13\, \gd^I{}_i$, then we produce the true geometrical standard embedding, and the interacting part of the theory possesses (2,2) supersymmetry; otherwise the model only has (2,0) supersymmetry.

Finally, we briefly comment on whether we can switch on various blow-up modes and possible Wilson lines simultaneously. One can check that it is impossible to combine the gauging associated with the blow-up mode $(\rep{1},\rep{1})_{\sm 2}$ with any of the other states, because the mixed anomalies never cancel. Also the untwisted state $\rep{3_R}(\rep{3},\rep{1})_{\sm 2}$ cannot be combined with any of the states in the orbifold spectrum. The other three states, $(\rep{1},\rep{26})_1$, $\rep{3_R}(\crep{3},\rep{1})_{0}$ and $\rep{3_R}(\rep{3},\rep{26})_{1}$,  can be used simultaneously to define GLSMs, since all the mixed anomalies vanish.

\subsubsection*{The $\Intr_3$ orbifold model with $\boldsymbol{n=0}$}

The orbifold model with $n=0$ has no orbifold embedding on the gauge degrees of freedom and hence the unbroken gauge group remains $SO(32)$. Given that conventional wisdom states that one always need to have a non-trivial gauge bundle when one has a non-trivial geometry, this orbifold model looks very odd. It is therefore interesting to investigate this model in light of the GLSM construction:

The $n=0$ orbifold model only has twisted states with integral internal shifted momenta, hence we will never be able to find any configuration of charges of the chiral-Fermi superfields, such that the worldsheet gauge anomalies cancel.\footnote{The fact that these twisted states have two oscillators does not seem to be relevant; we will encounter similar twisted states in the $\Cplx^3/\Intr_4$ orbifold model of Subsection \ref{sc:Z4ResModel}, which can be treated following the general procedures put forward in this paper.} We conclude that it is impossible to construct a resolved version of this orbifold model, as there seems to be no GLSM that could correspond to it. In other words the $\Intr_3$ model with $n=0$ is rigid.

\subsubsection*{The $\Intr_3$ orbifold models with $\boldsymbol{1 < n \leq 5}$}

The possible GLSMs corresponding to the orbifold models with higher $n$ can be analyzed in a similar fashion as model $n=1$. In Subsection \ref{sc:Z3spectra} we will use two of them (the models with $n=2$ and $4$) to demonstrate how we can compute the charge spectrum in target space directly from the GLSM worldsheet action. For this reason we do not describe these models here in any detail further.

\subsection{Resolutions of a $\boldsymbol{\Cplx^3/\Intr_4}$ orbifold model}
\label{sc:Z4ResModel}

\begin{table}
\[
\renewcommand{\arraystretch}{1.3} 
\arry{|  r | c  c | c |}{
\hline 
\multicolumn{1}{|c|}{\text{state}} & \text{representation} & \text{in 4D} & 
Q(\gL^A,\gL^3,\gL^\ga; \gL^{\!\sm 1}, \gL^{\!\sm 2}) 
\\ \hline\hline 
\multicolumn{4}{| l |}{\text{1st twisted sector:}~~ \tN_1=0,~~ P_1^2 = \frac{11}8}
\\ \hline \hline 
\big| p_1 \ \big( \frac 14^2\ \frac 12\ \undr{0^{12}\ \pm \!1} \big) \big\rangle 
&
(\rep{26},\rep{1})_{\sm \frac 12, \sm \frac18} 
& \checkmark & 
 \big( \frac 14^2\ \frac 12\ \undr{0^{12}\ \pm \!1};0\ 0 \big) 
\\
\big| p_1 \ \big( \sm\frac 34^2\ \sm\frac 12\ 0^{13} \big) \big\rangle 
&
(\rep{1},\rep{1})_{1, \sm \frac18} 
& \checkmark & 
 \big( \sm\frac 34^2\ \sm\frac 12\ 0^{13}; 0\ 0 \big) 
\\ \hline\hline 
\multicolumn{4}{| l |}{\phantom{\text{1st twisted sector:}}~~ \tN_1=\frac 14,~~ P_1^2 = \frac{7}8}
\\ \hline \hline 
\tga_{-\frac 14}^a 
\big| p_1 \ \big( \undr{\frac 14\ \sm\frac 34}\ \frac 12\ 0^{13} \big) \big\rangle 
&
\rep{2}_R(\rep{1},\rep{2})_{0, \sm \frac38} 
& \checkmark & 
 \big( \frac 14^2\ \frac 12\ 0^{13}; \sm1\ 0 \big) 
\\ \hline\hline 
\multicolumn{4}{| l |}{\phantom{\text{1st twisted sector:}}~~ \tN_1=\frac 12,~~ P_1^2 = \frac{3}8} 
\\ \hline \hline 
\tga_{-\frac 12}^{3} 
\big| p_1 \ \big( \frac 14^2\ \sm\frac 12\ 0^{13} \big) \big\rangle 
&
(\rep{1},\rep{1})_{0, \frac38} 
& \checkmark & 
 \big( \frac 14^2\ \frac 12\ 0^{13}; \sm1\ 0 \big) 
\\ 
\tga_{-\frac 12}^\undr{3} 
\big| p_1 \ \big( \frac 14^2\ \sm\frac 12\ 0^{13} \big) \big\rangle 
&
(\rep{1},\rep{1})'_{0, \frac38} 
& \checkmark & 
 \big( \frac 14^2\ \frac 12\ 0^{13}; \sm1\ 0 \big) 
\\ 
\tga_{-\frac 14}^a \tga^b_{-\frac 14} 
\big| p_1 \ \big( \frac 14^2\ \sm\frac 12\ 0^{13} \big) \big\rangle 
&
\rep{3}_R(\rep{1},\rep{1})_{0, \frac38} 
& \checkmark & 
 \big( \frac 14^2\ \frac 12\ 0^{13}; \sm1\ 0 \big) 
\\ \hline\hline 
\multicolumn{4}{| l |}{{\text{2nd twisted sector:}}~~ \tN_2=0,~~ P_2^2 = \frac{3}2}
\\ \hline \hline 
\big| p_2 \ \big( \frac 12^2\ 0\ \undr{0^{12}\ \pm\!1} \big) \big\rangle 
&
(\rep{26},\rep{1})_{\sm\frac 12, \frac 14} 
& \checkmark & 
 \big( \frac 12^2\ 0\ \undr{0^{12}\ \pm\!1}\ ;\ 0\  0 \big)  
\\  
\big| p_2 \ \big( \sm\frac 12^2\ 1\ 0^{13} \big) \big\rangle 
&
(\rep{1},\rep{1})_{0, \sm\frac34} 
& \checkmark & 
 \big( \sm\frac 12^2\ 1\ 0^{13}\ ;\ 0\ 0 \big)  
\\
\big| p_2 \ \big( \sm \frac 12^2\ \sm1\ 0^{13} \big) \big\rangle 
&
(\rep{1},\rep{1})_{1, \frac34} 
& \checkmark & 
 \big( \sm \frac 12^2\ \sm1\ 0^{13}\ ;\ 0\ 0 \big)  
\\ \hline  
\big| p_2 \ \big( \sm\frac 12^2\ 0\ \undr{0^{12}\ \pm\!1} \big) \big\rangle 
&
(\rep{26},\rep{1})_{\frac 12, \sm\frac 14} 
& \text{x} & 
 \big( \sm\frac 12^2\ 0\ \undr{0^{12}\ \pm\!1}\ ;\ 0\  0 \big)  
\\  
\big| p_2 \ \big( \frac 12^2\ \sm 1\ 0^{13} \big) \big\rangle 
&
(\rep{1},\rep{1})_{0, \frac34} 
& \text{x} & 
 \big( \frac 12^2\ \sm 1\ 0^{13}\ ;\ 0\ 0 \big)  
\\
\big| p_2 \ \big( \frac 12^2\ 1\ 0^{13} \big) \big\rangle 
&
(\rep{1},\rep{1})_{\sm1, \sm\frac34} 
& \text{x} & 
 \big( \frac 12^2\ 1\ 0^{13}\ ;\ 0\ 0 \big)  
\\ \hline\hline 
\multicolumn{4}{| l |}{\phantom{\text{2st twisted sector:}}~~ \tN_2=\frac 12,~~ P_2^2 = \frac{1}2}
\\ \hline \hline 
\tga_{-\frac 12}^a 
\big| p_2 \ \big( \undr{\frac 12\ \sm\frac 12}\ 0\ 0^{13} \big) \big\rangle 
&
\rep{2}_R(\rep{1},\rep{2})_{0, 0} 
& \checkmark & 
 \big( \frac 12^2\ 0\ 0^{13}\ ;\ 0\ \sm1 \big) 
\\ \hline 
\tga_{-\frac 12}^\ua 
\big| p_2 \ \big( \undr{\frac 12\ \sm\frac 12}\ 0\ 0^{13} \big) \big\rangle 
&
\crep{2}_R(\rep{1},\rep{2})_{0, 0} 
& \text{x} & 
 \big( \frac 12^2\ 0\ 0^{13}\ ;\ 0\ \sm1 \big) 
\\ \hline  
}
\renewcommand{\arraystretch}{1} 
\]
\captn{
This Table lists the twisted states of the non-compact $\Cplx^3/\Intr_4$ orbifold in the orbifold standard embedding, i.e.\ 
$V= \big(\frac 14^2\ \sm\frac 12\ 0^{13}\big)$. 
Here $p_1 = \tv_1 = \big(0\ \frac 14^2\ \frac 12\big)$ and 
$p_2 = \tv_2 = \big( 0\ \frac 12^2\ 0 \big)$ and the indices $a,b, A,B = 1,2$ and $\ga,\gb =1,\ldots 13$. 
The first charge on the target space states generated by $(\sm \frac 12^3\ 0^{13})$ is anomalous, while the second generated by $(\frac 14^2\ \sm \frac 12\ 0^{13})$ is non-anomalous. 
\label{tb:TwistedStatesZ4}
}
\end{table}

In the current subsection, we consider the heterotic $SO(32)$ theory on the non-compact orbifold $\Cplx^3/\Intr_4$. Here we restrict our attention to the orbifold standard embedding, i.e.\ take 
\equ{
v = \big( 0,\frac 14^2, -\frac 12 \big)~, 
\qquad 
V = \big( \frac 14^2, -\frac 12, 0^{13} \big)~. 
}
This orbifold has two distinct twisted sectors, $r=1$ and $2$. In the first twisted sector the states have at most two oscillator excitations, and in the second twisted sector at most a single one, see Table \ref{tb:TwistedStatesZ4}. The momenta $p_1$ and $p_2$ are uniquely given by 
$p_1 = \tv_1 = \big( 0, \frac 14^2, \frac 12 \big)$ and 
$p_2 = \tv_2 = \big( 0, \frac 12^2, 0 \big)$ 
for the target space complex bosons. On the orbifold, the theory possesses a residual SU(2)$_R$ symmetry, the oscillator states with indices $a,b = 1,2$ form non-trivial representations under this group. In addition we introduce the following indices for the chiral-Fermi multiplets: $A,B=1,2$ and $\ga,\gb = 1,\ldots,13$.

Next we want to investigate the consequences of selecting a set of these states as blow-up modes. As we have two sectors, and in each sector there could be a blow-up mode switched on, we need to introduce the two exceptional chiral superfields $\gPs^{\sm 1}$ and $\gPs^{\sm 2}$. This gives us the following table of charges:
\[
\arry{|c || c c c | c c |}{
\hline 
& \gPs^0 & \gPs^a & \gPs^3 & \gPs^{\sm 1} & \gPs^{\sm 2} 
\\ \hline \hline 
q_1  & 0 & \frac 14^2 & \frac 12 & -1 & 0 
\\ 
q_2 & 0 & \frac 12^2 & 0 & 0 & -1 
\\ \hline 
}
\]
where $a=1,2$.

In addition we need to give the corresponding charges of the chiral-Fermi multiplets $\gL^\ga$ and $\gL^{\!\sm r}$. This information has also been collected in Table~\ref{tb:TwistedStatesZ4}: The charges in the first and second twisted sectors are represented by $Q_1$ and $Q_2$, respectively. For the blow-up modes containing oscillators, we see that these charges do not distinguish these state uniquely. But since precisely these states required the introduction of the additional chiral-Fermi multiplets $\gL^{\!\sm r}$, there are compensating fermionic gauge transformations. For the oscillatory blow-up modes from the first twisted sector the charge vector always reads $Q_1 = (\frac 14^2,\frac 12, 0^{13}; \sm1, 0)$ and the fermionic gauging is given in general by 
\equ{
\gd_1 \gL^A = \gb_1{}^A{}_a\, \gPs^a\, \gX_1~, 
\quad 
\gd _1\gL^3 = (\gb_1\, \gPs^3  + \frac 12 \gb_{1\,ab}\, \gPs^a\gPs^b) \gX_1~, 
\quad 
\gd_1 \gL^{\!\sm 1} = - \gPs^{\sm r} \, \gX_1~, 
\label{fermiGauge1} 
}
and $\gd_1 \gL^{\ga} = \gd_1 \gL^{\!\sm 2} =  0$. For the oscillatory modes of the second twisted sector we have 
\equ{
\gd_2 \gL^A = \gb_2{}^A{}_a\, \gPs^a\, \gX_2~, 
\qquad 
\gd_2 \gL^{\!\sm 2} = - \gPs^{\sm 2} \, \gX_2~, 
\qquad 
\gd_2 \gL^{\ga} = \gd_2 \gL^3 = \gd_2 \gL^{\!\sm 1} = 0~, 
}
with $Q_2 = (\frac 12^2,0, 0^{13}; 0, \sm 1)$. 
We see that the coefficients of these transformations can be uniquely identified with the various twisted states with oscillators with one exception: In the first twisted sector we have 
\equ{
 \rep{2}_R(\rep{1},\rep{2})_{0, \sm \frac38} 
 \leftrightarrow  \gb_1{}^A{}_a~, 
\quad 
(\rep{1},\rep{1})_{0, \frac38} 
 \leftrightarrow  \gb_1~,
\quad
\rep{3}_R(\rep{1},\rep{1})_{0, \frac38} 
\leftrightarrow \gb_{1\,ab}~, 
}
and $\rep{2}_R(\rep{1},\rep{2})_{0, 0} \leftrightarrow \gb_2{}^a{}_b$ in the second twisted sector. There is only one state we do not account for here: The reason for this is that the two states with the oscillators $\tga_{-1/2}^3$ and $\tga_{-1/2}^\undr{3}$ in the first twisted sector are distinguished only by {\em our} choice of complex structure of the coordinate $z^3$.

Each of the states in Table \ref{tb:TwistedStatesZ4} can individually be used as a blow-up mode. If we consider combinations of blow-up modes, the mixed anomalies have to vanish. For example, they cancel 
%
when we use the first twisted state $(\rep{26},\rep{1})_{-\frac 12,-\frac 18}$ and a multiple of the first twisted states with oscillators as blow-up modes simultaneously.

If we take one blow-up mode from the first and one from the second twisted sector then the mixed anomaly constraint reads 
\equ{
Q_1 \cdot Q_2 = q_1 \cdot q_2 = \frac 14~. 
\label{MixedZ4Anom}
}
(One can check this condition and $Q_1^2 =11/8$ and $Q_2 = 3/2$ are equivalent to the constraints found in ref~\cite{Nibbelink:2007pn} provided one makes the change of notation 
$Q_1 = \frac 12 V_1 + \frac 14 V_2, Q_2 = \frac 12 V_2$, 
where $V_1$ and $V_2$ were used in that reference.) This condition is fulfilled if we take any of the states from the first twisted sector except $(\rep{1}, \rep{1})_{1,-\frac 18}$ and the state $(\rep{26},\rep{1})_{-\frac 12,\frac 14}$ or $\rep{2}_R(\rep{1},\rep{1})_{0,0}$ from the second twisted sector. One extra constraint arises when one uses both $\rep{26}$'s from the first and second twisted sector simultaneously: The $\pm 1$ entries defining this representation should be different. This means that different components of these $\rep{26}$'s should take VEVs.

Finally, we consider the issue of projected out twisted states in the context of resolutions of this heterotic $\Intr_4$ orbifold model. We argue that it is impossible to associate a GLSM with the second twisted states that got projected out from the four dimensional spectrum. These states are marked with an ''x" in Table \ref{tb:TwistedStatesZ4}: 
First of all, the state $\crep{2}_R(\rep{1},\rep{2})_{0, 0}$, 
or explicitly 
\(
\tga_{-\frac 12}^\ua 
\big| p_2 \ \big( \undr{\frac 12\ \sm\frac 12}\ 0\ 0^{13} \big) \big\rangle,
\)
cannot be associated with a GLSM, as it would induce a non-holomorphic fermionic gauging 
\(
\gd_2 \gL^A = \gb_2{}^{A\, a} \bgPs_a \, \gX_2, 
\) 
which is inadmissible by (2,0) supersymmetry. Furthermore, none of the other projected out states 
$(\rep{26},\rep{1})_{\frac 12, \sm\frac 14}$, 
$(\rep{1},\rep{1})_{0, \frac34}$ and 
$(\rep{1},\rep{1})_{\sm1, \sm\frac34}$ 
can ever satisfy the mixed anomaly cancellation condition \eqref{MixedZ4Anom} with any of the charges $Q_1$ associated with the states in the first twisted sector. Hence whenever we have switched on a first twisted state as blow-up and included its effect in the GLSM, these projected out states cannot be consistently associated to it.

\subsection{Resolutions of the $\boldsymbol{\Cplx^3/\ztwo}$ orbifolds}
\label{sc:Z2Z2top}

Our final example involves resolutions of the orbifold $\Cplx^3/\Intr_2\!\times\!\Intr_2$. The orbifold group contains four elements, apart from the identity we have 
\equ{
\gth_a ~:~ 
\left\{ 
\arry{ll}{
z^a \ra z^a~,
\\[1ex] 
z^b \ra - z^b~, & b \neq a~,
}
\right. 
}
for $a=1,2,3$. There exist six heterotic orbifold models distinguished by the choice of gauge shifts $V_a$, $a=1,2,3$.

\begin{table}
\[
\arry{| c || c | }{
\hline 
\text{triangulation} & \text{condition(s)} 
\\ \hline\hline 
``E_1" & b_1 > b_2 + b_3 
\\ 
``E_2" & b_2 > b_1 + b_3 
\\ 
``E_3" & b_3 > b_1 + b_2 
\\ \hline 
 & 
b_1 < b_2 + b_3 
\\ 
``S" & b_2 < b_1 + b_3 
\\ 
 & b_3 < b_1 + b_2 
\\ \hline 
} 
\qquad \qquad \qquad \qquad 
\raisebox{-12ex}{\scalebox{0.45}{\mbox{\input{triangZ22}}}}
\] 
\captn{
The values of the K\"ahler parameters that define the four phases that distinguish the four triangulations ``$E_1$",  ``$E_2$",  ``$E_3$" and ``$S$" of the completely resolved $\Cplx^3/\Intr_2\!\times\!\Intr_2$ orbifold, i.e.\ with $b_1, b_2, b_3 > 0$. 
The picture gives a sketch of these four regions defining $x = b_1/b$ and $y=b_2/b$ with $b = b_1+b_2+b_3$. (Outside the large triangle at least one of the K\"ahler parameters is negative and the resulting phase does not correspond to a triangulation of smooth toric variety.) 
\label{fg:TriangulationsZ22}}
\end{table}

Before we investigate the GLSMs that can be associated to this orbifold, we briefly recall the resolution of this orbifold in the language of toric geometry. The toric resolution of this orbifold has three ordinary divisors $D_a$ and three exceptional ones $E_r$. The resolution is not unique in toric geometry; there are four topologically distinct complete resolutions possible, corresponding to the four different triangulations of the toric diagram. In \cite{Nibbelink:2007pn} we referred to these triangulations as the ``$E_1$", ``$E_2$", ``$E_3$" and ``$S$" triangulations. In the triangulation ``$E_1$" the curve $D_1 E_1$ exists but the curve $E_2 E_3$ does not. The triangulations ``$E_2$" and ``$E_3$" have similar properties, while in the ``$S$" triangulation all the curves $E_1 E_2$, $E_2 E_3$ and $E_1 E_3$ exist, but none of the curves $D_r E_r$. In figure \ref{fg:TriangulationsZ22} we have given the conditions on the K\"ahler parameters that distinguishes these four triangulations, see e.g.\ \cite{Blaszczyk:2010db}.

\subsubsection{GLSM description of topological transitions}

These different topologies can be easily understood as some of the phases of the associated GLSM. For this we do not even need to specify exactly which twisted states take VEVs, as long as a VEV is switched on in each twisted sector. We now describe this in detail without requiring any detailed knowledge of toric geometry. (Of course all the information discussed here can be efficiently be read of from the toric fan associated with the resolution.)

The $\ztwo$ orbifold models have three distinct twisted sectors, with shifted right-moving momenta 
$p_1=(0,\sfrac 12,\sfrac 12)$, 
$p_2=(\sfrac 12,0,\sfrac 12)$, 
$p_3=(\sfrac 12,\sfrac 12,0)$,
respectively. Hence, in addition to the chiral superfields $\gPs^a$, $a=0,1,2,3$, any resolution model contains three exceptional chiral superfields $\gPs^{\sm r}$, $r=1,2,3$. We have the following charge assignment: 
\[
\arry{|c|| c c c c || c c c |}{
\hline 
\text{superfield} & \gPs^{0} & \gPs^{1} & \gPs^2 & \gPs^3 & \gPs^{\sm 1} & \gPs^{\sm 2} & \gPs^{\sm 3} 
\\ \hline\hline 
q_1 & 0 & 0 & 1/2 & 1/2 & -1 & 0 & 0 
\\ 
q_2 & 0 & 1/2 & 0 & 1/2 & 0 & -1 & 0 
\\
q_3 & 0 & 1/2 & 1/2 & 0 & 0 & 0 & -1
\\ \hline 
}
\]
Switching on a VEV of a state in each twisted sector, leads to the potential 
\equ{
V = 
\frac {e_1^2}2\, 
\Big( \sfrac{ |z^2|^2 + |z^3|^2}2  \!-\! |x^1|^2 \!-\! b_1 \Big)^2
+ 
\frac {e_2^2}2\, 
\Big( \sfrac{|z^1|^2 + |z^3|^2}2  \!-\! |x^2|^2 \!-\! b_2 \Big)^2
+ 
\frac {e_3^2}2\, 
\Big( \sfrac {|z^1|^2 + |z^2|^2}2  \!-\! |x^3|^2 \!-\! b_3 \Big)^2~, 
\label{DtermPotZ22} 
}
for the scalars $z^a$ and $x^r$ of the chiral superfields $\gPs^a$ and $\gPs^{\sm r}$, respectively. The divisors $D_a$ and $E_r$ are defined by the equation $D_a = \{z^a = 0\}$ and $E_r = \{x^r = 0\}$. By studying the phases defined by this potential we can get a lot of information about the topology of the corresponding geometries.

\begin{table}[]
\[
\arry{| l || c |}{
\hline 
\multicolumn{1}{|c||}{\text{divisor}} & \multicolumn{1}{|c|}{\text{exists when}} 
\\ \hline\hline 
D_1 & \text{always}
\\
D_2 & \text{always}
\\
D_3 & \text{always}
\\ \hline\hline  
E_1 & b_1 \geq 0 
\\
E_2 & b_2 \geq 0 
\\
E_3 & b_3 \geq 0 
\\ \hline 
}
\,
\arry{| l || r l |}{
\hline 
\multicolumn{1}{|c||}{\text{curve}} & \multicolumn{2}{|c|}{\text{exists when}} 
\\ \hline\hline 
E_1 E_2 & b_1, b_2 \geq 0~, & b_3 \leq b_1+b_2 
\\
E_2 E_3 & b_2, b_3 \geq 0~, & b_1 \leq b_2+b_3 
\\
E_1 E_3 & b_1, b_3 \geq 0~, & b_2 \leq b_1+b_3 
\\ \hline\hline 
D_1 D_2 & b_3 \leq 0 ~~ & 
\\ 
D_1 D_3 & b_2 \leq 0 ~~ & 
\\ 
D_2 D_3 & b_1 \leq 0 ~~ & 
\\ \hline 
}
\,
\arry{| l || r l |}{
\hline 
\multicolumn{1}{|c||}{\text{curve}} & \multicolumn{2}{|c|}{\text{exists when}} 
\\ \hline\hline 
D_1 E_1 & b_1 \geq 0~, & b_1 \geq b_2+b_3 
\\
D_2 E_2 & b_2 \geq 0~, & b_2 \geq b_1+b_3 
\\
D_3 E_3 & b_3 \geq 0~, & b_3 \geq b_1+b_2 
\\ \hline\hline 
D_1 E_{2,3} & b_2 \geq 0~, &  b_3 \geq 0 
\\ 
D_2 E_{1,3} & b_1 \geq 0~, & b_3 \geq 0 
\\ 
D_3 E_{1,2} & b_1 \geq 0~, & b_2 \geq 0 
\\ \hline 
}
\]
\captn{
This Table indicates under which restrictions of the K\"ahler parameters $b_r$ the various divisors and curves exist. 
\label{tb:CurveExistenceZ22}}
\end{table}

%

We start by investigating the existence of possible divisors depending on the K\"ahler parameters $b_r$. The divisors $D_a$ exist for any value of the K\"ahler parameters, as one can always find values of the other coordinates such that the potential vanishes identically. Contrary, the divisor $E_r$ only exists when $b_r \geq 0$: When $b_r < 0$ the square in which it occurs only contains positive or non-negative terms, so that it is impossible to make the potential to vanish. In order that we can talk about a smooth geometry, we need that all exceptional divisors exist, hence all three K\"ahler parameters $b_1,b_2, b_3 \geq 0$.

For curves the situation becomes more complicated, because there are many possible curves inside the geometry. In Table~\ref{tb:CurveExistenceZ22} we indicate under which conditions curves, that are intersection of two distinct divisors, exist. Here we only illustrate the principle by considering the curve $D_1 E_1$. By setting $z^1= x^1 = 0$ the potential \eqref{DtermPotZ22} reduces to 
\equ{ 
V\Big|_{D_1E_1} = 
\frac {e_1^2}2\, 
\Big( \sfrac{ |z^2|^2 + |z^3|^2}2 \!-\! b_1 \Big)^2
+ 
\frac {e_2^2}2\, 
\Big( \sfrac{|z^3|^2}2  \!-\! |x^2|^2 \!-\! b_2 \Big)^2
+ 
\frac {e_3^2}2\, 
\Big( \sfrac {|z^2|^2}2  \!-\! |x^3|^2 \!-\! b_3 \Big)^2~. 
}
For this potential to vanish, the terms in each of the three squares need to vanish separately. The first square can only vanish when $b_1 \geq 0$. By adding the terms in the second and third squares and subtracting the terms in the first square we find an additional condition on the K\"ahler parameters
\equ{
0 \leq |x^2|^2 + |x^3|^2 = b_1 - b_2 - b_3~.
} 
Hence the curve $D_1E_1$ exists when $b_1 \geq 0$ and $b_1\geq b_2+b_3$. These conditions are satisfied even when $b_2$ or $b_3$ are negative, i.e.\ when a smooth description does not apply. For a smooth geometry all K\"ahler parameters are positive, hence the first condition is then implied by the second. Using similar arguments one can confirm the results summarized Table~\ref{tb:CurveExistenceZ22}.

\begin{table}[]
\[
\arry{ccc}{
\arry{c c}{
\hline 
\multicolumn{2}{|c|}{\text{orbifold phase:}} \\ 
\multicolumn{2}{|c|}{\text{\small no exceptional divisors}}
\\ \hline \\[-2ex] 
\arry{c}{ b_1 \leq 0 \\ b_2 \leq 0 \\ b_3 \leq 0}
&  
\raisebox{-7ex}{\scalebox{0.25}{\mbox{\input{Z22orbi}}}}
\\ \\  \hline 
\multicolumn{2}{|c|}{\text{partial resolution:}} \\ 
\multicolumn{2}{|c|}{\text{\small one exceptional divisor}}
\\ \hline \\[-2ex] 
\arry{c}{ b_1 \geq 0 \\ b_2 \leq 0 \\ b_3 \leq 0}
&  
\raisebox{-7ex}{\scalebox{0.25}{\mbox{\input{Z22single}}}}
}
&\qquad & 
\arry{c c}{
\hline 
\multicolumn{2}{|c|}{\text{partial resolution:}} \\ 
\multicolumn{2}{|c|}{\text{\small two exceptional divisors}}
\\ \hline \\[-2ex] 
\arry{c}{ b_1 \geq b_2 \geq 0 \\[1ex] b_3 \leq 0}
&  
\raisebox{-7ex}{\scalebox{0.25}{\mbox{\input{Z22double12}}}}
\\ \\ 
\arry{c}{ b_2 \geq b_1 \geq 0 \\[1ex] b_3 \leq 0}
&  
\raisebox{-7ex}{\scalebox{0.25}{\mbox{\input{Z22double21}}}}
\\[13ex]
}
\\ \\ 
\hline 
\multicolumn{3}{|c|}{\text{full resolution: {\small three exceptional divisors}}} 
\\ \hline \\[-2ex] 
\arry{cc}{
\arry{c}{ b_1,~b_2 \geq 0 \\[1ex] b_3 \geq b_1+b_2}
&  
\raisebox{-7ex}{\scalebox{0.25}{\mbox{\input{Z22fullE3}}}}
\\ \\ 
\arry{c}{ b_1,~b_3 \geq 0 \\[1ex] b_2 \geq b_1+b_3}
&
\raisebox{-7ex}{\scalebox{0.25}{\mbox{\input{Z22fullE2}}}}
} & &
\arry{cc}{
\arry{c}{ b_2,~b_3 \geq 0 \\[1ex] b_1 \geq b_2+b_3}
& 
\raisebox{-7ex}{\scalebox{0.25}{\mbox{\input{Z22fullE1}}}}
\\ \\ 
\arry{c}{
b_1+b_2 \geq b_3 \geq 0 \\ 
b_1+b_3 \geq b_2 \geq 0 \\ 
b_2+b_3 \geq b_1 \geq 0
}
&  
\raisebox{-7ex}{\scalebox{0.25}{\mbox{\input{Z22fullS}}}}
}
}
\]
\captn{This table gives the 14 phases of the GLSM associated with the $\Cplx^3/\ztwo$ orbifold. We have not displayed all the partial resolutions: The others are obtained by cyclic permute of the labels 1,2,3.
\label{tb:Z22Phases}}
\end{table}

Not only the existence of divisors and curves can be determined from the scalar potential \eqref{DtermPotZ22}, it can even be used to compute the intersection numbers of three distinct divisors in a given phase: The existence of a given intersection is determined in the same way as the existence of a divisor or a curve: It exists when the potential restricted to this intersection can vanish. Suppose that this is possible, then the intersection number is obtained by the following considerations: If the vacuum is unique, then the intersection number equals unity. Alternatively, if there is a residual $\Intr_2$ gauge symmetry, then the intersection equals $1/2$. Consider the partial resolution with $b_1 > b_2 >0 > b_3$  as an illustration: As the corresponding toric diagram suggests, is the intersection $D_1E_1E_2 = 1$. Indeed the potential \eqref{DtermPotZ22} restricted to $D_1E_1E_2$ has as vacuum solution 
\equ{
\langle z_3 \rangle = \sqrt{2 b_2}~, 
\qquad 
\langle z_2 \rangle = \sqrt{2(b_1-b_2)}~, 
\qquad
\langle x_3 \rangle = \sqrt{b_1-b_2-b_3}~, 
}
on which no residual discrete gauge transformation acts. On the contrary, the vacuum solution of the potential \eqref{DtermPotZ22} restricted to $D_1D_2E_1$ 
\equ{ 
\langle z_3 \rangle = \sqrt{2 b_2}~, 
\qquad 
\langle x_2 \rangle = \sqrt{b_1-b_2}~, 
\qquad
\langle x_3 \rangle = \sqrt{-b_3}~, 
}
has a residual $\Intr_2$ gauge symmetry: The gauge symmetry $q_3$ does not act on $z_3$ hence fixing $x_2$ and $x_3$ only fixes this $U(1)$ gauge symmetry up to a sign, the $\Intr_2$ phase. Consequently, the intersection number $D_1D_2E_1$ equals $1/2$.

These considerations combined imply that the GLSM associated to the $\Cplx^3/\ztwo$ orbifold has 14 phases in total. The distinctions between the different phases can be visualized by the toric diagrams of the (partially) resolved $\ztwo$ singularity. This information has been collected in Table \ref{tb:Z22Phases}. The orbifold phase has all K\"ahler parameters negative; when we take them to tend to $-\infty$ we recover the orbifold CFT. Other regions in parameter space, where we have some computational controle have all K\"ahler parameters positive and large. In this case we find ourselves in a supergravity regime and we can analyze the theory using geometrical means. But as noted at the beginning of this Section, this geometry is not unique, but depends on the triangulation of the toric diagram. In previous works \cite{Nibbelink:2007xq,Blaszczyk:2009in,Blaszczyk:2010db} we have tried to go from the orbifold regime directly to the regimes where supergravity can be trusted. From the results found here, we conclude that these analyses only took into account five of the 14 possible phases. The orbifold phase and the four geometrical phases are connected to each other only by a single point:  $b_1=b_2=b_3=0$. The other phases have neither a fully geometrical interpretation nor are describable as an exactly solvable CFT.

\subsubsection{GLSM anomalies versus Bianchi identities}

One can consider heterotic supergravity on one of the toric resolutions. The fundamental consistency relations in geometrical compactification are the Bianchi identities. For resolutions of the $\Cplx^3/\Intr_2\!\times\!\Intr_2$ orbifold these conditions strongly depend on the triangulation \cite{Nibbelink:2007pn}, because the intersection numbers do. As our general discussion showed, from the worldsheet perspective the role of the Bianchi identities is played by the anomaly cancellation conditions of the GLSM. Let us investigate their relation in more detail:

When we use a single blow-up mode in each of the three twisted sectors (without oscillators), the anomaly cancellation conditions read
\equ{
Q_1^2 = Q_2^2 = Q_3^2 = \frac 32~, 
\qquad 
Q_1\cdot Q_2 = Q_2\cdot Q_3 = Q_3 \cdot Q_1 = \frac 14~,
\label{anomaliesZ22}
}
where $Q_1$, $Q_2$ and $Q_3$ are the charges of the chiral-Fermi multiplets of the GLSM.

It is instructive to compare this with the Bianchi identities on the three exceptional divisors $E_1$, $E_2$ and $E_3$ within the geometries defined by the four different triangulations. In the symmetric triangulation ``$S$" we find 
\equ{
Q_1^2 + 2\, Q_2\cdot Q_3 = 2~,
\qquad 
Q_2^2 + 2\, Q_1\cdot Q_3 = 2~, 
\qquad 
Q_3^2 + 2\, Q_1\cdot Q_2 = 2~,
}
while in asymmetric triangulation ``$E_1$" we obtain 
\equ{
Q_2^2 + Q_3^2 = 3~, 
\qquad 
Q_2^2 - 2\, Q_1\cdot Q_3 = 1~, 
\qquad 
Q_3^2 - 2\, Q_1\cdot Q_2 = 1~. 
}
(The results in the other two asymmetric triangulations are obtained by cyclic permutations of the labels.)  Each of these sets of conditions separately are weaker than \eqref{anomaliesZ22}. However, when we combine all four sets of Bianchi identities, we have a set of equations that are equivalent to the anomaly cancellation conditions \eqref{anomaliesZ22} of the GLSM.

The fact that the anomaly conditions \eqref{anomaliesZ22} of the GLSM contains all the possible Bianchi identities of the supergravity models on the four resolutions, should not come as a surprise: The GLSM formalism allows us to smoothly move in moduli space between the various phases, including the four geometrical phases described by these triangulations. Therefore a consistent string model should at least produce consistent supergravity models in each of these full resolutions. Even from the supergravity perspective alone, one can argue that the various Bianchi identities that arise in the different triangulations should be superimposed: Even if a flop-transition itself is beyond the range of validity of supergravity, in each of the resolutions associated with the different triangulations supergravity should be valid. Since neither the flux nor any of the exceptional divisors have disappeared during the flop transition, the Bianchi identities on the different resolutions have to be imposed on the same gauge flux.

In ref.\ \cite{Nibbelink:2007pn} line bundle models on orbifold resolutions were also investigated (using the parameterization $V_i = -Q_i/2$ there). In addition to the Bianchi identities within a certain triangulation, also three Bianchi identities, 
\equ{
Q_1^2 = Q_2^2 = Q_3^2 = \frac 32~, 
}
were enforce on the three resolutions of $C^2/\Intr_2$ contained in any $C^3/\ztwo$  resolution. This gave in total a maximum of six conditions. However only for resolution ``$S$" these six conditions are all independent and equivalent to the GLSM anomaly cancellation \eqref{anomaliesZ22}. For the asymmetric triangulations this procedure leaves one condition out.

In the same paper\ \cite{Nibbelink:2007pn} it was argued that none of the asymmetric triangulations can be used for model building purposes, as the flux quantization conditions and the Bianchi identities are incompatible. On non-compact spaces it is not strictly necessary to satisfy all flux quantization conditions. On resolutions of the compact orbifold $T^6/\ztwo$ it was shown that the flux quantization conditions are automatically satisfied, when the fluxes are identified in the proper way from the orbifold shifts and Wilson lines \cite{Blaszczyk:2010db}. Since we identify the superfield charges with shifted momenta up to integers, our formalism also automatically ensure that the relevant quantization conditions are fulfilled.

\section{Charged spectrum} 
\label{sc:Spectrum}

\subsection{Marginal kinetic deformations}

An important and difficult question is how to determine the charged spectrum of a resolution model described by a GLSM directly. There are two regimes where computational methods are available: the orbifold point and large volume limits. At the orbifold point in moduli space the theory is described by a free CFT, so that the complete spectrum can be obtained. In section \ref{sc:Orbifolds} we reviewed how the massless spectrum can be computed. Also in the large volume regime, where all cycles are large, one can use cohomology methods and index theorems to compute the spectrum. These methods are essentially field theoretical in nature. These methods might therefore not take all string effects into account and are not obviously applicable in phases other than the large volume regimes.

In this section we would like to propose a method to compute the charge massless spectrum which is intrinsically stringy, i.e.\ that uses properties of the worldsheet theory. The idea behind this method is motivated by how one can identify the massless states that the heterotic string describes in ten dimensions. Recall that the free action \eqref{FreeHetAc} can be extended with various marginal deformations giving a formulation of the heterotic string in a general background. Each of these deformations corresponds to some states in the target space theory. E.g.\ the kinetic terms of the worldsheet coordinate fields define the metric and the Kalb-Ramond field, while marginal deformations of the kinetic terms of the left-moving fermions gives rise to the gauge fields. Inspired by this we investigate the marginal deformations of GLSMs.

A certain part of such deformations have already been identified. First of all, the FI-parameters for the various bosonic gaugings are holomorphic functions of the chiral superfields, hence correspond to massless states in the target space theory, which may be interpreted as complexified axions. (In this work we ignore possible non--perturbatively generated superpotentials by e.g.\ worldsheet instantons.) In addition, when we consider twisted states with oscillator excitations as blow-up modes, we were forced to introduce fermionic gaugings \eqref{gXgauging} on the worldsheet. The coefficients in these fermionic gauge transformations are again holomorphic and could be identified with the twisted states with oscillator modes. Therefore, we consider here only marginal deformations of the kinetic terms of the chiral and chiral-Fermi superfields in the GLSM.

To this end we start with a generic GLSM, and consider all possible kinetic deformations of it. This produces a gauged non-linear sigma model (GNLSM). Insisting that these deformations respect (2,0) supersymmetry, the worldsheet action\footnote{Here we made the assumption that there are no fermionic gaugings; including fermionic gaugings requires the introduction of the Fermi-gauge superfields $\gS_r$ in the chiral-Fermi multiplet kinetic terms, see \eqref{(2,0)kinactions} of Appendix \ref{sc:(2,0)actions}.}
\equ{
S_{GNLSM} =  \int\d^2\gs\d^2\gth^+\, 
\Big\{ 
\frac i4\,\Big(
k_a\, \bcD \gPs^a - \bcD\, \bgPs_a\, \bk^a
\Big)
-\frac 12\, N^\ga{}_\gb\, \bgL_\ga \gL^\gb 
- \frac 14\, n^{\ga\gb}\, \bgL_\ga \bgL_\gb 
- \frac 14\, \bn_{\ga\gb}\, \gL^\ga \gL^\gb 
\Big\} 
} 
is encoded in a few functions $k_a$, $\bk^a$, $N^\ga{}_\gb$, $n^{\ga\gb}$ and $\bn_{\ga\gb}$ of the chiral superfields $\gPs^a$ and their conjugates $\bgPs_a$.  In equations \eqref{SgPs} and \eqref{SgL} of  Appendix \ref{sc:(2,0)actions} we give the full component worldsheet action for the non-gauged non-linear sigma model (NLSM) up to auxiliary field contributions.

To determine the massless states in the four dimensional target space theory one considers deformations that are generic functions of $\gPs^0$ and $\bgPs_0$. In analogy to ten dimensions, the deformations that are only functions of these coordinates only constitute the four dimensional metric, Kalb-Ramond field and the massless gauge fields in four dimensions. There are also deformations possible that involve the internal chiral superfields $\gPs^a$ and $\bgPs_a$, $a\neq 0$. One might think that any deformation, that depends on the internal coordinates, gives rise to states that are massive in the target space theory. However, this is not always the case: Massless state can still arise because some of these internal coordinates take a non-vanishing VEV in certain phases of the GLSM. The analysis of the spectrum is therefore strongly dependent on the phase of the GLSM one is considering.

This method does not capture all the massless states in all phases. In partially resolved (or hybrid) phases, where the VEVs do not specify the vacuum uniquely, the corresponding twisted sectors have to be added by hand. Fortunately, by standard orbifold CFT techniques they can be computed. Hence in particular in the orbifold phase, the deformations only give us the untwisted sector.

Since we are primarily interested in the charged target space spectrum we describe this expansion for the deformations of the kinetic terms of the chiral-Fermi multiplets in more detail. From \eqref{SgL} and \eqref{gaugeComp} in Appendix \ref{sc:(2,0)actions} we read off that the scalars of the four dimensional chiral multiplets are identified by 
\equ{ 
N^\ga{}_{\gb,a}~, 
\qquad 
n^{\ga\gb}{}_{,a}~, 
\qquad 
\bn_{\ga\gb}{}_{,a}~, 
\label{perturbations} 
}
with $a \neq 0$ in leading order, that do not vanish in the phase in question. (As observed above $a=0$ corresponds to the massless four dimensional gauge fields.) Since we would like to consider only the massless modes, we ignore any contributions in these functions that depend on $\bgPs_a\gPs^a$ or fully gauge invariant (anti-)holomorphic combinations. Because of worldsheet gauge anomaly cancellation the charges of the Fermi multiplets are bounded.  Consequently so are the possible charges of the functions $N^\ga{}_\gb$ and $n^{\ga\gb}$. By restricting only to a derivative w.r.t.\ the holomorphic coordinates $z^a$, we have selected a given four dimensional helicity. This allows us to identify holomorphic derivatives of these functions with the scalar components of chiral multiplets in the effective four dimensional theory (anti--holomorphic derivatives are associated to their complex conjugates). To summarize, not from the functions $N^\ga{}_\gb$ and $n^{\ga\gb}$ themselves, but from their holomorphic derivatives we can read off the four dimensional spectrum of chiral multiplets.

It may happen that a state might be present in a resolved phase, while it seems to be absent in the orbifold phase. In this case this state can be interpreted as a twisted state which remained massless in the resolution process: In the orbifold phase it is not really absent, but has to be included to the spectrum via the twisted worldsheet boundary conditions. When a state exists both in the orbifold and resolution phases, it should be thought of as an untwisted state. In the examples discussed in the next Subsection we encounter these effects explicitly.

\begin{table}[]
\[
\renewcommand{\arraystretch}{1.3}
\arry{| c || c | c | c | c | c | c | c | c | }{
\hline 
Q & 1/3 & \sm 1/3 & 2/3 & \sm 2/3 & 1  & \sm 1  & 4/3 & \sm 4/3 
\\\hline\hline
\text{Monomial} & 
\gPs^i & \!\gPs^i \gPs^j \gPs^{\sm 1}\! & 
\gPs^i \gPs^j & \gPs^i \gPs^{\sm 1} &  
\gPs^i \gPs^j \gPs^k & \gPs^{\sm 1} &
 \bgPs_{\sm 1} \gPs^i  & \bgPs_i \gPs^{\sm 1}
\\ \hline 
\langle z^{\sm 1} \rangle \neq 0 & 
\checkmark  & \text{x} & 
\text{x} & \checkmark & 
\text{x} & \text{x} & 
\checkmark & \text{x}
\\\hline 
\langle z^l \rangle \neq 0 & 
\hspace{3ex}\checkmark\hspace{4ex} &  
\hspace{3ex}\text{x}\hspace{4ex} &
\hspace{3ex}\text{x}\hspace{4ex} & 
\hspace{3ex}\checkmark\hspace{4ex} & 
\hspace{3ex}\text{x}\hspace{4ex} & 
\hspace{3ex}\checkmark\hspace{4ex} & 
\hspace{3ex}\text{x}\hspace{4ex} & 
\hspace{3ex}\checkmark\hspace{4ex} 
\\\hline 
\text{Type} & 
\text{U} & - & 
- & \text{U} & 
- & \text{T} & 
\text{U'} & \text{T'}
\\\hline\hline 
\text{Multiplicity} &
1/9 & - &
- & 1/9 & 
- & 1 & 
1/9 & 26/9 
\\\hline 
}
\renewcommand{\arraystretch}{1}
\]
\captn{
Possible monomials that can build up the functions $N^\ga{}_\gb$, $n^{\ga\gb}$ and $\bn_{\ga\gb}$. The third row indicates whether a monomial contributes to massless states in the orbifold phase with $\langle z^{\sm 1}\rangle \neq 0$.
 The fourth row whether a monomial contributes to the massless spectrum in at least one of the three resolution patches with $\langle z^l \rangle \neq 0$. 
Based on this one can interpret the corresponding state as an untwisted (U) or twisted (T) state as displayed in the fifth row. (With U' we indicate that the monomial gives an untwisted contribution in the orbifold phase, while with T' twisted contributions that partially compensate the untwisted states.) The final row gives the resulting multiplicity. 
\label{tb:Monomials}
}
\end{table}

\subsection{Example: spectra of $\Intr_3$ orbifold resolutions}
\label{sc:Z3spectra}

We now illustrate how one can determine the spectrum of a GLSM which describes a resolution of an orbifold model. Concretely, we consider again the non-compact orbifold $\Cplx^3/\Intr_3$ as in Subsection \ref{sc:Z3resolutions}. Because of worldsheet gauge anomaly cancellation the non-vanishing charges of Fermi multiplet bilinears $\bgL_\ga \gL^\gb$, $\bgL_\ga\bgL_\gb$ and $\gL^\ga\gL^\gb$ lie between $-4/3$ and $4/3$. It follows that there are eight monomials out of which the functions  $N^\ga{}_\gb$, $n^{\ga\gb}$ and $\bn_{\ga\gb}$ can be build, which are given in Table \ref{tb:Monomials}.

As Table \ref{tb:Monomials} contains important information, let us briefly explain how to read it. The eight monomials are unique by the following requirements: 1) They involve at least a single holomorphic superfield $\gPs^a$, and  2) they do not involve any gauge invariant factors, like $\bgPs_a \gPs^a$ (not necessarily summed over $a$) or $\gPs^1\gPs^2\gPs^3 \gPs^{\sm 1}$, etc. In order that one of the combinations \eqref{perturbations} corresponds to a massless state in a given phase, it should not vanish when the corresponding VEV(s) are inserted. The $\checkmark$'s and $\text{x}$'s specify precisely whether the monomial exists or vanishes in that phase. The orbifold phase has a $\Intr_3$-degenerate vacuum with $\langle z^{\sm 1} \rangle \neq 0$, hence in this phase we can only identify untwisted states. These are marked as $\text{U}$ in Table \ref{tb:Monomials}. Since the monomials corresponding to these untwisted states always involve $\gPs^i$, they have a three-fold degeneracy resulting in a multiplicity $3/27 = 1/9$. As we can see from this Table most untwisted states survive also in the resolved phases with $\langle z^l\rangle \neq 0$. In addition, derivatives of some other monomials do not vanish. These novel contributions can be interpreted to represent the twisted states that remain present in the resolution phases. The monomial $\gPs^{\sm 1}$ gives rise to a single twisted state. The monomial $\bgPs_i \gPs^{\sm 1}$ gives rise to three resolutions states, i.e.\ their multiplicity is 3. However, when going from the orbifold phase to one of the resolution patches, the untwisted deformation corresponding to the monomial $\bgPs_{\sm 1} \gPs^i$ (denoted by $\text{U'}$) is removed, hence the net multiplicity is $3-3/27 = 26/9$. (We refer to this contribution as $\text{T'}$.)

We have applied this prescription to compute the resolution spectrum of each of the seven line bundle resolution models described in \cite{Nibbelink:2007rd} and found agreement in each case. We display the analysis here only for the following two cases:

\subsubsection{A $\boldsymbol{\Intr_3}$ orbifold resolution model with 
$\boldsymbol{n=4}$}

First consider the resolution of the orbifold model ($n=4$ see \eqref{Z3class}) with twist and shift given by 
\equ{
v = \big(0, \frac 13^2,-\frac 23\big)~, 
\qquad 
V = \big(\frac 13^{8}, - \frac 23^4,0^4\big)~. 
}
The corresponding orbifold model has gauge group $U(12)\times SO(8)$ and spectrum
\equ{ 
\frac 1{27} \Big[ \rep{3_R}(\rep{12},\rep{8})_1 + \rep{3_R}(\crep{66},\rep{1})_{\sm 2} \Big] +  (\rep{1},\rep{1})_4 +(\rep{8_+},\rep{1})_{\sm 2}~. 
} 
Hence we can choose either the singlet $(\rep{1},\rep{1})_4$ or the spinor representation $(\rep{8_+},\rep{1})_{\sm 2}$ of $SO(8)$ as a blow-up mode. If we take the singlet, we have $Q=(\frac 13^{12},0^4)$, and we obtain a GLSM with the following charge assignment
\equ{
\arry{c|| c|c|c|| c|c}{
\text{Field} & \gPs^0 & \gPs^i & \gPs^{\sm 1} & \gL^I  & \gL^\ga 
\\ \hline 
q, Q & 0 & \frac 13 & -1 & \frac 13 & 0
}
\label{chargesZ34} 
}
with $i=1,2,3$, $I = 1,\ldots, 12$ and $\ga = 1,\ldots, 4$.

We can make the following expansion of the kinetic terms of the chiral-Fermi multiplets: 
\equ{
S_\text{Fermi} = - \frac 14 \, \int \d^2\gs\d^2 \gth^+\, 
\Big\{
N^I{}_J\, \bgL_I \gL^J + N^\ga{}_\gb\, \bgL_\ga \gL^\gb  
+ n^{\ga\gb}\, \bgL_\ga \bgL_\gb 
\non \\[1ex] 
+ N^I{}_\ga\, \bgL_I \gL^\ga
+ n^{IJ}\, \bgL_I \bgL_J 
 + n^{I\ga}\, \bgL_I \bgL_\ga 
\Big\} + \text{h.c.}~.  
\label{FermiSinglet}
}
The functions $N^I{}_J$, $N^\ga{}_\gb$, $n^{\ga\gb}$ and $\bn_{\ga\gb}$ are all neutral and therefore correspond to the gauge sector of the theory. We read off that the resolution gauge group is $U(12)\times SO(8)$; the same as on the orbifold.

Given the charge assignment in equation \eqref{chargesZ34}, we infer that the remaining functions $N,n$ carry non-vanishing charges 
\equ{
\arry{c|| 
c|c| c|c| c|c}{
\text{Function} & 
N^I{}_\ga & N^\ga{}_I & 
n^{IJ} & \bn_{IJ} & 
n^{I\ga} & \bn_{I\ga} 
\\ \hline 
Q & 
\frac 13 & - \frac 13 & 
\frac 23 & - \frac 23 & 
\frac 13 & - \frac 13 
\\ \hline 
\text{Type} & 
~~\text{U}  & - &
- & ~~\text{U} &
\text{U} & - 
}
}
We do not give the monomials explicitly as the charges of the monomials determine them uniquely, see Table \ref{tb:Monomials}.  
From this result we can immediately read off the charged spectrum in the resolution phase: The function $\bn_{IJ}$ is anti-symmetric in $I$ and $J$, hence it gives the untwisted state $(\crep{66},\rep{1})$. The functions $N^I{}_\ga$ and $n^{I\ga}$ are also untwisted and form the $(\rep{12},\crep{4})$ and $(\rep{12},\rep{4})$ representations of $U(12)\times U(4)$, respectively. Together they combine to $(\rep{12},\rep{8})$ of $U(12)\times SO(8)$. In summary we have found the charged spectrum 
\equ{ 
\frac 1{9}\, (\rep{12},\rep{8}) + \frac 19\, (\crep{66},\rep{1})
}
This agrees with the resolution spectrum of Table 2 in \cite{Nibbelink:2007rd} computed via a generalization of an index theorem.

\subsubsection{A $\boldsymbol{\Intr_3}$ Resolution model with 
$\boldsymbol{n=2}$}

In the previous example we only found untwisted states in the resolution phase. With our next example we illustrate that we can also correctly determine twisted states in the four dimensional spectrum. We start from the $\Intr_3$ orbifold model ($n=2$) with shift 
\equ{ 
V = \big(\frac 13^{4}, -\frac 23^2, 0^{10}\big)~, 
} 
so that the orbifold gauge group equals $U(6)\times SO(20)$. The orbifold spectrum of this model is 
\equ{ 
\frac 1{27}\Big[ 
\rep{3_R}(\rep{6},\rep{20})_1 + \rep{3_R}(\crep{15},\rep{1})_{\sm 2}  
\Big] 
+ \rep{3_R}(\rep{1},\rep{1})_2 + (\crep{15},\rep{1})_0~. 
} 

As blow-up mode we choose a mode in the $(\crep{15},\rep{1})_0$. Concretely we take 
\equ{
Q = P_1 = 
\big( \frac 13^4, -\frac 23^2, 0^{10} \big)~. 
}
This gives the following charge assignment
\equ{
\arry{c|| c|c|c|| c|c|c}{
\text{Field} & 
\gPs^0 & \gPs^i & \gPs^{\sm 1} & 
\gL^I & \gL^A & \gL^\ga 
\\ \hline 
q, Q & 
0 & \frac 13 & -1 & 
\frac 13 & -\frac 23 & 0
}
\label{chargesZ32}
}
with $i =1,2,3$, $I = 1,\ldots,4$, $A = 1,2$, and $\ga = 1,\ldots, 10$. This results in the following expansion of the kinetic terms of the chiral-Fermi multiplets: 
\equa{
S_\text{Fermi} = &- \frac 14 \int \d^2\gs\d^2 \gth^+\, 
\Big\{
N^I{}_J\, \bgL_I \gL^J 
+ N^A{}_B\, \bgL_A \gL^B 
+ N^\ga{}_\gb\, \bgL_\ga \gL^\gb  
+ n^{\ga\gb}\, \bgL_\ga \bgL_\gb 
\non \\[0ex] &
+ N^I{}_A\, \bgL_I \gL^A 
+ N^A{}_I\, \bgL_A \gL^I 
+ N^I{}_\ga\, \bgL_I \gL^\ga
+ N^\ga{}_I \, \bgL_\ga \gL^I 
+ N^A{}_\ga\, \bgL_A \gL^\ga 
+ N^\ga{}_A\, \bgL_\ga \gL^A 
\non \\[1ex] & 
+ n^{IJ}\, \bgL_I \bgL_J 
+ n^{AB}\, \bgL_A \bgL_B 
+ n^{IA}\, \bgL_I \bgL_A 
+ n^{I\ga}\, \bgL_I \bgL_\ga 
+ n^{A\ga}\, \bgL_A \bgL_\ga 
\Big\} + \text{h.c.}~.  
}
The functions on the first line, $N^I{}_J$, $N^A{}_B$, $N^\ga{}_\gb$, $n^{\ga\gb}$ and $\bn_{\ga\gb}$, are neutral, hence they correspond to the unbroken four dimensional gauge group $U(4) \times U(2) \times SO(20)$ in the resolution phase. Given the charge assignment of equation \eqref{chargesZ32}, we infer that the charges of the remaining functions are: 
\equ{
\arry{l}{\dsp 
\arry{c|| c|c|c|c|c|c }{
\text{Function} & 
N^A{}_I & N^I{}_A & 
N^I{}_\ga & N^\ga{}_I & 
N^A{}_\ga & N^\ga{}_A 
\\ \hline 
Q & 1 & -1 & 
\frac 13 & - \frac 13 & 
- \frac 23 & \frac 23 
\\ \hline 
\text{Type} & 
- & \text{T} & 
\text{U} & - & 
\text{U} & - 
}
\\[5ex] 
\arry{c||  c|c|c|c|c|c| c|c|c|c}{
\text{Function} & 
n^{IJ} & \bn_{IJ}& 
n^{AB} & \bn_{AB} & 
n^{IA} & \bn_{IA} & 
n^{I\ga} & \bn_{I\ga} & 
n^{A\ga} & \bn_{A\ga}  
\\ \hline 
Q &
\frac 23 & -\frac 23 & 
-\frac 43 & \frac 43 & 
-\frac 13 & \frac 13 & 
\frac 13 & -\frac 13 & 
-\frac 23 & \frac 23 
\\ \hline 
\text{Type} & 
- & \text{U} &
\text{T'} & \text{U'} & 
- & \text{U} & 
\text{U} & - & 
\text{U} & - 
}
}
}
By Table \ref{tb:Monomials} we have determined how we should interpret the resulting states in the effective four dimensional theory. Putting all this information together we obtain the following resolution spectrum: 
\equ{ 
\frac 19 \, (\crep{6},\rep{1},\rep{1}) + 
\frac 19 \, (\rep{4},\rep{1},\rep{20}) + 
\frac 19 \, (\rep{1},\rep{2},\rep{20}) + 
\frac 19 \, (\crep{4},\crep{2},\rep{1}) + 
(\crep{4},\rep{2},\rep{1}) + 
\frac {26}9 \, (\rep{1},\rep{1},\rep{1})~. 
}
This again agrees with the resolution spectrum of Table 2 in \cite{Nibbelink:2007rd} up to charge conjugation.

\section{Conclusions}
\label{sc:concl}

The aim of this paper was to consider a framework that can smoothly interpolate between heterotic orbifold models and supergravity on their resolutions with gauge bundles. As gauged linear sigma models (GLSMs) reduce to orbifolds, when taking a certain limit of the K\"ahler parameters, and allow for a large volume description in another limit, they provide an ideal setting for this investigation. However, before we could use this formalism, we first needed to address the following question: Given a heterotic orbifold model with a certain number of twisted states with non--vanishing VEVs, what is the corresponding GLSM? We proposed that the shifted momenta that characterize these twisted modes essentially determine the charge assignment of the superfields of the GLSM.

To be precise, we made the following identifications between the shifted momenta and the GLSM charges: By adding integers to the entries of the right--moving shifted momenta of each blow--up mode we arranged them to all lie between 0 and 1, and to sum up to unity. These entries are the charges of the chiral superfields that contain the target space coordinates. To ensure that the total sum of charges vanishes, an exceptional chiral superfield with charge $-1$ was added for each blow--up mode. Furthermore, for the twisted modes without any oscillator excitations the left--moving shifted momenta are identical to the charges of the chiral--Fermi multiplets that generate the gauge field and matter representations in target space. When twisted modes contain oscillators, it was necessary to modify some components of the left--moving shifted momenta by an integral amount before they could be identified with the chiral--Fermi superfield charges. Consequently, twisted states with oscillatory excitations induce fermionic gauging and result in non--Abelian gauge bundles, like the standard embedding.

This proposal leads to some intriguing relations between the orbifold theory and the GLSM description: The level matching and massless conditions of the twisted states ensured that the pure GLSM anomalies cancel automatically. The $\Intr_N$ projection conditions on higher twisted states is closely related to the absence of mixed gauge anomalies in the GLSM. In particular for a $\Intr_4$ orbifold model we showed that none of the second--twisted states, that were projected out, could be associated with a consistent GLSM: Either the mixed gauge anomalies did not cancel, or the fermionic gauging was not holomorphic as it should. Moreover, irrespectively of whether a twisted mode with non--vanishing VEV contains oscillator excitations or not, it is associated with an exceptional cycle inside the resolved geometry: The volume of this cycle is determined by a K\"ahler parameter, which in the GLSM appears as a FI--parameter for the worldsheet gauge multiplet. This justifies calling such a twisted state with non--zero VEV a blow--up mode.

The GLSM framework provides a very powerful environment to study different regions of the string moduli space: Different triangulations of geometrical phases and the resulting intersections of divisors can be easily understood by analyzing the supersymmetric minima of $D$--term potentials on the worldsheet. Moreover, even in regions that neither allow for an orbifold nor for a large volume description, the GLSM nevertheless makes perfect sense. We have illustrated this by investigating the 14 different phases associated with (partial) resolutions of the $\Cplx^3/\Intr_2\!\times\!\Intr_2$ orbifold. Beside the orbifold phase and the four different triangulations of the fully resolved geometry in the large volume limit, there are nine additional hybrid phases, where some K\"ahler parameters are positive while some others are negative. This corresponds to partially resolved geometries for which both orbifold and  supergravity descriptions presumably fail, yet for the GLSM this does not seem to pose any serious obstruction.

In order to confirm that in the large volume limit the GLSM gives the same results as certain gauge bundles (line bundles in the cases we considered) on the resolved orbifold, we developed a method to read off the effective four dimensional charged massless spectrum directly from the GLSM. The method is inspired by how one identifies the ten dimensional target space metric, B--field and gauge fields in the sigma model on the string worldsheet: We considered marginal deformations of the kinetic terms of the chiral--Fermi multiplets that induce the gauge degrees of freedom. For all seven line bundle models on the resolved $\Intr_3$ orbifold, identified in \cite{Nibbelink:2007rd}, we found exact agreement.

\subsection*{Outlook}

As we have seen GLSMs associated to more complicated orbifolds, like the $\Cplx^3/\Intr_2\!\times\!\Intr_2$, may possess many quasi--geometrical phases. In these phases even the approximate validity of both supergravity and the orbifold CFT description are questionable to say the least. Our hope is that this work improves the understanding of how to obtain physical statements in these hybrid and therefore intrinsically stringy regimes of the moduli space. Heterotic orbifolds can be equipped with (generalized) discrete torsion \cite{Vafa:1986wx,Vafa:1994,Ploger:2007iq}. Recent works \cite{Adams:2006kb,Adams:2009av,Adams:2009zg,Adams:2009tt} suggest how discrete torsion may be implemented in a GLSM description. It would be intriguing to extend these ideas to GLSM resolutions.

It would be very interesting to extend our GLSM methods to compact orbifold resolutions as well. The recent description of the resolved $T^6/\Intr_2\!\times\!\Intr_2$ in ref.\ \cite{Blaszczyk:2010db} might be a good starting point for such an investigation. For the non--compact $\Cplx^3/\ztwo$ we have shown that the pure and mixed GLSM anomaly cancellation conditions encode the Bianchi identities of all four triangulations of the fully resolved orbifold combined. This might suggest that we should combine all possible Bianchi identities on compact orbifold resolutions to ensure that the resolution model has a proper string lift. Probably part of these conditions are identical to the mass--shell conditions of the twisted blow--up modes. In refs.\ \cite{Nibbelink:2009sp,Blaszczyk:2010db} we have already used such conditions to find solutions to the Bianchi identities in a given triangulation. However, with our current knowledge we cannot exclude that there are additional conditions that need to be fulfilled in order that a given heterotic supergravity in a given triangulation has a full string lift.


Another question, which needs to be clarified further, what is the scale $M_r$ in the $\ln(\langle T_r\rangle/M_r)$ in equation \eqref{logdependence}. The precise value of $M_r$ of course depends on the precise parameterization of the moduli space described by the K\"ahler (or FI--)parameter $\gr_r$ and the definition and normalization of the twisted state $|T_r\rangle$ in the effective target space theory associated to the orbifold CFT. The arguments presented here only suggests that $M_r$ is of the order of the string scale, but its precise value is unknown. Does it depend on $r$, or is it universal for all twisted sectors? Knowledge of the parameter $M_r$ is of phenomenological relevance as it is probably related to the scale where treatment of the orbifold CFT with ``small" perturbations induced by the VEV $\langle T_r\rangle$ breaks down.

As mentioned in the Introduction GLSMs do not describe true string dynamics as their dimensionful coupling constant break conformal invariance. Therefore one should really study GLSMs in the strong coupling limit. This means that non--perturbative quantum effects, like instantons, can be very important. It has been shown that worldsheet instantons do not destabilize these models \cite{Silverstein:1995re,Silverstein:1994ih,Basu:2003bq,Adams:2003zy}. More recently, there have been investigations of the resulting non--perturbative superpotential of GLSMs \cite{McOrist:2008ji,Aspinwall:2010ve,Kreuzer:2010ph}. Such investigations could be extended to the GLSMs discussed in the present work.

The main reason to restrict ourselves to GLSMs has been simplicity. True string backgrounds are described by non--linear sigma models (NLSMs). As we have seen in the computation of the effective four dimensional spectra the study of gauged NLSMs (GNLSMs) could be very useful. It would therefore be very appealing to investigate gauged NLSMs in more detail. One might think that once one is considering NLSMs, gauged NLSMs are essentially obsolete, because one can remove the gauge degrees of freedom in the strong coupling limit. However, the gauged description makes a useful distinction between topological properties that characterize the gauge bundles involved and the precise forms of the metric, $B$--field, and gauge field backgrounds.

Finally the successful approach of calculating part of the spectrum by investigating the kinetic terms of the chiral-Fermi multiplets may be somewhat surprising, because such non--holomorphic terms are not protected against renormalization effects. However, given that these terms give rise to the charged chiral spectrum in the effective target space theory in four dimensions, the computation may nevertheless be reliable in general. Further investigations are needed to understand exactly under which conditions what part of the four dimensional spectrum can be determined by such techniques.

\subsection*{Acknowledgements}

First of all I would like to thank Filipe Paccetti Correia and Michele Trapletti for collaboration at the initial stage of this project. I would further like to thank Allan Adams, Ralph Blumenhagen and Ilka Brunner for very helpful discussions on GLSMs, and Wilfried Buchm\"uller and 
Michael Ratz for discussions on orbifold models and their possible blow--ups. In addition, I am very grateful to Michael Blaszczyk, Fabian R\"uhle, and Patrick Vaudrevange for careful reading the manuscript and many interesting discussions at various occasions. 
This research has been supported by the "LMUExcellent" Programme.

\appendix 
\def\theequation{\thesection.\arabic{equation}} 
\setcounter{equation}{0}

\section{Reduction of 4D superspace to 2D (2,2) and (2,0) superspaces} 
\label{sc:superspace}

In four dimensions the $\cN=1$ superspace has complex Grassmann coordinate $\gth^\ga$ and $\bgth^\dga$, with $\ga, \dga =+,-$. The anti-symmetric tensor $\ge^{\ga\gb}$ carries spinor indices with $\ge^{+-}=-\ge_{+-}=1$. The (2,2) superspace in two dimensions is identical to the four dimensional superspace. In the (2,0) superspace one only keeps $\gth^+$ and $\bgth^+$ as complex Grassmann variables \cite{Dine:1986by}. The supercovariant derivatives read
\equ{
D_\ga = \der_\ga + i \gs^m_{\ga\dga} \bgth^\dga \der_m~, 
\qquad 
\bD_\dga = - \der_\dga - i \gs^m_{\ga\dga} \gth^\ga \der_m~, 
}
and satisfy the algebra 
\equ{
\{ D_\ga, \bD_\dga \} = -2i \gs^m_{\ga\dga} \der_m~.  
}
Here $m = 0,1,2,3$ are the four dimensional space-time indices. In the reduction to two dimensions we only keep the coordinates $\gs^0$ and $\gs^3$. Left- and right-moving or holomorphic and anti-holomorphic coordinates are introduced via
$\gs = (\gs^0 + \gs^3)/2$ and $\bgs = (\gs^0 - \gs^3)/2$, such that 
\equ{ 
\der = \der_0 + \der_3~, 
\qquad 
\bder = \der_0 - \der_3~. 
}
The non-vanishing part of the algebra of super covariant derivatives becomes
\equ{
\{ D_+, \bD_+ \} = -2i \der~, 
\qquad 
\{ D_-, \bD_- \} = -2i \bder~. 
} 
Note that $D^2 = -2 D_+ D_-$ and $\bD^2 = 2 \bD_+ \bD_-$. The integration over Grassmann variables is defined as differentiation 
\equ{
\int d^4\gth = \int d^2\gth^+ \int\d^2\gth^-~, 
\qquad 
\int\d^2\gth^\pm A = \frac 12 [\bD_\pm, D_\pm] A\big|_\pm~. 
}

\section{(2,2) supersymmetric field theories}
\label{sc:(2,2)theories}

\subsection{(2,2) superfields}
\label{sc:(2,2)superfields}

There are three basic bosonic superfields or multiplets of (2,2) supersymmetry in two dimensions: the chiral, the vector and the twisted-chiral multiplet. The first two have an analog in four dimensional superspace, the last one is a curiosity of two dimension.

Chiral and anti-chiral multiplets, $\cC$ and $\cC^\dag$, are defined by the constraints
\equ{
\bD_+ \cC = \bD_- \cC = 0~, 
\qquad 
D_+ \cC^\dag = D_- \cC^\dag = 0~. 
}
A vector superfield $\cV$ is defined as a real superfield with a gauge transformation 
\equ{
\cV^\dag = \cV~, 
\qquad 
\cV \ra \cV - \cL - \bcL~, 
}
where $\cL, \bcL$ are chiral and anti-chiral multiplets. The vector field itself is not gauge invariant, but out of it one can construct gauge invariant field strengths 
\equ{ 
\cT = \frac 1{\sqrt 2} \bD_+ D_- \cV~, 
\qquad 
\bcT = \frac 1{\sqrt 2} \bD_- D_+ \cV~. 
}
These are twisted-chiral and twisted-anti-chiral superfields, defined by the constraints 
\equ{
\bD_+ \cT = D_- \cT = 0~, 
\qquad 
D_+ \bcT = \bD_- \bcT = 0~. 
\label{Superfieldstrengths} 
}

If the chiral multiplets are charged under the vector field, they transform under gauge transformations as:
\equ{
\cC \ra e^{2\cL\cdot q}\, \cC~, 
\qquad
\cC^\dag \ra  \cC^\dag\, e^{2\bcL\cdot q}~, 
}
where we allow for a number of gauge fields and chiral multiplets, so that the gauge transformations are encoded by a charge matrix $q$. It is then convenient to introduce gauge covariant derivatives 
\equ{
\cD_\ga = D_\ga~, 
\qquad  
\bcD_\dga = e^{2\cV\cdot q}\bD_\dga e^{-2\cV\cdot q}
= \bD_\dga - 2 q\cdot (\bD_\dga \cV)~. 
} 
Using them we can define gauge covariant (anti-)chiral superfields, defined by 
\equ{
\bcD_+ \cC = \bcD_- \cC = 0~, 
\qquad 
\cD_+ \bcC = \cD_- \bcC = 0~. 
}
This amounts to defining $\bcC =  \cC^\dag e^{2\cV\cdot q}$ while keeping $\cC$ as before. Their gauge transformations read 
\equ{
\cC \ra e^{2\cL\cdot q}\, \cC~, 
\qquad 
\bcC \ra \bcC\,  e^{-2\cL\cdot q}~.
}
The anti-commutation algebra of the gauge covariant derivatives is more involved
\equ{
\arry{lcl}{\dsp 
\{ \cD_+, \bcD_+ \} = -2i\, \cD~, 
&\qquad & 
\{ \cD_-, \bcD_- \} = -2i \,\bcD~,  
\\[2ex] 
 \{ \bcD_+, \cD_- \} = 2\sqrt 2\,  q\cdot \cT~, 
&\qquad&
 \{ \bcD_-, \cD_+ \} = 2\sqrt 2\, q\cdot \bcT~, 
} 
}
where $\cD$ and $\bcD$ denote the gauge covariant versions of the derivatives $\der$ and $\bder$, respectively.

\subsection{(2,2) actions}
\label{sc:(2,2)actions}

A general action of a (2,2) theory can be decomposed in the following parts: a K\"ahler potential, a superpotential, a twisted-superpotential and a gauge kinetic action. They read: 
\equ{
S_\text{K\"ahler} = -\frac 14\, \int \d^2\gs \d^4\gth\, 
K\big(\bcC, \cC\big)~, 
\label{(2,2)Kahler}
}
\equ{
S_\text{super} =  \frac 12\int \d^2\gs \d^2\gth\, W(\cC) + \hc~, 
\qquad 
S_\text{twisted} = \frac 12 \int \d^2\gs \d\gth^+\d \bgth^-\, \gr\cdot\cT + \hc~. 
\label{(2,2)supertwisted} 
}
The dimensionality of the twisted superfield $\cT$ only allows linear terms to appear in the twisted superpotential. The quantities $\gr$ can be thought of as Fayet-Iliopoulos parameters. And finally the gauge kinetic action can be written as  
\equ{
S_\text{gauge} = - \frac 1{4 e^2}\int \d^2\gs \d^4\gth\, \bcT \cT~, 
\label{(2,2)gauge} 
}
where $e$ is the gauge coupling.

\section{(2,0) supersymmetric field theories}
\label{sc:(2,0)theories}

\subsection{(2,0) superfields} 
\label{sc:(2,0)superfields}

In this Appendix we describe the basic (2,0) multiplets in terms of the (2,0) superspace language reviewed above. These multiplets are the vector multiplet, the chiral multiplet, the chiral-Fermi multiplet and the Fermi-gauge (or unconstraint Fermi) multiplet.

The vector multiplet consists of two real superfields $V$ and $A$. Their bosonic gauge transformations read 
\equ{
V \ra V - \gTh - \bgTh~, 
\qquad 
A \ra A +i \bder (\gTh- \bgTh)~, 
\label{(2,0)gauge} 
}
where $\gTh$ and $\bgTh$ are chiral and anti-chiral super gauge parameter, i.e.\ $\bD_+\gTh = 0$ and $D_+ \bgTh=0$, respectively. 
These transformation can be used to reach the Wess-Zumino (WZ) gauge. In this gauge the only non-vanishing components of $V$ and $A$ are 
\equ{
\arry{c}{\dsp 
A_\gs = \frac 12[\bD_+, D_+]V\big|_+~,
\qquad 
A_\bgs = A\big|_+~,
\qquad 
\tD = \frac 12[\bD_+, D_+] A\big|_+~, 
\\[2ex] \dsp 
\gf = -\frac 1{\sqrt 2}\, D_+ A \big|_+~,
\qquad 
\bgf =  \frac 1{\sqrt 2}\, \bD_+ A\big|_+~,
}
\label{compVectorM}
}
by restricting (derivatives of) it to $\gth^+=\bgth^+=0$ indicated by $|_+$. This WZ gauge is left invariant by residual transformations for $\gTh=-i\ga/2= - \bgTh$ (where $\ga$ is a real function not a superfield), which induces an Abelian gauge transformation $A_\gs \ra A_\gs - \der \ga$ and $A_\bgs\ra A_\bgs - \bder \ga$. We define the super field strengths
\equ{
F = - \frac 12 \bD_+ \big( A - i \bder V \big)~, 
\qquad 
\bF = \frac 12 D_+ \big( A + i \bder  V \big)~ 
\label{(2,0)superfieldstrengths}
}
for the pair $V$ and $A$. The super field strengths fulfill the constraints 
\equ{
\bD_+ F = D_+ \bF = 0~, 
}
hence they are examples of chiral Fermi multiplets which are introduced below. The components of the super field strengths are given by 
\equ{
F|_+ =  \frac 1{\sqrt 2}\, \gf~, 
\quad 
\bF|_+ = \frac 1{\sqrt 2}\, \bgf~, 
\qquad 
D_+ F|_+ = \frac 12( \tD + i\, F_{\gs\bgs} )~, 
\quad 
\bD_+ \bF|_+ = \frac 12( \tD - i\,  F_{\gs\bgs} )~, 
}
where $F_{\gs\bgs} = \der A_\bgs - \bder A_\gs$ is the gauge field strength.

Next we introduce gauge covariant chiral and anti-chiral superfields $\gPs, \bgPs = \gPs^\dag e^{2V\cdot q}$ that obey the constraints 
\equ{ 
\bcD_+ \gPs = \cD_+ \bgPs = 0~, 
}
respectively, where the gauge covariant derivatives are defined as 
\equ{
\cD_+ = D_+~, 
\qquad  
\bcD_+ = e^{2 V\cdot q}\bD_+ e^{-2V\cdot q}
= \bD_+ - 2 q\cdot (\bD_+ V)~, 
\label{covDers} 
}
in the chiral basis. Consequently they transform under bosonic gauge transformations as 
\equ{
\gPs \ra e^{2\gTh \cdot q}\, \gPs~, 
\qquad 
\gPs^\dag \ra  \gPs^\dag\, e^{2\bgTh \cdot q}
\qquad 
\bgPs \ra \bgPs\, e^{-2\gTh \cdot q}~. 
}
The notation $\gTh \cdot q$ allows for the existence of a set of chiral superfield covariant w.r.t.\ the set of bosonic gauge transformations. The gauge covariant versions of the derivatives $\der$ and $\bder$ read 
\equ{
\cD = \frac i2\, \{ \cD_+, \bcD_+ \}~, 
\qquad 
\bcD = \bder + 2i (A -i \bder V)\cdot q~, 
}
on covariant chiral superfields, respectively. The components of the gauge covariant chiral and anti-chiral superfields are defined as 
\equ{
z = \gPs\big|_+~, 
\quad 
\gps = \frac 1{\sqrt 2} \cD_+ \gPs\big|_+~, 
\qquad 
\bz = \bgPs\big|_+~, 
\quad 
\bgps = \frac 1{\sqrt 2} \bcD_+ \gPs\big|_+~. 
\label{compChiral} 
}

The chiral-Fermi multiplets are fermionic superfields that fulfill the following constraints
\equ{ 
 \bcD_+ \gL = \cD_+ \bgL  =0~,
\label{FermiConstraint} 
}
where the gauge covariant derivative $\bcD_+ \bgL = (\bD_+ -2 \bD_+ V\cdot Q)\bgL$ on $\bgL$ is given by charges $Q$, as in \eqref{covDers} for the chiral superfields. The components of chiral-Fermi and anti-chiral-Fermi superfields are given by 
\equ{
\gl = \gL\big|_+~, 
\quad 
h = \frac 1{\sqrt 2} \cD_+ \gL\big|_+~, 
\qquad 
\bgl = \bgL\big|_+~, 
\quad 
\bh = \frac 1{\sqrt 2} \bcD_+ \bgL\big|_+~, 
\label{compChiralFermi} 
}
The super field strengths \eqref{(2,0)superfieldstrengths} are chiral-Fermi multiplets that are inert under gauge transformations.

The fourth and final type of (2,0) superfields we need to introduce are Fermi-gauge superfields, $\gS$ and $\bgS$. They transform with shifts under fermionic gauge transformations
\equ{
\gS \ra \gS + 2\, \gX~, 
\qquad 
\bgS \ra \bgS + 2\, \bgX~, 
\label{FermionicGauge} 
} 
where $\gX$ and $\bgX$ are neutral chiral-Fermi and anti-chiral-Fermi superfields. These transformations allow us to gauge some components of $\gS$ and $\bgS$ away; the remaining physical components are 
\equ{
s = \frac 1{\sqrt 2} \bD_+ \gS \big|_+~,
\quad 
\gvf = \frac 1{2} [D_+,\bD_+] \gS \big|_+~,
\qquad 
\bs = \frac 1{\sqrt 2} D_+ \bgS \big|_+~,
\quad 
\bgvf = \frac 1{2} [\bD_+,D_+] \bgS \big|_+~.
\label{compFermiGauge}
}

\subsection{(2,0) actions}
\label{sc:(2,0)actions}

A general (2,0) action can be decomposed into kinetic actions for the various (2,0) multiplets, a superpotential and a Fayet-Illiopoulos (FI) term. We first give their kinetic actions 
\equ{
S_\text{gauge} = 
\frac 1{2e_I^2}\, \int\d^2\gs \d^2\gth^+ \, \bF_I F_I~,
\qquad 
S_\text{chiral} =
\frac i2\, \int\d^2\gs \d^2\gth^+ \, \bgPs_a \bcD \gPs^a~, 
\\[2ex] 
S_\text{Fermi} = 
- \frac 12\,  \int\d^2\gs \d^2\gth^+ \, 
\big(\bgL_\ga - \bgS_j \, \bM^j{}_\ga(\bgPs) \big) 
 \big( \gL^\ga - M^\ga{}_i(\gPs)\, \gS^i \big)~, 
\\[2ex] 
S_\text{fer.\,gauge} =
-\frac 1{2e_i^2}\,  \int\d^2\gs \d^2\gth^+\, \bD_+\gS_i \bder D_+\bgS^i 
\label{(2,0)kinactions}
}
for sets of gauge superfields $\{(V,A)^I\}$, chiral superfields $\{\gPs^a\}$, chiral-Fermi superfields $\{\gL^\ga\}$ and Fermi-gauge superfields $\{\gS^i\}$, respectively, with the fermionic gauging given in~\eqref{gXgauge}. The (2,0) superpotential and FI actions are given by 
\equ{ 
S_\text{super} = 
 \int\d^2\gs \d\gth^+ \, m\, N_\ga(\gPs)\, \gL^\ga~, 
\qquad 
S_\text{FI} = 
 \int\d^2\gs \d\gth^+ \, \gr_I(\gPs) \, F^I~, 
 \label{super&FI} 
}
in terms of holomorphic functions $N_\ga(\gPs)$ and $\gr_I(\gPs)$ of the chiral superfields $\gPs^a$. The parameters $e_I$, $e_i$ and $m$ have dimension of mass.

\subsection{Non-linear sigma models}

A non-linear sigma model (NLSM) of the heterotic string can be described by the action $S_\text{het}  = S_{\gPs} + S_{\gL}$ with \cite{Hull:1985jv}
\begin{subeqns}
\equa{
\qquad 
& S_{\gPs} = 
\frac i4\, \int \d^2\gth^+\,
 \Big(  k_a \bder \gPs^a  - \bk^a \bder \bgPs_a \Big) 
~, 
\\[2ex] 
& S_{\gL} = 
- \frac 12 \, \int \d^2 \gth^+\, 
 \Big\{ 
 N^\gb{}_\ga \, \bgL_\gb \gL^\ga 
+\frac 12 \, \bn_{\ga\gb} \, \gL^\ga \gL^\gb
+\frac 12 \, n^{\ga\gb} \, \bgL_\ga \bgL_\gb
\Big\}~, 
\label{HetAc} 
}
\end{subeqns} 
where $k_a$ and $n_{\ga\gb}= -n_{\gb\ga}$ are arbitrary complex functions of the chiral superfields $\gPs$ and $\bgPs$. Their complex conjugates are denoted as $\bk^a$ and $\bn^{\ga\gb}$. Finally, $N^\gb{}_\ga$ is an Hermitean function of $\gPs$ and $\bgPs$.

We introduce the following short-hand notations
for the background metric 
\equ{ 
G^{a}{}_b = \frac 12 \big( K^a{}_b + \bK^a{}_b \big)~, 
\qquad\quad  
K^a{}_b = k_b{},{}^c~, 
\qquad 
\bK^a{}_b = \bk^a{}_{,b}
}
and the Kalb-Ramond two-form 
$B_2 = B^a{}_{b}\,\d \bz_a \d z^b + \frac 12\, b_{ab}\, \d z^a \d z^b + \bb^{ab} \d \bz_a \d \bz_b$ 
given by \cite{Hull:1985jv}
\equ{ 
B^{a}{}_b = \frac 12 \big( K^a{}_b -\bK^a{}_{b}\big)~, 
\qquad 
b_{ab} = \frac 12 \big( k_{a,b}-k_{b,a} \big)~, 
\qquad 
\bb^{ab} = \frac 12 \big( \bk^{a,b}-\bk^{b,a} \big)~. 
}
The one-form $K_1 = i(k_1 - \bk_1)$, with $k_1= k_a \, \d z^a$ and $\bk_1 = \bk^a \, \d \bz_a$, can be thought of a prepotential for both the metric and the Kalb-Ramond field simultaneously. In particular, one has 
\equ{
B_2 = i( \der_1 - \bder_1)K_1 = 
\der_1 \bk_1 + \bder_1 k_1 - \der_1 k_1 - \bder_1 \bk_1 
~,
}
with $\der_1 = \d z^a \der_a$ and $\bder_1 = \d \bz_a \bder^a$. This does not define the most general possible torsion classes as only the following components of the three-form field strength $H_3 = \d B_2$ are switched on 
\equ{
H_{abc} = H^{abc} = 0~, 
\qquad 
H_{ab}{}^c = (k_{a,b} - k_{b,a})_,{}^c~,
\qquad 
H^{ab}{}_c = (\bk^{a}{},{}^b - \bk^{a}{},{}^b)_{,c}~. 
}
In the presence of torsion the Christoffel connections are modified 
\equ{
\gG_{\!\! +}{}^d_{ca} = (G\inv)^d{}_b 
\big( G^b{}_{c,a} + \sfrac 12\, H_{ca}{}^b \big)~,
\qquad  
\bgG_{\!\! +}{}^{cb}_d = 
\big(G^c{}_{a,}{}_b  +\sfrac 12\, H^{cb}{}_a \big)
(G\inv)^a{}_d~. 
}
The component action $S_\gPs$ can be written either as 
\equ{
\arry{rl}{ \dsp
S_\gPs = & - \frac 12\, G^a{}_b\,  \big(\der \bz_a \bder z^b + \bder \bz_a \der z^b \big)
 - \frac 12\, B^a{}_b\,  \big(\der \bz_a \bder z^b - \bder \bz_a \der z^b \big) 
 \\[2ex] \dsp  
&+ \frac i2 \, \bgps_b \, G^b{}_d\, 
\Big( \bder \gps^d  + \gG_{\!\! +}{}_{ca}^d\, \bder z^c\,\gps^a  \Big) 
- \frac i2 \, 
\Big( 
\bder \bgps_d +  \bgps_b \bder \,\bz_c \,\bgG_{\!\!+}{}^{cb}_d  
\Big) \, G^d{}_a\, \gps^a~, 
}
\label{SgPs}
}
or as 
\equ{
\arry{rl}{\dsp 
S_\gPs = & - \frac 12\, G^a{}_b\,  \big(\der \bz_a \bder z^b + \bder \bz_a \der z^b \big)
 +\frac 12\, b_{ab}\,  \bder z^a \der z^b 
 +\frac 12\, \bb^{ab}\,  \bder \bz_a \der \bz_b 
 \\[2ex] \dsp 
 &+ \frac i2 \, \bgps_b \, G^b{}_d\, 
\Big( \bder \gps^d  + \gG_{\!\! +}{}_{ca}^d\, \bder z^c\,\gps^a  \Big) 
- \frac i2 \, 
\Big( 
\bder \bgps_d +  \bgps_b \bder \,\bz_c \,\bgG_{\!\!+}{}^{cb}_d  
\Big) \, G^d{}_a\, \gps^a~, 
}
}
In the first form the purely holomorphic and anti-holomorphic components of the $B$-field are absent, while in the second form the $B^a{}_b$ does not appear. These two actions describe the same physics as they differ by a total derivative only. On the level of the two-form $B_2$ this can be understood as a gauge transformation 
$B_2' - B_2 = (\der_1 + \bder_1)(l_1 + \bl_1)$: 
\equ{ 
B_2' = 
\der_1 (\bk_1+\bl_1) + \bder_1 (k_1+l_1) - \der_1 (k_1-l_1) - \bder_1 (\bk_1 - \bl_1)~. 
}
We can therefore find a gauge in which either $B^a{}_b$ vanishes or where $b_{ab} = \bb^{ab} =0$, but we cannot set all of them zero at the same time.

The component form of the Fermi multiplet action (ignoring the auxiliary field  contributions) reads
\equ{ 
\arry{rl}{S_\gL =  & \dsp 
+ \frac i2\, \bgl_\ga N^\ga{}_\gb \, 
\Big( \der \gl^\gb  + A_a{}^\gb{}_\gg \, \der z^a\, \gl^\gg \Big)  
-  \frac i2\, A_a{}^{\ga\gb} \,\der z^a\, \bgl_\ga \bgl_\gb 
\\[2ex] & \dsp 
- \frac i2\, 
\Big( \der \bgl_\ga + \bgl_\gg \, \der \bz_a\, \bA^a{}^\gg{}_\gb \Big) N^\ga{}_\gb\, \gl^\gb 
+ \frac i2\, \bA^a{}_{\ga\gb} \,\der \bz_a\, \gl^\ga \gl^\gb 
}
\label{SgL}
}
where we have introduced an $SO(32)$ gauge field $A_a$ decomposed in the adjoint and the anti-symmetric tensor (and conjugate) representations of $U(16)$, respectively, as 
\equ{
A_a{}^\ga{}_\gb = (N\inv)^\ga{}_\gg \,N^\gg{}_{\gb,a}~, 
\quad
\bA^a{}^\ga{}_\gb = N^\ga{}_{\gg,}{}^{a}\, (N\inv)^\gg{}_\gb~, 
\qquad 
A_a{}^{\ga\gb} =  n^{\ga\gb}{}_{,a}~, 
\quad
\bA^a{}_{\ga\gb} =  \bn_{\ga\gb,}{}^{a}~.
\label{gaugeComp} 
}

\section{Reduction of (2,2) to (2,0) supersymmetric field theories}
\label{sc:Reduction}

\subsection{Decomposition of (2,2) superfields in (2,0) superspace}
\label{sc:ReductionSuperfields}

The chiral and chiral-Fermi (2,0) superfields defined in Appendix \ref{sc:(2,0)theories} can be recovered from the (2,2) chiral superfield by reducing the (2,2) superspace to (2,0) by making an expansion in the Grassmann coordinates $\gth^-, \bgth^-$: 
\equ{
\gPs = \cC\big|_-~,
\quad 
\gL = \frac 1{\sqrt 2} D_- \cC \big|_-~, 
\qquad 
\bgPs = \bcC\big|_-~,
\quad 
\bgL = \frac 1{\sqrt 2} \bD_- \bcC \big|_-~. 
}
The (2,2) chirality of $\cC$ is therefore inherited by $\gPs$ and $\gL$.  
These expansions are by definition (2,0) gauge covariant, but not fully (2,2) gauge covariant. The reason is that we use the Distler-Kachru convention \cite{Distler:1993mk,Distler:1995mi} in this decomposition which uses the ordinary super covariant derivatives $D_-$ and $\bD_-$ and not their gauge covariant versions $\cD_-$ and $\bcD_-$: This ensures that the chiral-Fermi multiplet $\gL$ is indeed chiral.\footnote{If one instead, following Witten \cite{Witten:1993yc}, uses gauge covariant derivates to define this (2,0) decomposition, a Fermi multiplet is chiral up to a holomorphic function: $\bcD_+ \gL = E(\gPs)$.}

Consequently, the (2,2) gauge parameters $\cL$ and $\bcL$ are restricted to (2,0) superspace: 
\equ{
\gTh = \cL\big|_-~, 
\quad 
\gX = \frac 1{\sqrt 2} D_- \cL\big|_-~, 
\qquad
\bgTh = \bcL|_-~, 
\quad 
\bgX = \frac 1{\sqrt 2} \bD_-\bcL\big|_-~, 
}
with $\gTh$ and $\gX$ chiral and chiral-Fermi superfields, respectively. The induced transformations act on the (2,0) components of (2,2) chiral superfields as 
\equ{
\gPs \ra e^{2\gTh\cdot q} \gPs~, 
\qquad 
\gL \ra e^{2\gTh\cdot q}\Big( 2 \,\gX\cdot q \, \gPs + \gL \Big)~. 
}
This reproduces the bosonic and the fermionic gauge transformations.

The (2,2) vector superfield can be reduced in a similar fashion. The (2,0) vector multiplet is obtained by the restrictions   
\equ{ 
V = \cV\big|_-~, 
\quad 
A = \frac 12 \, [\bD_-,D_-] \cV\big|_-~, 
\qquad 
\gS = \frac 1{\sqrt 2}\, D_-\cV \big|_-~, 
\quad 
\bgS = \frac 1{\sqrt 2}\, \bD_- \cV \big|_-.
}
Using the fermionic transformations \eqref{FermionicGauge} allows us to gauge some components of $\gS$ away, however, given that $\cT|_- = \bD_+ \gS$ and $\bcT|_- = D_+ \bgS$ given in~\eqref{Superfieldstrengths} define gauge invariant superfield strengths, their components are physical.  The other fermionic parts of $\cT$ and $\bcT$ 
\equ{
F = \frac 1{\sqrt 2} \bD_- \cT\big|_-~, 
\qquad 
\bF = \frac 1{\sqrt 2} D_- \bcT\big|_-~, 
}
define the superfield strengths for the pair $V$ and $A$ defined in~\eqref{(2,0)superfieldstrengths}.

\subsection{(2,0) Reduction of (2,2) actions}
\label{sc:ReductionAction}

In section~\ref{sc:(2,2)superfields} three basic actions for (2,2) superfields have been defined. With the technology presented in the previous subsection, the three actions \eqref{(2,2)Kahler}-\eqref{(2,2)gauge} can be rewritten in the (2,0) language.

The superpotential action becomes  
\equ{
S_\text{super} =  \frac 1{\sqrt 2} \int \d^2\gs\d \gth^+\, W_,{}_a(\gPs) \gL^a + \hc~,
}
where the subscript $a$ on $W_,{}_a$ denotes differentiation w.r.t.\ $\gPs^a$. Similarly for the twisted-superpotential action we find
\equ{
S_\text{twisted} =  \frac 1{\sqrt 2} \int \d^2\gs\d \gth^+\, \gr_I F^I + \hc~,
}
where the index $I$ labels the Abelian vector multiplets. The reduction of the K\"ahler potential action reads 
\equ{ 
S_\text{K\"ahler} = \int \d^2\gs\d^2\gth^+\, 
\Big[ 
\frac i4 \, K_,{}_a \bcD \gPs^a 
- \frac i4 \,K_,{}^b \bcD \bgPs_b 
- \frac 12 \, K_,{}^b{}_a (\bgL - \bgS\cdot q \,\bgPs)_b (\gL - \gS\cdot q\, \gPs)^a
\Big]~.
}
Finally, the gauge action takes the form 
\equ{
S_\text{gauge} = \frac 1{2e^2} \int \d^2\gs\d^2\gth^+\, 
\Big[ \bF F + i\, \bD_+\gS \bder D_+\bgS \Big]~. 
}
%

%
%

\providecommand{\href}[2]{#2}\begingroup\raggedright\endgroup

\end{document}

%% file: Limits.tex
\begin{picture}(0,0)%
\epsfig{file=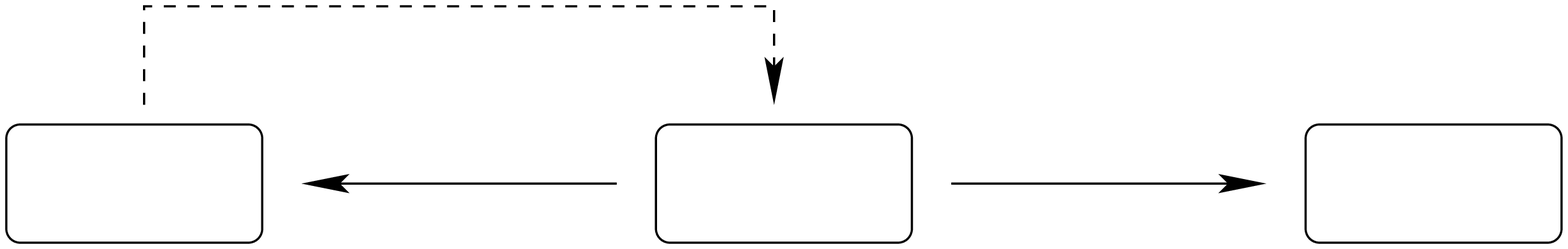}%
\end{picture}%
\setlength{\unitlength}{3947sp}%
\begingroup\makeatletter\ifx\SetFigFont\undefined%
\gdef\SetFigFont#1#2#3#4#5{%
  \reset@font\fontsize{#1}{#2pt}%
  \fontfamily{#3}\fontseries{#4}\fontshape{#5}%
  \selectfont}%
\fi\endgroup%
\begin{picture}(11894,2188)(429,-1433)
\put(2701,-61){\makebox(0,0)[lb]{\smash{{\SetFigFont{20}{24.0}{\rmdefault}{\mddefault}{\updefault}{\color[rgb]{0,0,0}$|p_r, P_r\rangle \rightarrow (q_r, Q_r)$}%
}}}}
\put(3301,-811){\makebox(0,0)[lb]{\smash{{\SetFigFont{20}{24.0}{\rmdefault}{\mddefault}{\updefault}{\color[rgb]{0,0,0}blow down}%
}}}}
\put(7801,-811){\makebox(0,0)[lb]{\smash{{\SetFigFont{20}{24.0}{\rmdefault}{\mddefault}{\updefault}{\color[rgb]{0,0,0}large volume}%
}}}}
\put(8251,-1261){\makebox(0,0)[lb]{\smash{{\SetFigFont{20}{24.0}{\rmdefault}{\mddefault}{\updefault}{\color[rgb]{0,0,0}$b_r \rightarrow \infty$}%
}}}}
\put(3451,-1261){\makebox(0,0)[lb]{\smash{{\SetFigFont{20}{24.0}{\rmdefault}{\mddefault}{\updefault}{\color[rgb]{0,0,0}$\sm\infty \leftarrow b_r$}%
}}}}
\put(2701,539){\makebox(0,0)[lb]{\smash{{\SetFigFont{20}{24.0}{\rmdefault}{\mddefault}{\updefault}{\color[rgb]{0,0,0}charge assignment}%
}}}}
\put(10501,-811){\makebox(0,0)[lb]{\smash{{\SetFigFont{20}{24.0}{\rmdefault}{\mddefault}{\updefault}{\color[rgb]{0,0,0}Line Bundle}%
}}}}
\put(10951,-1261){\makebox(0,0)[lb]{\smash{{\SetFigFont{20}{24.0}{\rmdefault}{\mddefault}{\updefault}{\color[rgb]{0,0,0}on CY}%
}}}}
\put(5851,-811){\makebox(0,0)[lb]{\smash{{\SetFigFont{20}{24.0}{\rmdefault}{\mddefault}{\updefault}{\color[rgb]{0,0,0}GLSM}%
}}}}
\put(1201,-1261){\makebox(0,0)[lb]{\smash{{\SetFigFont{20}{24.0}{\rmdefault}{\mddefault}{\updefault}{\color[rgb]{0,0,0}CFT}%
}}}}
\put(901,-811){\makebox(0,0)[lb]{\smash{{\SetFigFont{20}{24.0}{\rmdefault}{\mddefault}{\updefault}{\color[rgb]{0,0,0}Orbifold}%
}}}}
\end{picture}%

%% file: triangZ22.tex
\begin{picture}(0,0)%
\epsfig{file=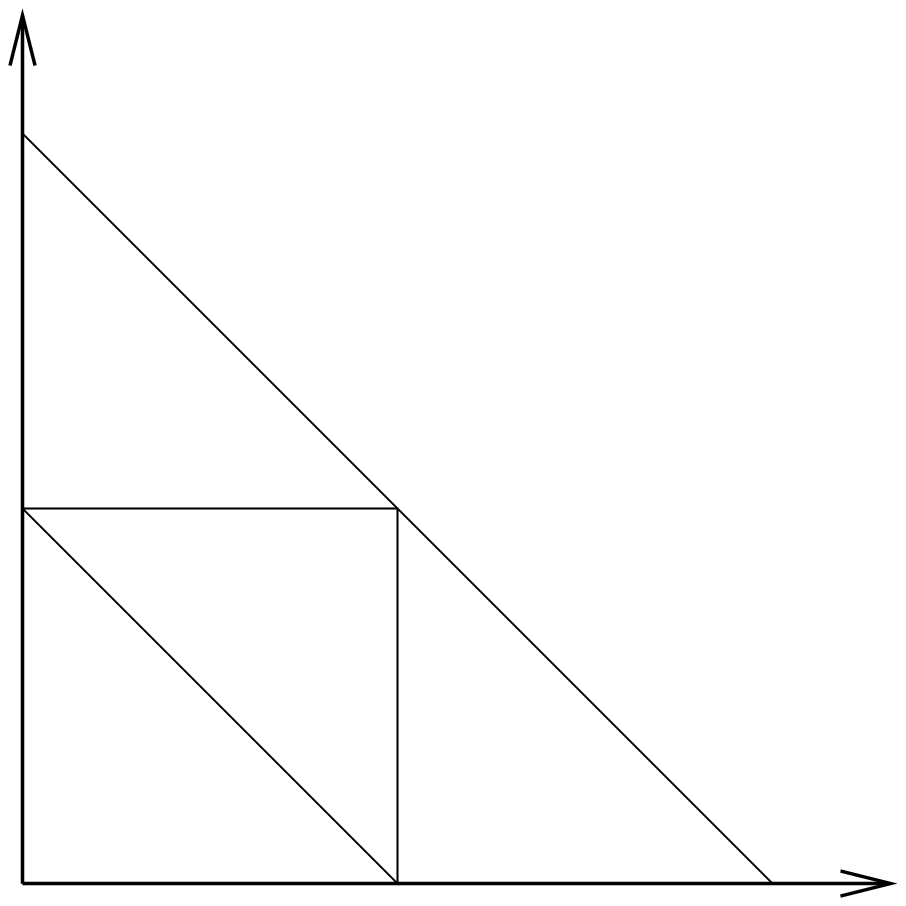}%
\end{picture}%
\setlength{\unitlength}{3947sp}%
\begingroup\makeatletter\ifx\SetFigFont\undefined%
\gdef\SetFigFont#1#2#3#4#5{%
  \reset@font\fontsize{#1}{#2pt}%
  \fontfamily{#3}\fontseries{#4}\fontshape{#5}%
  \selectfont}%
\fi\endgroup%
\begin{picture}(4522,4598)(2101,-6737)
\put(3301,-5161){\makebox(0,0)[lb]{\smash{{\SetFigFont{20}{24.0}{\rmdefault}{\mddefault}{\updefault}{\color[rgb]{0,0,0}``$S$"}%
}}}}
\put(2551,-2461){\makebox(0,0)[lb]{\smash{{\SetFigFont{20}{24.0}{\rmdefault}{\mddefault}{\updefault}{\color[rgb]{0,0,0}$y$}%
}}}}
\put(6151,-6061){\makebox(0,0)[lb]{\smash{{\SetFigFont{20}{24.0}{\rmdefault}{\mddefault}{\updefault}{\color[rgb]{0,0,0}$x$}%
}}}}
\put(2101,-2761){\makebox(0,0)[lb]{\smash{{\SetFigFont{20}{24.0}{\rmdefault}{\mddefault}{\updefault}{\color[rgb]{0,0,0}$1$}%
}}}}
\put(2101,-6661){\makebox(0,0)[lb]{\smash{{\SetFigFont{20}{24.0}{\rmdefault}{\mddefault}{\updefault}{\color[rgb]{0,0,0}$0$}%
}}}}
\put(5851,-6661){\makebox(0,0)[lb]{\smash{{\SetFigFont{20}{24.0}{\rmdefault}{\mddefault}{\updefault}{\color[rgb]{0,0,0}$1$}%
}}}}
\put(2701,-4111){\makebox(0,0)[lb]{\smash{{\SetFigFont{20}{24.0}{\rmdefault}{\mddefault}{\updefault}{\color[rgb]{0,0,0}``$E_2$"}%
}}}}
\put(2701,-5911){\makebox(0,0)[lb]{\smash{{\SetFigFont{20}{24.0}{\rmdefault}{\mddefault}{\updefault}{\color[rgb]{0,0,0}``$E_3$"}%
}}}}
\put(4501,-5911){\makebox(0,0)[lb]{\smash{{\SetFigFont{20}{24.0}{\rmdefault}{\mddefault}{\updefault}{\color[rgb]{0,0,0}``$E_1$"}%
}}}}
\end{picture}%

%% file: Z22orbi.tex
\begin{picture}(0,0)%
\includegraphics{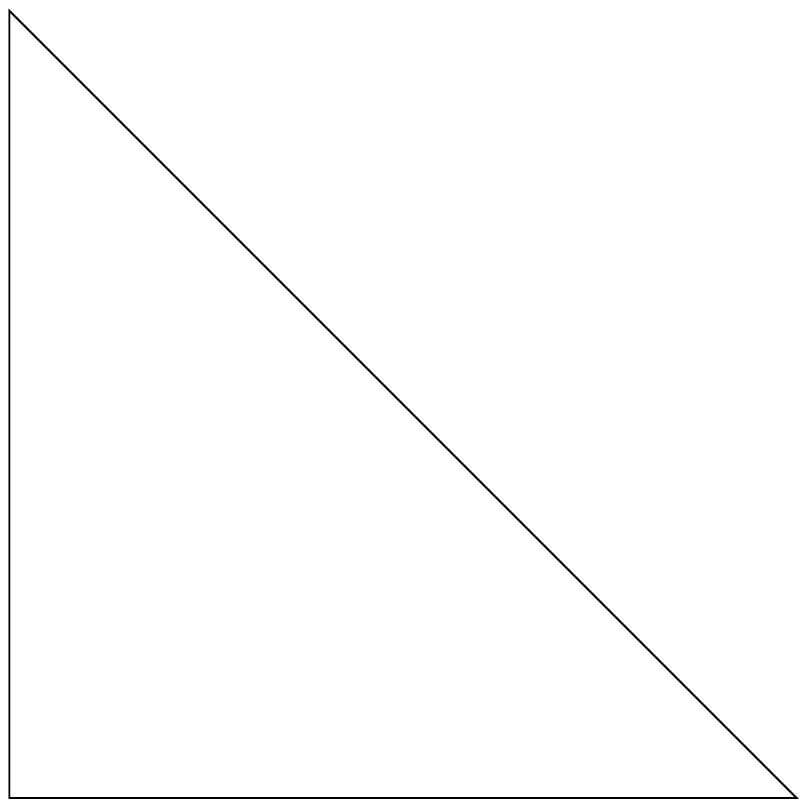}%
\end{picture}%
\setlength{\unitlength}{4144sp}%
\begingroup\makeatletter\ifx\SetFigFont\undefined%
\gdef\SetFigFont#1#2#3#4#5{%
  \reset@font\fontsize{#1}{#2pt}%
  \fontfamily{#3}\fontseries{#4}\fontshape{#5}%
  \selectfont}%
\fi\endgroup%
\begin{picture}(3852,4690)(2461,-7442)
\put(2926,-3211){\makebox(0,0)[lb]{\smash{{\SetFigFont{34}{40.8}{\rmdefault}{\mddefault}{\updefault}{\color[rgb]{0,0,0}$D_3$}%
}}}}
\put(6076,-7261){\makebox(0,0)[lb]{\smash{{\SetFigFont{34}{40.8}{\rmdefault}{\mddefault}{\updefault}{\color[rgb]{0,0,0}$D_2$}%
}}}}
\put(2476,-7261){\makebox(0,0)[lb]{\smash{{\SetFigFont{34}{40.8}{\rmdefault}{\mddefault}{\updefault}{\color[rgb]{0,0,0}$D_1$}%
}}}}
\end{picture}%

%% file: Z22single.tex
\begin{picture}(0,0)%
\includegraphics{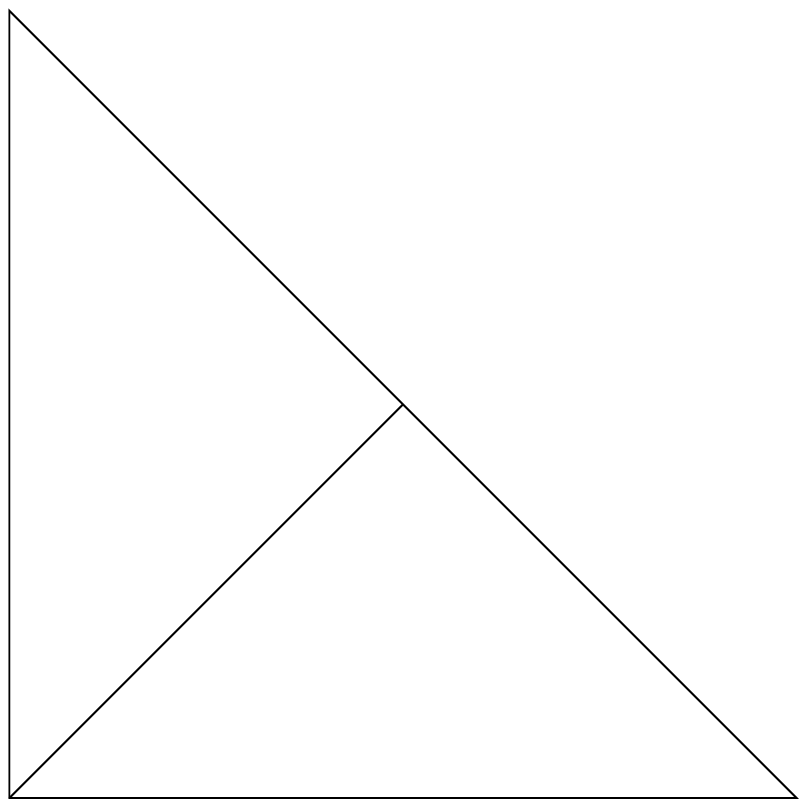}%
\end{picture}%
\setlength{\unitlength}{4144sp}%
\begingroup\makeatletter\ifx\SetFigFont\undefined%
\gdef\SetFigFont#1#2#3#4#5{%
  \reset@font\fontsize{#1}{#2pt}%
  \fontfamily{#3}\fontseries{#4}\fontshape{#5}%
  \selectfont}%
\fi\endgroup%
\begin{picture}(3852,4690)(2461,-7442)
\put(4726,-4786){\makebox(0,0)[lb]{\smash{{\SetFigFont{34}{40.8}{\rmdefault}{\mddefault}{\updefault}{\color[rgb]{0,0,0}$E_1$}%
}}}}
\put(2926,-3211){\makebox(0,0)[lb]{\smash{{\SetFigFont{34}{40.8}{\rmdefault}{\mddefault}{\updefault}{\color[rgb]{0,0,0}$D_3$}%
}}}}
\put(6076,-7261){\makebox(0,0)[lb]{\smash{{\SetFigFont{34}{40.8}{\rmdefault}{\mddefault}{\updefault}{\color[rgb]{0,0,0}$D_2$}%
}}}}
\put(2476,-7261){\makebox(0,0)[lb]{\smash{{\SetFigFont{34}{40.8}{\rmdefault}{\mddefault}{\updefault}{\color[rgb]{0,0,0}$D_1$}%
}}}}
\end{picture}%

%% file: Z22double12.tex
\begin{picture}(0,0)%
\epsfig{file=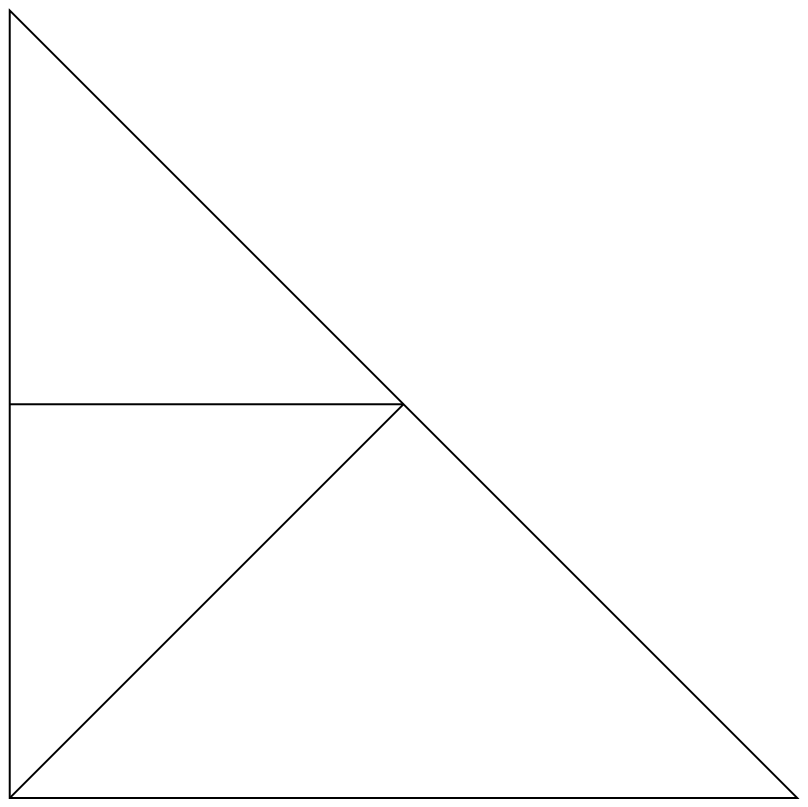}%
\end{picture}%
\setlength{\unitlength}{4144sp}%
\begingroup\makeatletter\ifx\SetFigFont\undefined%
\gdef\SetFigFont#1#2#3#4#5{%
  \reset@font\fontsize{#1}{#2pt}%
  \fontfamily{#3}\fontseries{#4}\fontshape{#5}%
  \selectfont}%
\fi\endgroup%
\begin{picture}(5681,4611)(2071,-7414)
\put(2071,-4741){\makebox(0,0)[lb]{\smash{{\SetFigFont{34}{40.8}{\rmdefault}{\mddefault}{\updefault}{\color[rgb]{0,0,0}$E_2$}%
}}}}
\put(4726,-4786){\makebox(0,0)[lb]{\smash{{\SetFigFont{34}{40.8}{\rmdefault}{\mddefault}{\updefault}{\color[rgb]{0,0,0}$E_1$}%
}}}}
\put(2926,-3211){\makebox(0,0)[lb]{\smash{{\SetFigFont{34}{40.8}{\rmdefault}{\mddefault}{\updefault}{\color[rgb]{0,0,0}$D_3$}%
}}}}
\put(6076,-7261){\makebox(0,0)[lb]{\smash{{\SetFigFont{34}{40.8}{\rmdefault}{\mddefault}{\updefault}{\color[rgb]{0,0,0}$D_2$}%
}}}}
\put(2476,-7261){\makebox(0,0)[lb]{\smash{{\SetFigFont{34}{40.8}{\rmdefault}{\mddefault}{\updefault}{\color[rgb]{0,0,0}$D_1$}%
}}}}
\end{picture}%

%% file: Z22double21.tex
\begin{picture}(0,0)%
\epsfig{file=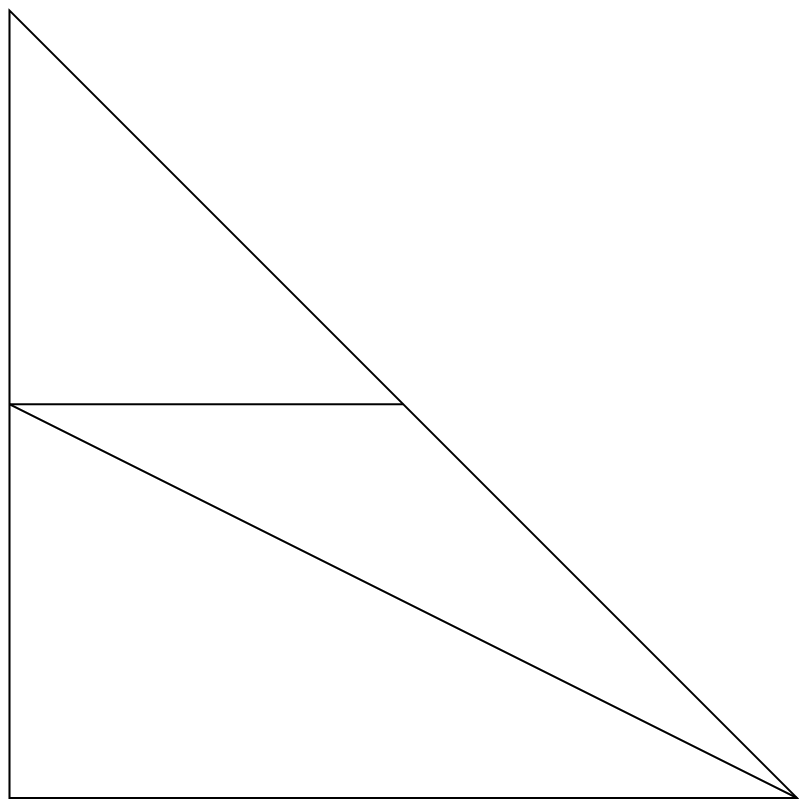}%
\end{picture}%
\setlength{\unitlength}{4144sp}%
\begingroup\makeatletter\ifx\SetFigFont\undefined%
\gdef\SetFigFont#1#2#3#4#5{%
  \reset@font\fontsize{#1}{#2pt}%
  \fontfamily{#3}\fontseries{#4}\fontshape{#5}%
  \selectfont}%
\fi\endgroup%
\begin{picture}(5771,4611)(1981,-7414)
\put(1981,-4741){\makebox(0,0)[lb]{\smash{{\SetFigFont{34}{40.8}{\rmdefault}{\mddefault}{\updefault}{\color[rgb]{0,0,0}$E_2$}%
}}}}
\put(4726,-4786){\makebox(0,0)[lb]{\smash{{\SetFigFont{34}{40.8}{\rmdefault}{\mddefault}{\updefault}{\color[rgb]{0,0,0}$E_1$}%
}}}}
\put(2926,-3211){\makebox(0,0)[lb]{\smash{{\SetFigFont{34}{40.8}{\rmdefault}{\mddefault}{\updefault}{\color[rgb]{0,0,0}$D_3$}%
}}}}
\put(6076,-7261){\makebox(0,0)[lb]{\smash{{\SetFigFont{34}{40.8}{\rmdefault}{\mddefault}{\updefault}{\color[rgb]{0,0,0}$D_2$}%
}}}}
\put(2476,-7261){\makebox(0,0)[lb]{\smash{{\SetFigFont{34}{40.8}{\rmdefault}{\mddefault}{\updefault}{\color[rgb]{0,0,0}$D_1$}%
}}}}
\end{picture}%

%% file: Z22fullE3.tex
\begin{picture}(0,0)%
\epsfig{file=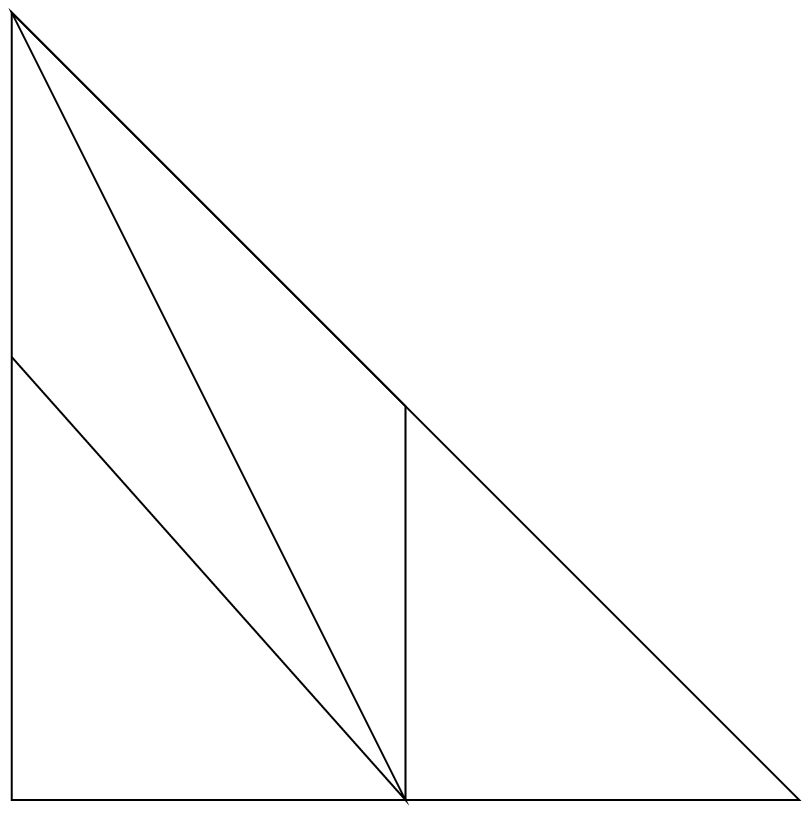}%
\end{picture}%
\setlength{\unitlength}{4144sp}%
\begingroup\makeatletter\ifx\SetFigFont\undefined%
\gdef\SetFigFont#1#2#3#4#5{%
  \reset@font\fontsize{#1}{#2pt}%
  \fontfamily{#3}\fontseries{#4}\fontshape{#5}%
  \selectfont}%
\fi\endgroup%
\begin{picture}(5681,4611)(2071,-7414)
\put(2071,-4741){\makebox(0,0)[lb]{\smash{{\SetFigFont{34}{40.8}{\rmdefault}{\mddefault}{\updefault}{\color[rgb]{0,0,0}$E_2$}%
}}}}
\put(4276,-7261){\makebox(0,0)[lb]{\smash{{\SetFigFont{34}{40.8}{\rmdefault}{\mddefault}{\updefault}{\color[rgb]{0,0,0}$E_3$}%
}}}}
\put(4726,-4786){\makebox(0,0)[lb]{\smash{{\SetFigFont{34}{40.8}{\rmdefault}{\mddefault}{\updefault}{\color[rgb]{0,0,0}$E_1$}%
}}}}
\put(2926,-3211){\makebox(0,0)[lb]{\smash{{\SetFigFont{34}{40.8}{\rmdefault}{\mddefault}{\updefault}{\color[rgb]{0,0,0}$D_3$}%
}}}}
\put(6076,-7261){\makebox(0,0)[lb]{\smash{{\SetFigFont{34}{40.8}{\rmdefault}{\mddefault}{\updefault}{\color[rgb]{0,0,0}$D_2$}%
}}}}
\put(2476,-7261){\makebox(0,0)[lb]{\smash{{\SetFigFont{34}{40.8}{\rmdefault}{\mddefault}{\updefault}{\color[rgb]{0,0,0}$D_1$}%
}}}}
\end{picture}%

%% file: Z22fullE2.tex
\begin{picture}(0,0)%
\epsfig{file=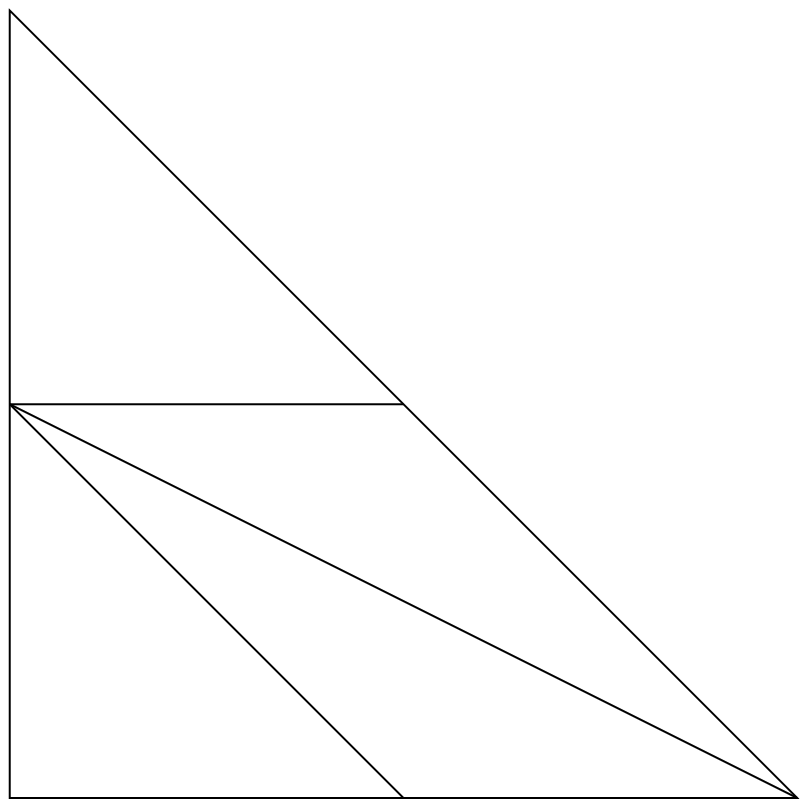}%
\end{picture}%
\setlength{\unitlength}{4144sp}%
\begingroup\makeatletter\ifx\SetFigFont\undefined%
\gdef\SetFigFont#1#2#3#4#5{%
  \reset@font\fontsize{#1}{#2pt}%
  \fontfamily{#3}\fontseries{#4}\fontshape{#5}%
  \selectfont}%
\fi\endgroup%
\begin{picture}(5681,4611)(2071,-7414)
\put(2071,-4741){\makebox(0,0)[lb]{\smash{{\SetFigFont{34}{40.8}{\rmdefault}{\mddefault}{\updefault}{\color[rgb]{0,0,0}$E_2$}%
}}}}
\put(4276,-7261){\makebox(0,0)[lb]{\smash{{\SetFigFont{34}{40.8}{\rmdefault}{\mddefault}{\updefault}{\color[rgb]{0,0,0}$E_3$}%
}}}}
\put(4726,-4786){\makebox(0,0)[lb]{\smash{{\SetFigFont{34}{40.8}{\rmdefault}{\mddefault}{\updefault}{\color[rgb]{0,0,0}$E_1$}%
}}}}
\put(2926,-3211){\makebox(0,0)[lb]{\smash{{\SetFigFont{34}{40.8}{\rmdefault}{\mddefault}{\updefault}{\color[rgb]{0,0,0}$D_3$}%
}}}}
\put(6076,-7261){\makebox(0,0)[lb]{\smash{{\SetFigFont{34}{40.8}{\rmdefault}{\mddefault}{\updefault}{\color[rgb]{0,0,0}$D_2$}%
}}}}
\put(2476,-7261){\makebox(0,0)[lb]{\smash{{\SetFigFont{34}{40.8}{\rmdefault}{\mddefault}{\updefault}{\color[rgb]{0,0,0}$D_1$}%
}}}}
\end{picture}%

%% file: Z22fullE1.tex
\begin{picture}(0,0)%
\epsfig{file=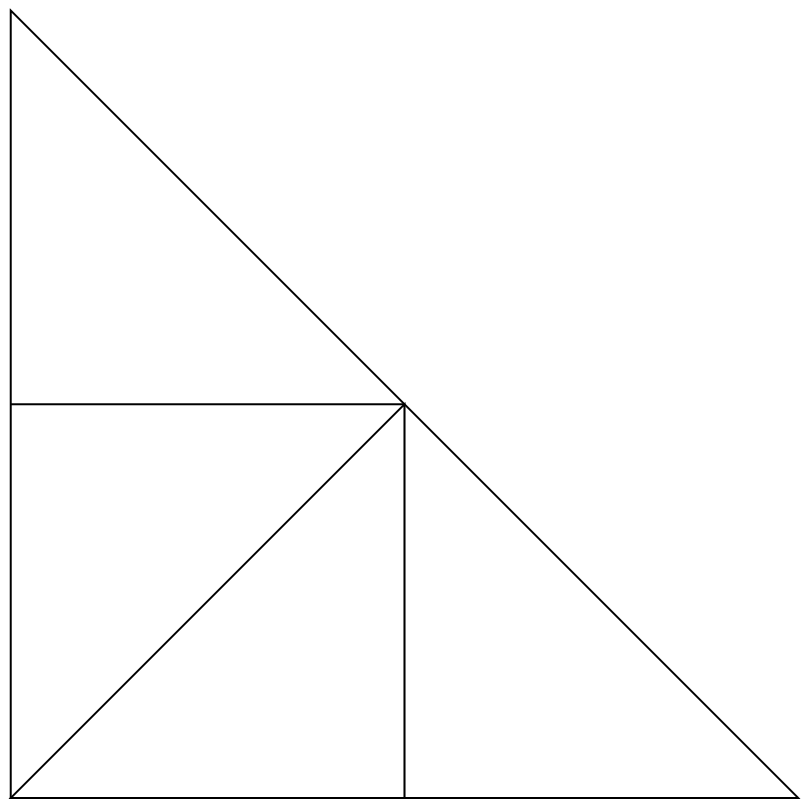}%
\end{picture}%
\setlength{\unitlength}{4144sp}%
\begingroup\makeatletter\ifx\SetFigFont\undefined%
\gdef\SetFigFont#1#2#3#4#5{%
  \reset@font\fontsize{#1}{#2pt}%
  \fontfamily{#3}\fontseries{#4}\fontshape{#5}%
  \selectfont}%
\fi\endgroup%
\begin{picture}(5681,4611)(2071,-7414)
\put(2071,-4831){\makebox(0,0)[lb]{\smash{{\SetFigFont{34}{40.8}{\rmdefault}{\mddefault}{\updefault}{\color[rgb]{0,0,0}$E_2$}%
}}}}
\put(4276,-7261){\makebox(0,0)[lb]{\smash{{\SetFigFont{34}{40.8}{\rmdefault}{\mddefault}{\updefault}{\color[rgb]{0,0,0}$E_3$}%
}}}}
\put(4726,-4786){\makebox(0,0)[lb]{\smash{{\SetFigFont{34}{40.8}{\rmdefault}{\mddefault}{\updefault}{\color[rgb]{0,0,0}$E_1$}%
}}}}
\put(2926,-3211){\makebox(0,0)[lb]{\smash{{\SetFigFont{34}{40.8}{\rmdefault}{\mddefault}{\updefault}{\color[rgb]{0,0,0}$D_3$}%
}}}}
\put(6076,-7261){\makebox(0,0)[lb]{\smash{{\SetFigFont{34}{40.8}{\rmdefault}{\mddefault}{\updefault}{\color[rgb]{0,0,0}$D_2$}%
}}}}
\put(2476,-7261){\makebox(0,0)[lb]{\smash{{\SetFigFont{34}{40.8}{\rmdefault}{\mddefault}{\updefault}{\color[rgb]{0,0,0}$D_1$}%
}}}}
\end{picture}%

%% file: Z22fullS.tex
\begin{picture}(0,0)%
\epsfig{file=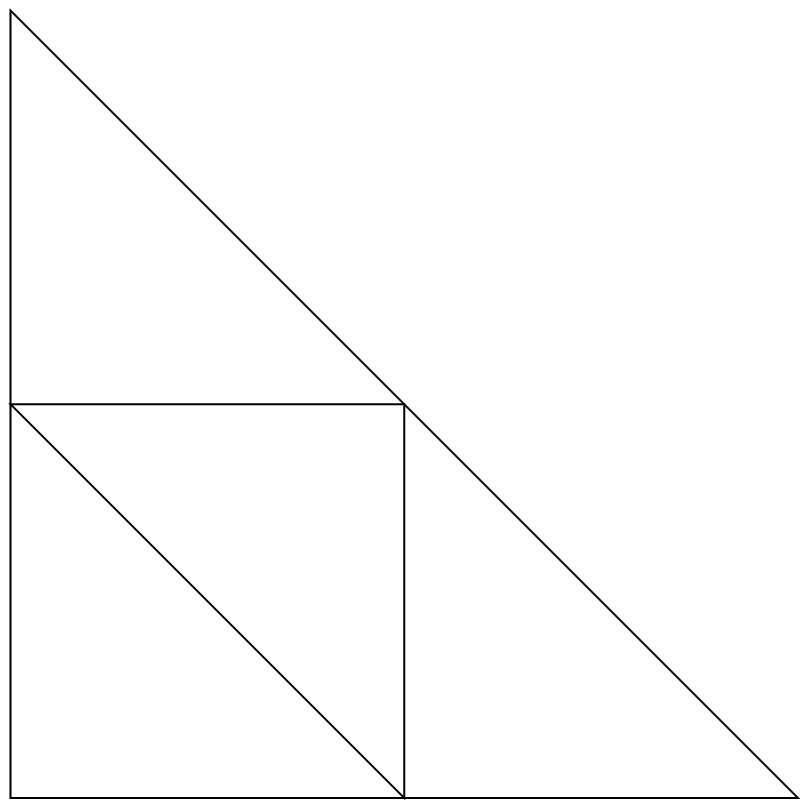}%
\end{picture}%
\setlength{\unitlength}{4144sp}%
\begingroup\makeatletter\ifx\SetFigFont\undefined%
\gdef\SetFigFont#1#2#3#4#5{%
  \reset@font\fontsize{#1}{#2pt}%
  \fontfamily{#3}\fontseries{#4}\fontshape{#5}%
  \selectfont}%
\fi\endgroup%
\begin{picture}(5771,4611)(1981,-7414)
\put(1981,-4741){\makebox(0,0)[lb]{\smash{{\SetFigFont{34}{40.8}{\rmdefault}{\mddefault}{\updefault}{\color[rgb]{0,0,0}$E_2$}%
}}}}
\put(4276,-7261){\makebox(0,0)[lb]{\smash{{\SetFigFont{34}{40.8}{\rmdefault}{\mddefault}{\updefault}{\color[rgb]{0,0,0}$E_3$}%
}}}}
\put(2476,-7261){\makebox(0,0)[lb]{\smash{{\SetFigFont{34}{40.8}{\familydefault}{\mddefault}{\updefault}{\color[rgb]{0,0,0}$D_1$}%
}}}}
\put(2926,-3211){\makebox(0,0)[lb]{\smash{{\SetFigFont{34}{40.8}{\rmdefault}{\mddefault}{\updefault}{\color[rgb]{0,0,0}$D_3$}%
}}}}
\put(4726,-4786){\makebox(0,0)[lb]{\smash{{\SetFigFont{34}{40.8}{\rmdefault}{\mddefault}{\updefault}{\color[rgb]{0,0,0}$E_1$}%
}}}}
\put(6076,-7261){\makebox(0,0)[lb]{\smash{{\SetFigFont{34}{40.8}{\rmdefault}{\mddefault}{\updefault}{\color[rgb]{0,0,0}$D_2$}%
}}}}
\end{picture}%